\documentclass{egpubl}
\usepackage{egsr2025}

\JournalSubmission    
\usepackage[T1]{fontenc}
\usepackage{dfadobe}  
\usepackage{algorithm}
\usepackage{amsmath} 
\usepackage{algpseudocode}

\usepackage{cite}  
\BibtexOrBiblatex
\electronicVersion
\PrintedOrElectronic
\ifpdf \usepackage[pdftex]{graphicx} \pdfcompresslevel=9
\else \usepackage[dvips]{graphicx} \fi

\usepackage{egweblnk} 

\usepackage{amsmath,amsfonts}
\usepackage{algorithm}
\usepackage{array}
\usepackage{textcomp}
\usepackage[table,dvipsnames]{xcolor}
\usepackage{stfloats}
\usepackage{url}
\usepackage{verbatim}
\usepackage{graphicx}
\usepackage{cite}
\usepackage{listings}
\usepackage{pgfplots}
\usepackage{comment}
\usepackage{tikz}
\usepackage{bm} 
\usepackage{subcaption} 
\usepackage{amsmath}
\usepackage{caption} 
\usepackage{overpic}
\usepackage{microtype}
\usepackage{placeins}
\usepackage{xr}
\usepackage{pgfplotstable}
\pgfplotsset{compat}
\usepackage{wrapfig}
\usepackage[utf8]{inputenc}

\usepackage{adjustbox}
\usepackage{multirow}
\usepackage{booktabs}
\usepackage{ulem}
\usepackage{xspace} 

\usepackage{color}          
\definecolor{gray}{rgb}{0.8,0.8,0.8}
\definecolor{lightgray}{rgb}{0.9,0.9,0.9}
\definecolor{cyan}{cmyk}{1,0,0,0}
\definecolor{blue}{rgb}{0,0,0.7}
\definecolor{darkblue}{rgb}{0,0,0.4}
\definecolor{darkred}{rgb}{0.4,0.0,0}
\definecolor{red}{rgb}{1.0,0.0,0}
\definecolor{darkgreen}{rgb}{0,0.6,0}
\definecolor{darkergreen}{rgb}{0,0.3,0}
\definecolor{orange}{rgb}{1,0.5,0}
\definecolor{magenta}{cmyk}{0,1,0,0}
\definecolor{darkyellow}{cmyk}{0,0,0.5,0}
\definecolor{purple}{cmyk}{0.5,1,0,0}

\newcommand{\FLIP}{\protect\reflectbox{F}LIP\xspace}

\title[]%
      {Procedural Multiscale Geometry Modeling using Implicit Functions}

\author[Bojja Venu, Adam Bosak, and Juan Raul Padron Griffe]
{\parbox{\textwidth}{\centering Bojja Venu$^{1}$\orcid{}, Adam Bosak$^{1}$\orcid{}
        and Juan Raul Padron-Griffe$^{2}$\orcid{} 
        }
        \\
{\parbox{\textwidth}{\centering $^1$Technical University of Denmark, Denmark \\
         $^2$University of Zaragoza, Spain
       }
}
}

\begin{document}

\teaser{  
    \centering
\begin{tikzpicture}
    \node[anchor=north west] (main) at (-3,8.5) {\includegraphics[width=0.3\linewidth]{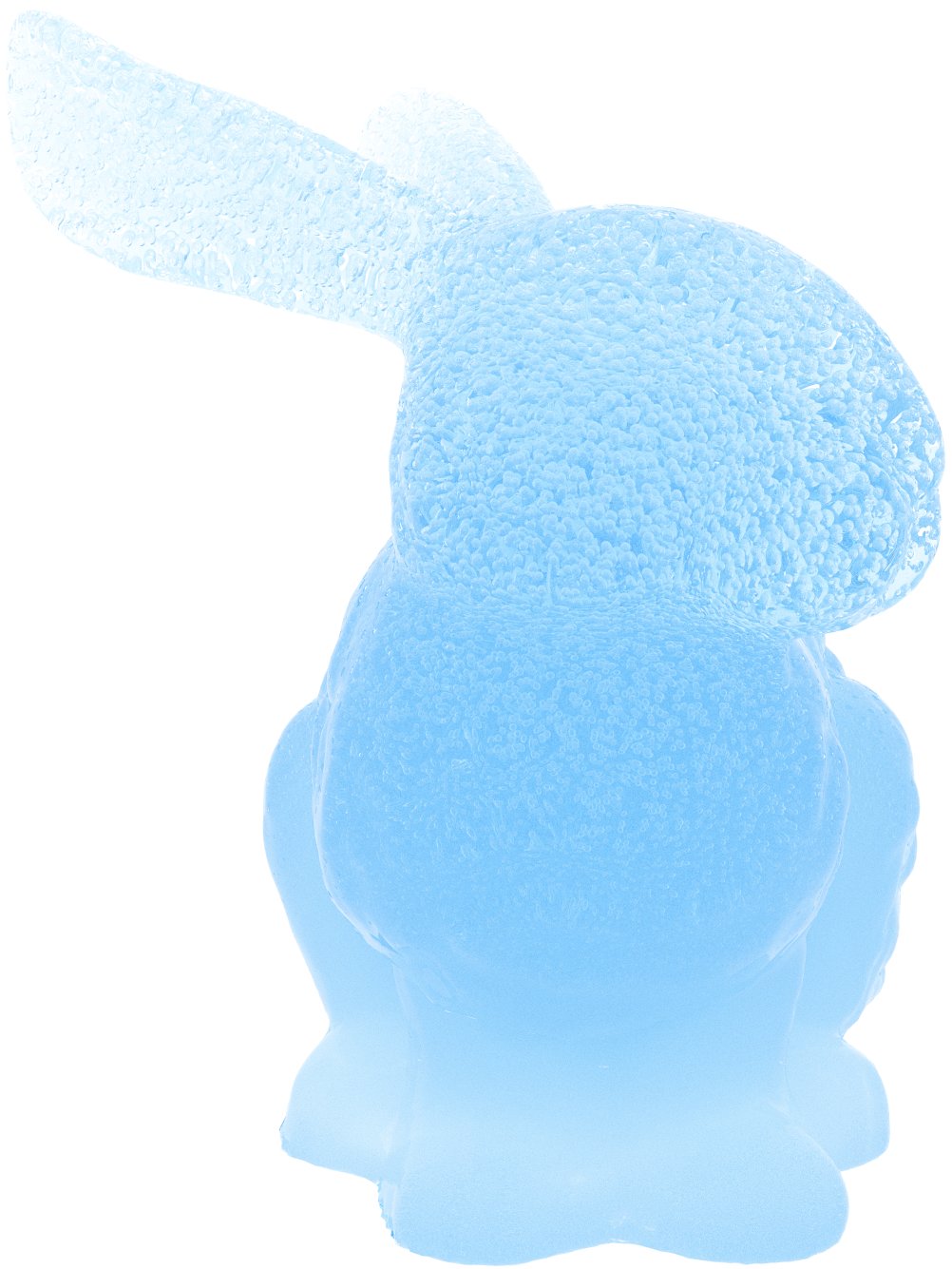}};
    
    \draw[thick, ->] ([xshift=-5pt,yshift=5pt]main.north west) 
        -- ([xshift=-5pt,yshift=-5pt]main.south west);

    \node[right=5pt] at ([xshift=-5pt, yshift=-02.5cm]main.north west) {Macroscale};
    \node[right=5pt] at ([xshift=-5pt, yshift=-4.4cm]main.north west) {Mesoscale};
    \node[right=5pt] at ([xshift=-5pt, yshift=1.2cm]main.south west) {Microscale};

    \node[draw, thick] (inset1) at (3.6,5.75) {\includegraphics[width=0.12\linewidth]{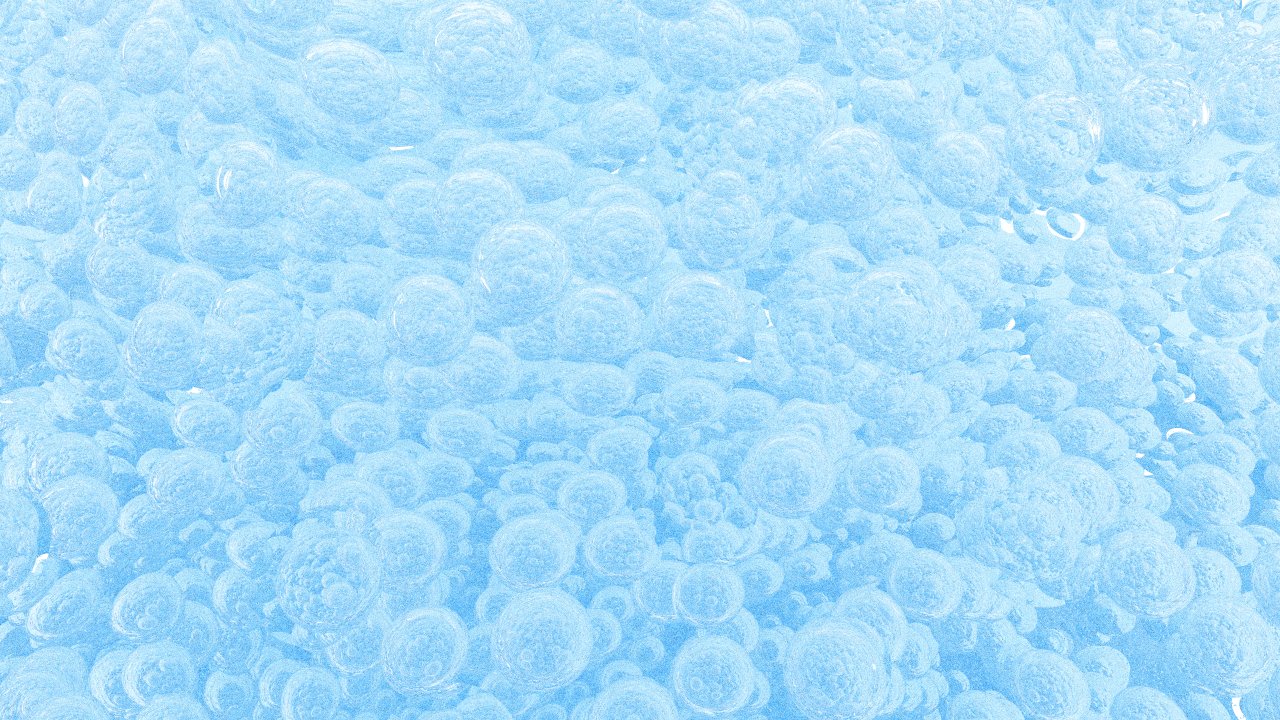}};
    \node[draw, thick] (inset2) at (3.6,4.0) {\includegraphics[width=0.12\linewidth]{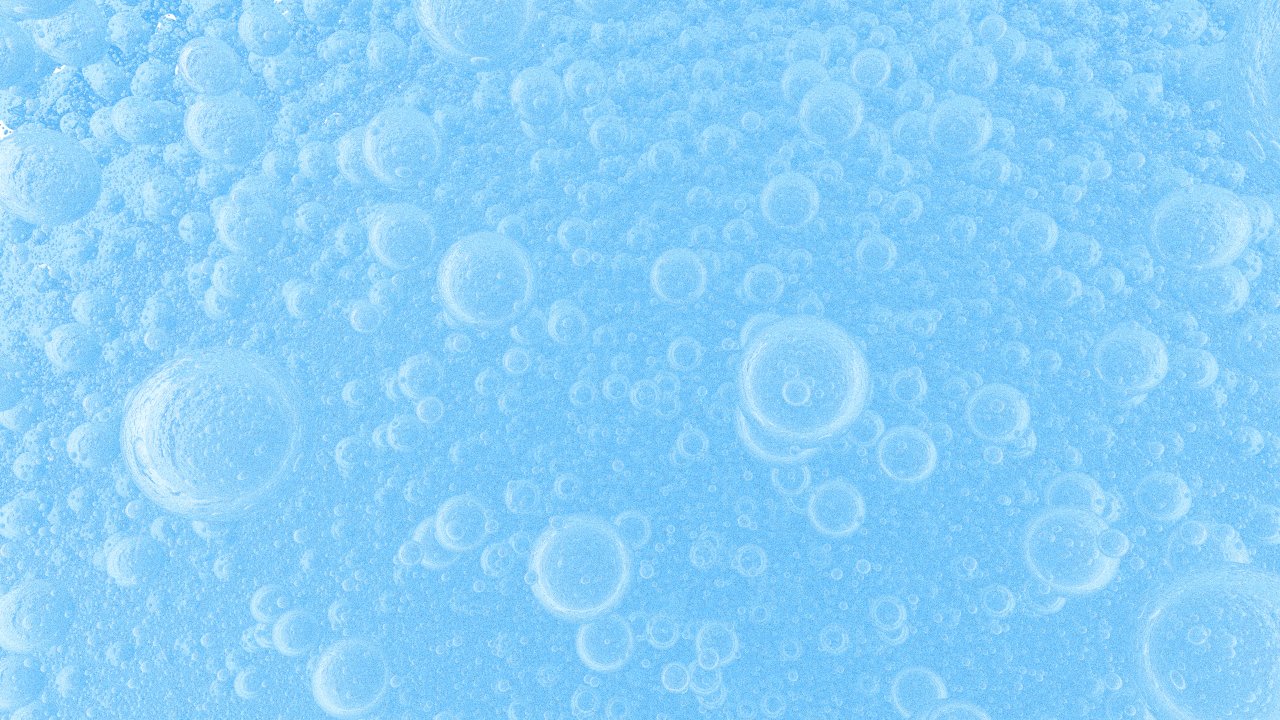}};
    \node[draw, thick] (inset3) at (3.6,2.25) {\includegraphics[width=0.12\linewidth]{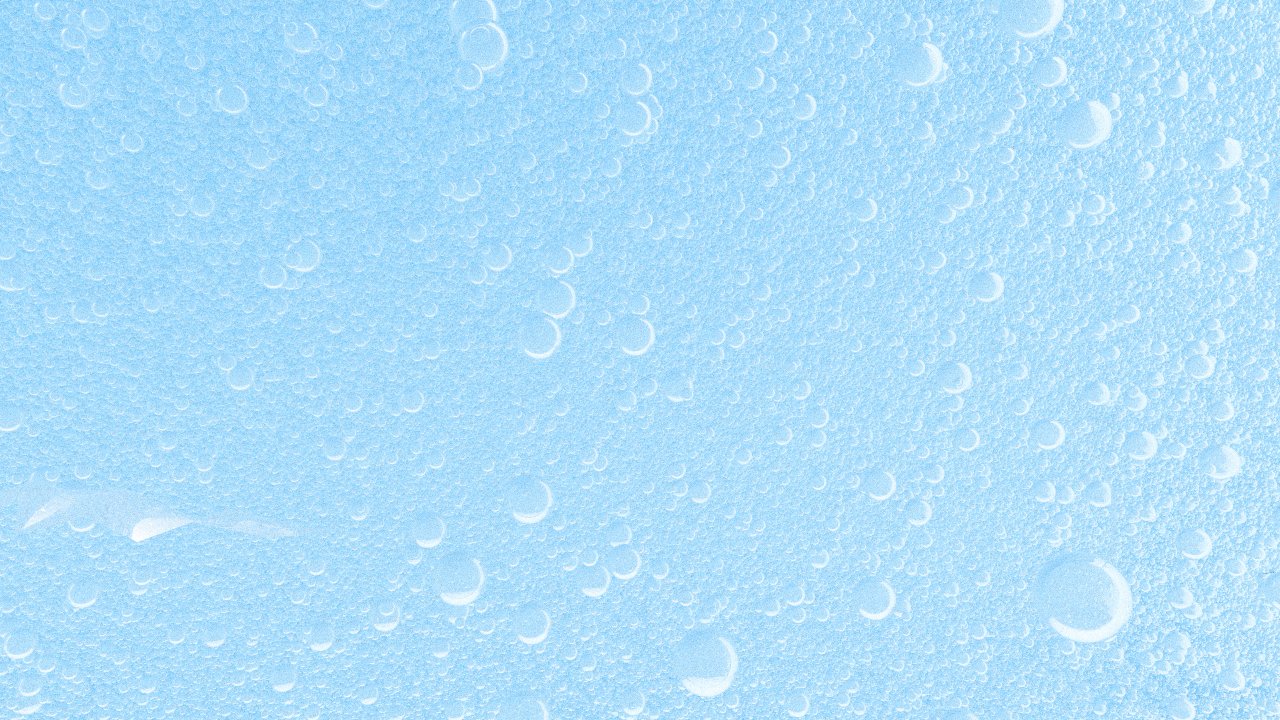}};
    \node[draw, thick] (inset4) at (6.68,6.75) {\includegraphics[width=0.18\linewidth]{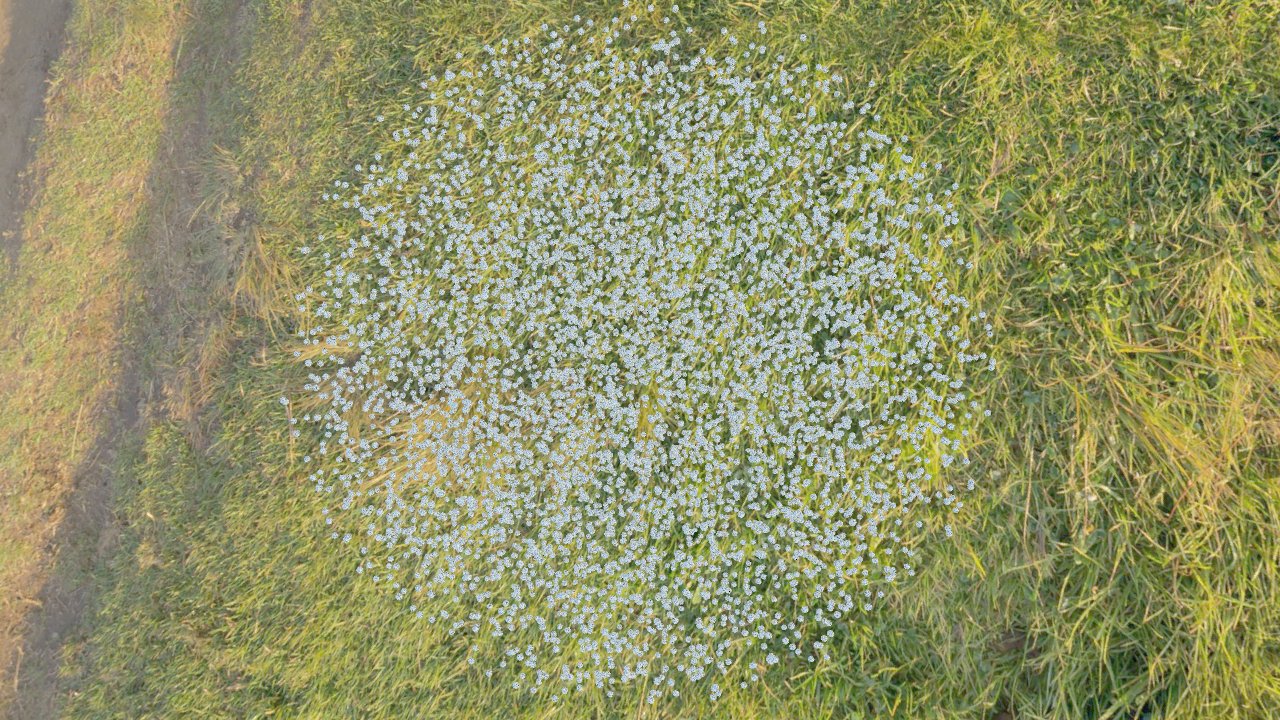}};
    \node[draw, thick] (inset5) at (6.68,4.75) {\includegraphics[width=0.18\linewidth]{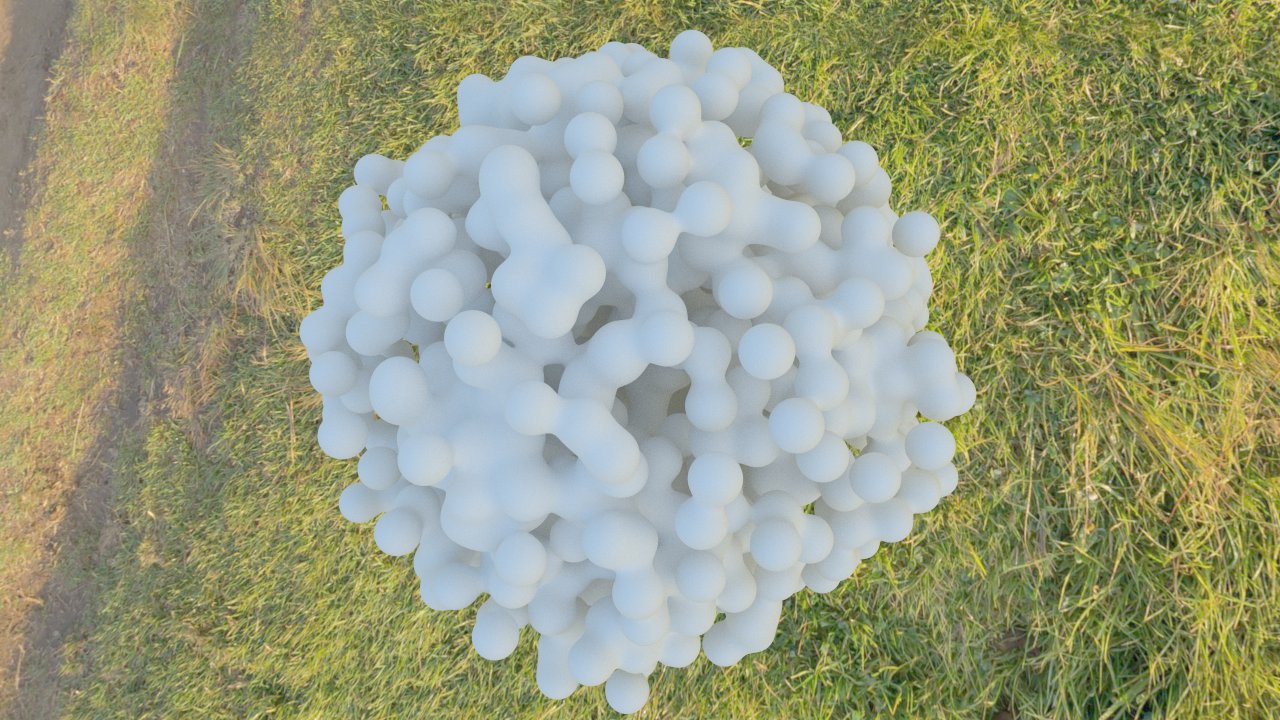}};
    \node[draw, thick] (inset6) at (6.68,2.75) {\includegraphics[width=0.18\linewidth]{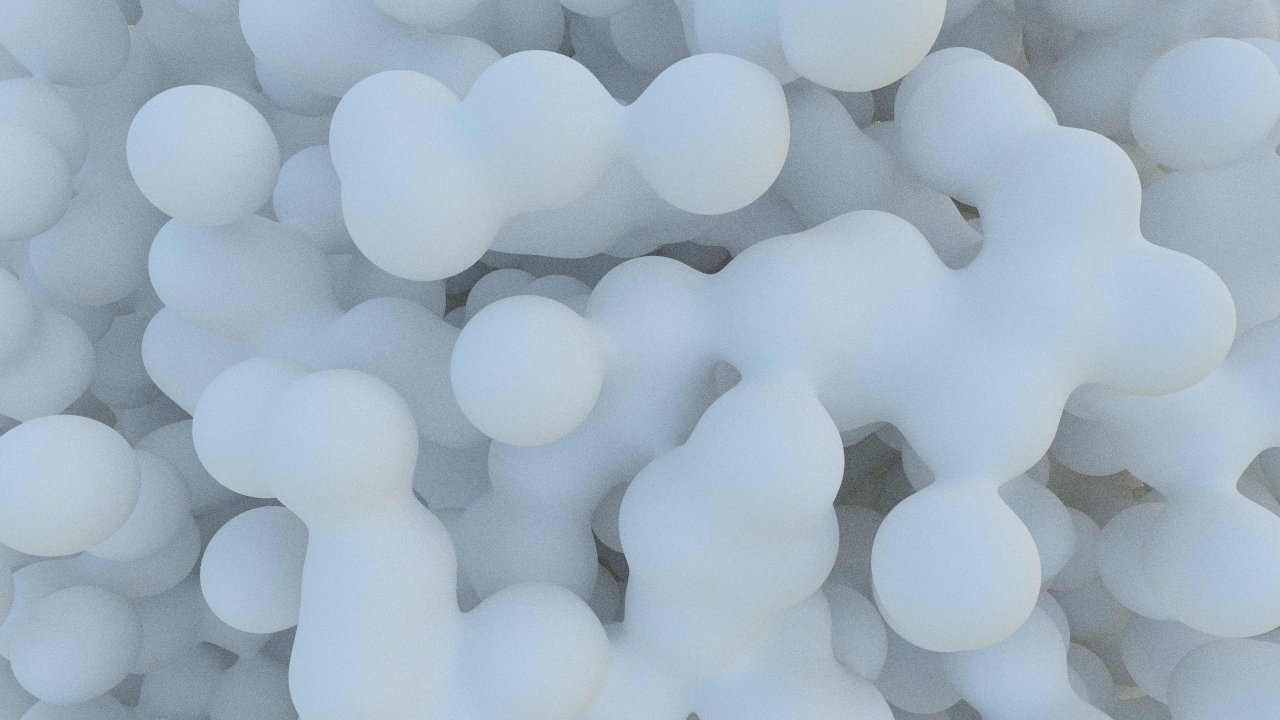}};
    \node[draw, thick] (inset7) at (11.5,6.25) {\includegraphics[width=0.315\linewidth]{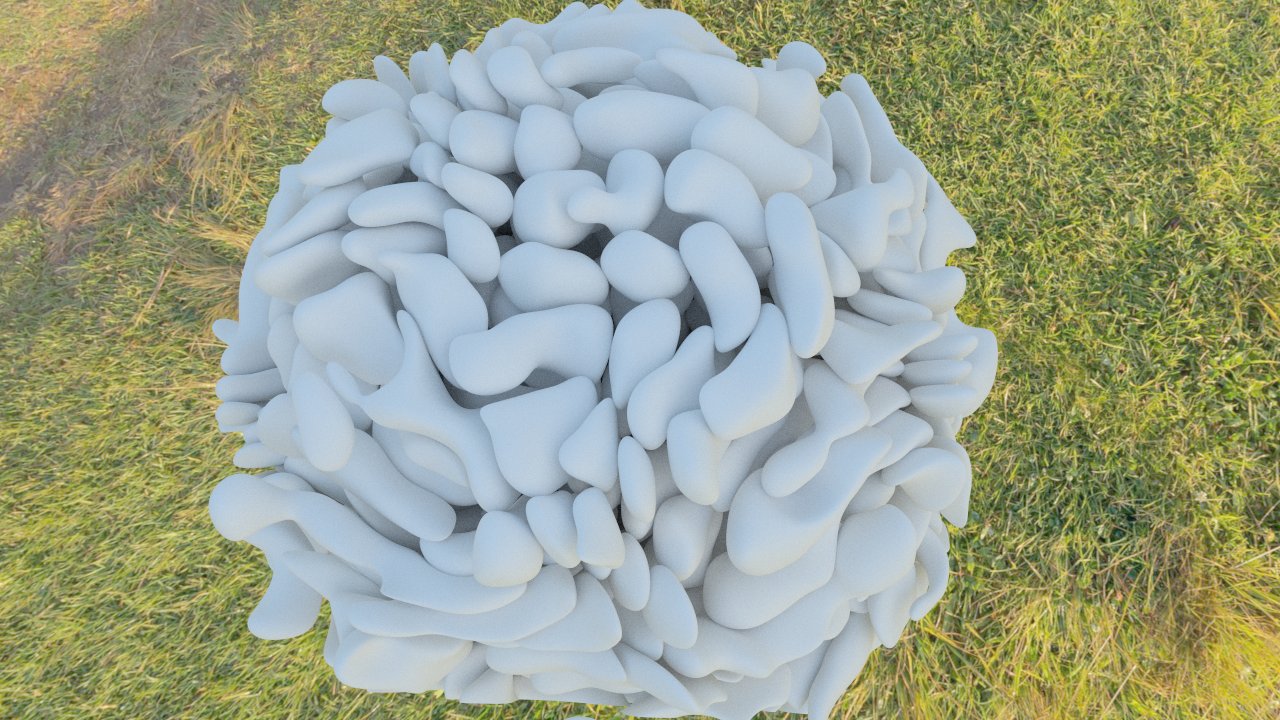}};
    \node[draw, thick] (inset8) at (11.5,2.75) {\includegraphics[width=0.315\linewidth]{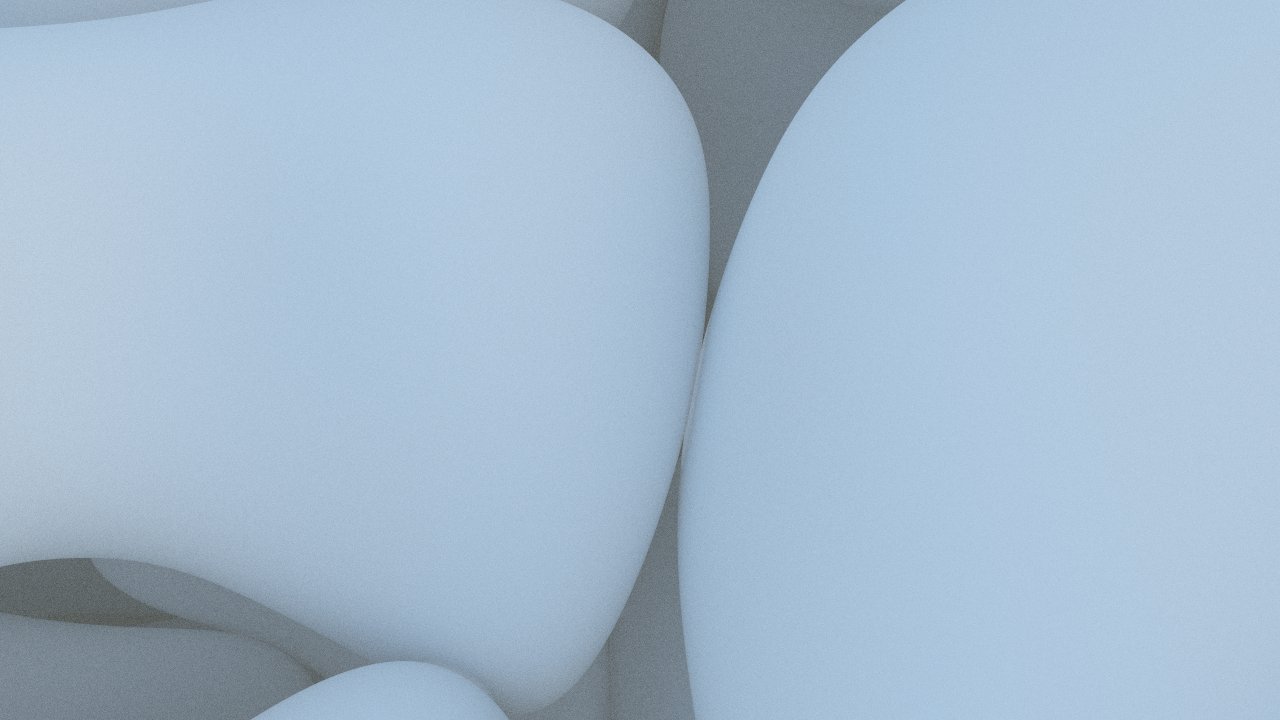}};
\end{tikzpicture}
 suspended particulate media \hspace{2.8cm}  particle agglomeration \hspace{2.5cm}  particle piling \\[-0.5ex]
    
    \caption{
Suspended particulate media: We generate spatially varying multiscale particulate media using a procedural particle representation without precomputation or tiling. A Stanford bunny made of ice demonstrates this, with large air bubbles sparsely in the distribution in the upper section transitioning to smaller ones and densely distributed below, affecting its macroscopic appearance. It shows macroscale, mesoscale, and microscale particles from top to bottom with smooth trasisitions. Our method efficiently computes the signed distance field (SDF) to nearby particles and employs simple surface rendering using adaptive sphere tracing. Unlike volume path tracing, individual particles become visible when large enough or viewed up close. Insets next to bunny show close-ups of different regions. Particle agglomeration: Particles from a suspended cloud (top) are selected using polynomial functions and grow in size to form agglomerates (middle and bottom). Particle piling: Illustrates on-the-fly granular media like rock piles generation using SDF grid deformation, even grains in concave shape, and including a zoomed-in contact point between grains.}
    \label{fig:svbunny}
}

\maketitle
\begin{abstract}

  Materials exhibit geometric structures across mesoscopic to microscopic scales, influencing macroscale properties such as appearance, mechanical strength, and thermal behavior. Capturing and modeling these multiscale structures is challenging but essential for computer graphics, engineering, and materials science. We present a framework inspired by hypertexture methods, using implicit functions and adaptive sphere tracing to synthesize multiscale structures on the fly without precomputation. This framework models volumetric materials with particulate, fibrous, porous, and laminar structures, allowing control over size, shape, density, distribution, and orientation. We enhance structural diversity by superimposing implicit periodic functions while improving computational efficiency. The framework also supports spatially varying particulate media, particle agglomeration, and piling on convex and concave structures, such as rock formations (mesoscale), without explicit simulation.  We show its potential in the appearance modeling of volumetric materials and explore how spatially varying properties influence perceived macroscale appearance. Our framework enables seamless multiscale modeling, reconstructing procedural volumetric materials from image and signed distance field (SDF) synthetic exemplars using first-order and gradient-free optimization.
\end{abstract}  
\section{Introduction}
Real-world materials exhibit various geometric structures at various length scales classified as particulate, fibrous, laminar, porous, or mixed, and each influences the properties of the material \cite{zallen2008physics, callister2020materials}. Understanding these classifications is essential for applications in appearance modeling, engineering, and material science. The forward and inverse modeling of the multiscale geometry can be helpful in various applications, such as appearance modeling and multiphysics modeling in chemical, thermal, mechanical, and electrical applications. We provide a framework for modeling and visualizing multiscale geometric structures, emphasizing geometry generation principles rather than rendering algorithms. This framework is flexible for generating various geometric structures, particularly useful for applications like fabrication and 3D printing of multiscale geometry.

 Materials consist of discrete particles that may be suspended, agglomerated, or in contact without merging. These particles can be crystalline or amorphous, affecting optical properties such as translucency \cite{frisvad2007computing, jarabo2018radiative, guo2021beyond}. While traditional volume rendering struggles with fine particle details, multiscale modeling preserves mesoscale complexity to enhance realism \cite{meng2015multi}.

A general approach to modeling particle-based materials involves arranging defined particle types within a tiled volume \cite{meng2015multi}. However, this tiling leads to regularity artifacts and density inconsistencies, lacks continuous spatial variation modeling. To address these issues, we extend hypertexture methods with procedural modeling to generate suspended particles, agglomerates, and granular media without pre-generation. By generating point distributions from implicit functions, we model spatially varying geometric structures with smooth transitions. Our method avoids conventional volume rendering by computing the SDF for nearby particles, enabling seamless zoom-in exploration and multi-scale density transitions \cite{muller2016efficient, michel2020real, zingsheim2024learning}. Using a 3D grid with pseudo-random seeding, we maintain controlled overlap within \(2^D\) nearest cells for particle instantiation. This approach supports multiscale particle distributions across varying grid scales, densities, and sizes, ensuring smooth transitions between scales, see Fig. \ref{fig:svbunny}. For larger particles exceeding half the grid cell width, SDF-based union operations across \(3^D\) cells are employed for modeling agglomeration, allowing particles to merge and form structures like gels, as shown in Fig. \ref{fig:svbunny}. Our particle agglomeration method, inspired by blobby object modeling \cite{muraki1991volumetric, jin2005blob, kanamori2008gpu}, is applicable to various materials, including ice, milk, foam, and composites.

For granular media such as rock piles and sand, our approach procedurally generates convex and even concave grains and their interactions using grid deformation with polynomial functions, see Fig. \ref{fig:svbunny},  and overcoming tiling artifacts and eliminating the need for collision detection between grains. \cite{paris2020modeling, peytavie2009arches}.

Additionally, we model volumetric geometric structures using space-filling other implicit primitives in discrete grids (cylinders, tubes, and laminates, etc.) and periodic functions without grid discretization, particularly triply periodic minimal surfaces (TPMS) \cite{han2018overview, dolan2015optical, inakage1987frequency, al2019multifunctional}. Although grid-based methods offer greater spatial variation, periodic functions enable efficient high-speed evaluation. Our contributions include:
\begin{itemize}
\item Procedural modeling of multiscale geometric structures using SDFs in discrete grids and periodic functions for volumetric material synthesis.
\item We propose an efficient method for calculating adaptive step lengths in sphere tracing using SDF gradients and polynomial functions to model geometry based on multiscale grid variations.
\item We propose optimization-based reconstruction methods to reconstruct total volumetric SDFs from 2D slices of the volume containing SDFs.
\end{itemize}


\section{Related Work} \label{sec:related}

In this section, we review the related research on modeling multiscale materials, emphasizing procedural methods using implicit surfaces.

\paragraph*{Implicit Representations}
Ricci~\cite{ricci1973constructive} defines solid surfaces using level sets within a distance field, employing $\max$ and $\min$ for union and intersection. Hart et al.\cite{hart1989ray} optimized rendering for such implicit objects, leading to sphere tracing\cite{hart1993sphere, hart1996sphere}, now widely used for rendering SDF-based objects. We leverage sphere tracing for Monte Carlo ray tracing of microgeometries.

Function-based representation of objects, along with the relations and operations defined by Pasko~\cite{pasko1995function}, has been used for blending, offsetting, and morphing geometric shapes and objects. SDFs enable Boolean operations, surface offsets, and blending via interpolation~\cite{payne1992distance}. Our particulate SDF modeling integrates these methods with spatially varying offsets.

Precomputed SDFs benefit from adaptive grids for scalability~\cite{frisken2000adaptively}. Sparse voxel octrees improve multi-scale representation and interpolation~\cite{heitz2012representing}, while displacement fields reduce voxelization artifacts~\cite{brunton2021displaced}. Our procedural approach avoids voxelization artifacts and precomputation constraints, ensuring smooth scale transitions, though high sample counts are needed for antialiased rendering. Adaptive ray marching~\cite{wu2022real} improves ray casting efficiency by utilizing precomputed binary density grids to reconstruct particle-based surfaces. However, these techniques struggle with multiscale grid variations. We propose an extension of gradient-based adaptive sphere tracing that accommodates multiscale grid geometry variations through polynomial-based gradient checks for step lengths, eliminating the need for precomputed grid scale data.


\paragraph*{Procedural Modeling}
Modeling particulate materials often involves procedural color variation. The bombing method~\cite{schachter1979random} distributes color bubbles, later adapted for 3D textures by Peachey~\cite{peachey1985solid}, storing bubble properties in a table for efficient lookup. This eliminates the need for surface parametrization in 2D textures.

Sparse convolution noise~\cite{lewis1984texture,lewis1989algorithms} translates point distributions into particulate color variations using a convolution kernel. Efficient generation methods for kernels with compact support~\cite{frisvad2007fast,luongo2020microstructure} employ space-filling arrangements of randomly positioned impulses, simplifying neighbor identification. We use this approach to create a space-filling particle distribution within a bounding sphere radius. Sparse convolution with Gabor noise kernels~\cite{lagae2009procedural} can be used to generate high-quality surface-level textures that are spectrally controlled. In this framework, we efficiently generate high-quality volumetric hypertextures using space-filling implicit object distributions.  Stolte~\cite{1003534} presented a method for efficiently filling space and voxelizing infinite implicit objects using recursive subdivision algorithms, which work best for repetitive structures and capture intricate details, such as cyclical surfaces.

Hypertexture~\cite{perlin1989hypertexture} extends color variation to density variations in a volume, enabling procedural surface rendering via ray marching~\cite{tuy1984direct,kajiya1984ray}. Worley~\cite{worley1996cellular} demonstrated that distances to nearest points effectively model surface irregularities in particulate materials. We apply this to transform our particle distribution into an SDF for physically based rendering.


Periodic surface modeling~\cite{wang2007periodic,lu2022structural} with triply periodic minimal surfaces (TPMS) enables intricate topological patterns across scales, including composite metamaterials. Leveraging trigonometric functions, we generate seamless multiscale microstructures with minimal parameters and computation.

Stereology estimates 3D particle number density from 2D exemplars, aiding tileable color variation in particulate materials~\cite{jagnow2004stereological,shu2014efficient}. The Worley method~\cite{dorsey1999modeling,soulie2007modeling} generates compact particle aggregates using a precomputed Poisson point distribution. Randomized particle instantiation within a volumetric tile disrupts tiling periodicity~\cite{meng2015multi} while space-filling point distributions eliminate the need for tiling or precomputation~\cite{baerentzen2023curl}. We leverage this to create spatially variable particulate materials with an SDF representation.

Most previous work relies on precomputation and lacks spatially-varying support. In contrast, our framework generates multiple structures on the fly and supports a spatially-varying definition.

Procedural models for porous structures have been proposed in the past with 3D textures~\cite{baravalle2017realistic}, Voronoi foams for fabrication~\cite{martinez2016procedural} with adaptive grids. We generate SDFs with adaptive grids for multiscale particle distributions efficiently.

High-frequency details in the mesoscale can be achieved with volume pertubations~\cite{kniss2002interactive}, generalized displacement maps~\cite{wang2004generalized}, volumetric spot noise for shell textures~\cite{turkay2016volumetric}  and recently aperiodically on meshes~\cite{michel2023mesogen}. We employ volumetric perturbations to enhance microgeometry details in volume rendering.


\paragraph*{Reconstruction}
Perlin noise~\cite{perlin1985image,perlin2002improving} combined with an encoder-decoder neural network can generate realistic 3D textures from 2D references~\cite{henzler2020learning}. Implicit periodic field networks use periodic encoding in MLPs to model particulate materials~\cite{chen2022exemplar}, but their reliance on precomputed 3D subvolumes makes them impractical for large-scale rendering. Generative models for spatially varying noise~\cite{maesumi2024one} are limited to surface noise while learning volumetric noise remains challenging. Our work explores SDF-based volume reconstruction, enabling full-volume reconstruction from examples.
\section{Multiscale Geometry Modeling}

An SDF is a scalar field representing the minimum distance from a point \(\bm{p}\) to a surface defined by an implicit function. It is negative inside the surface, positive outside, and zero on the isosurface~\cite{frisken2000adaptively}.


This section details the synthesis of multiscale material structures. Sec. \ref{sec:spm} addresses suspended particulate media via primitive space-filling SDFs, Sec. \ref{sec:pa} covers geometric structure formation through particle agglomeration, Sec. \ref{sec:pp} discusses granular media modeling through particle piling, and Sec. \ref{sec:pip} covers at periodic implicit functions for geometric structures.


\subsection{Modeling Suspended Particulate Media} \label{sec:spm}
Suspended particulate media consist of discrete particles in 3D space with no contact between them. In this section, we introduce the dual-grid-based approach for generating suspended particulate media, along with various principles for creating spatially varying particulate media with variations in density, distribution, shape, size, correlations, anisotropy, noise particles and more.

\subsubsection{Point Distribution} \label{sec:pointdistrib}

To generate particulate microgeometry, we require a space-filling point distribution that facilitates random instantiation of particles and provides a straightforward method to find any nearest particles that might overlap with a given point in space $\bm{p}$. Taking inspiration from methods used to produce sparse convolution noise~\cite{frisvad2007fast,luongo2020microstructure}, we accomplish this by utilizing a regular cubic grid in conjunction with a seeded pseudo-random number generator (PRNG). The grid cell index $\bm{q} = (q_x, q_y, q_z)$ is incorporated into a multilinear spatial hash function and then multiplied by the number of random numbers needed for a cell, denoted as $N$, to create a seed $t$ for the PRNG. By setting $N = 4$, we can generate a particle of random size $s$ and position $\bm{x}$ within each grid cell:
\begin{eqnarray}
t & = & N \sum_{i=0}^1\sum_{j=0}^1\sum_{k=0}^1a_{ijk} q_x^i q_y^j q_z^k \label{eq:seed}\\
(\xi_1,\dots,\xi_N) & = & \mathrm{rnd}(t, N) \label{eq:rnd}\\
\left[\bm{x}, s\right] & = & [\bm{q} + (\xi_1, \xi_2, \xi_3), \xi_4] \,, \label{eq:instance}
\end{eqnarray}

The pseudorandom number generator (PRNG), denoted as $\mathrm{rnd}$, produces $N$ random numbers that are uniformly distributed within the interval $[0,1)$. The coefficients $a_{ijk}$ are chosen to remove regularity artifacts in the distribution of particles that result in the process of generating particle clouds. Choosing prime numbers as coefficients eliminates the regularity artifacts in a better way.
To accurately identify particles that might overlap with a point \(\bm{p}\), we control the radius of the particle's bounding sphere to no more than half the width \(w\) of a grid cell. As a result, we can calculate the indices of the eight grid cells that could potentially contain particles overlapping the point \(\bm{p}\).

\begin{equation}
\bm{q}_i = \left\lfloor \bm{p}_w - \tfrac{1}{2} \right\rfloor + (i\&1,\ (i\gg1)\&1,\ (i\gg2)\&1),
\label{eq:cellid}
\end{equation}

\begin{figure}[ht]
\centering
\begin{tabular}{@{}c@{}}
\includegraphics[height=9.5em]{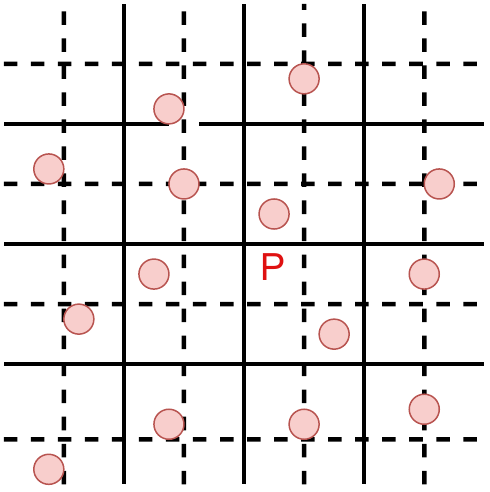}\hspace{\stretch{1}} 
    \end{tabular}
\caption{To consider the particles that may overlap an arbitrary query point $\bm{p}$, we check that the bounding sphere radius of a particle is at most half a grid cell width. Once we identify the cell in the stippled dual grid to which $\bm{p}$ belongs, the particles that could potentially overlap must be found in the $2^D$ grid cells that intersect with the stippled dual-grid cell (where $D$ represents the number of dimensions).}
\label{fig:distrib}
\end{figure}

The value of $i$ in the above expression goes from 0 to 7. The expression \(\bm{p}_w = \frac{1}{w}(\bm{p} + \bm{h}(\bm{p}))\) scales the coordinates based on the width of the grid cell and incorporates domain warping through an arbitrary vector function \(\bm{h}\). The adjustment of subtracting \(\frac{1}{2}\) from each scaled coordinate, followed by flooring, effectively utilizes a dual grid with a half-cell width offset. This design ensures that a point \(\bm{p}\) located within a cell of the dual grid is influenced solely by particles in the \(2^D\) neighboring cells, where \(D\) represents the number of dimensions (refer to Figure~\ref{fig:distrib}).

\subsubsection{Particle SDF in a Grid Cell} \label{sec:particle}

\begin{figure}[ht]
 \centering
  \begin{tabular}{@{}c@{}c@{}}
   \includegraphics[width=0.487\linewidth]{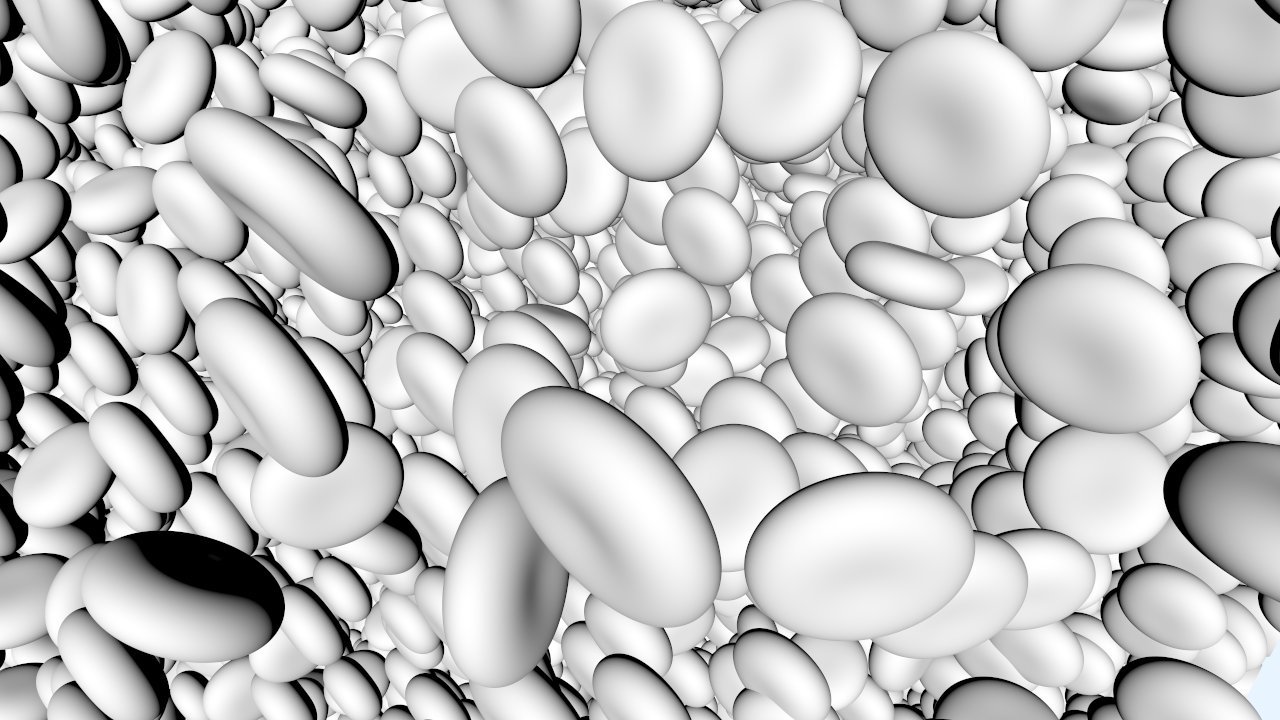} 
 \includegraphics[width=0.5\linewidth]{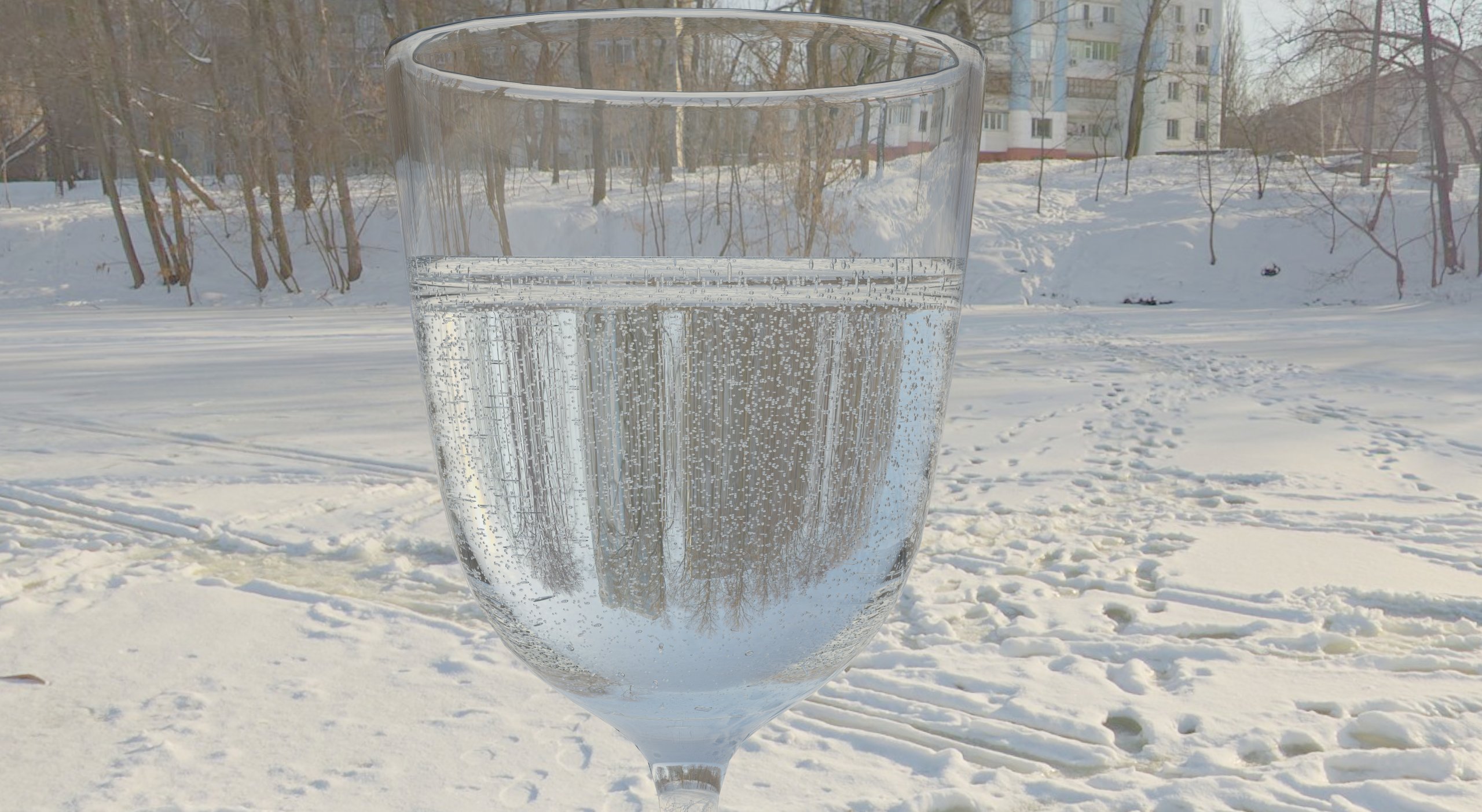} 
  \\[-0.5ex]
  \end{tabular}
\caption{Left: Red blood cell (RBC) particle cloud modeled using our method and an implicit function available for the shape of these cells in~\cite{kuchel2021surface}. Right: A rendering of sparkling water with spatially varying air bubble particle sizes. The bubbles are smaller and sparser at the bottom, gradually becoming larger and denser towards the top of the liquid.}
  \label{fig:rbc}
\end{figure}

Any particle shape can be used within a grid cell as long as its radius of the bounding sphere does not exceed \( w/2 \). The simplest SDF is that of a sphere. However, we could use any kind of particulate shape, see Fig.~\ref{fig:rbc} for particle shape as a red blood cell (RBC). Let \( [\bm{x}, s] \) represent the random placement of the particle within the grid cell indexed by \( \bm{q} \), as outlined in Eq.~\ref{eq:instance}. We denote \( d_{\bm{q}} \) as the signed distance function that describes this particle. For a sphere that is randomly placed with a radius that varies between \( 0 \) and \( w/2 \), we have
\begin{equation}
d_{\bm{q}}(\bm{p}_w) = \|\bm{p}_w - \bm{x}\| - \frac{1}{2}s \,.
\label{eq:p-sdf}
\end{equation}

To make things easier, we make the function accept a scaled point of interest as an argument, enabling the particle's signed distance function to be defined in a 3D space where the grid cell width is 1.

We can easily control the center of the particle at \(\bm{x}\) to a smaller portion of the grid cell or constrain its radius to a narrower range of values. By adjusting \(N\) in Eqs.~\ref{eq:seed}--\ref{eq:rnd}, we can define the particle using fewer or more random numbers. For example, we might set a fixed radius for all particles (\(N=3\)), or we could employ two random numbers to achieve random orientations of fixed-size ellipsoids (\(N=5\)). Spherical particles often serve as an excellent foundation for generating various particle shapes, while ellipsoids can be created through directional scaling, see Fig.~\ref{fig:anisotopic_ice}. As mentioned in Sec.~\ref{sec:related}, there are many SDFs of different shapes available in the literature and online. Furthermore, we can utilize Boolean operations to merge various particle clouds with different shapes together. Many SDFs applied in practice are non-Euclidean, such as those ellipsoids generated by directional scaling.


\subsubsection{Particulate Material SDF and Spatial Variation} \label{sec:particulate}

We now introduce the SDF for the particle suspensions. In this context, the subscript \(i\) refers to one of the $k$ nearest grid cell neighbors, indicated by the index \(\bm{q}_i\), as outlined in Eq.~\ref{eq:cellid}. The SDF for the particulate material with many particles is then defined as follows:
\begin{equation}
d_g(\bm{p}, w) = f(\bm{p}) + \min_{i=0,\dots,k} d_{\bm{q}_i}\left(\frac{1}{w}(\bm{p} + \bm{h}(\bm{p}))\right) \,, \label{eq:g-sdf}
\end{equation}
where $f$ is an arbitrary function for spatially varying offsetting of the surface. For modeling the suspended particulate media the value of the $k$ in the above equation is 7.

\begin{figure}[ht]
 \centering
  \begin{tabular}{@{}c@{}c@{}}
   \includegraphics[width=0.5\linewidth]{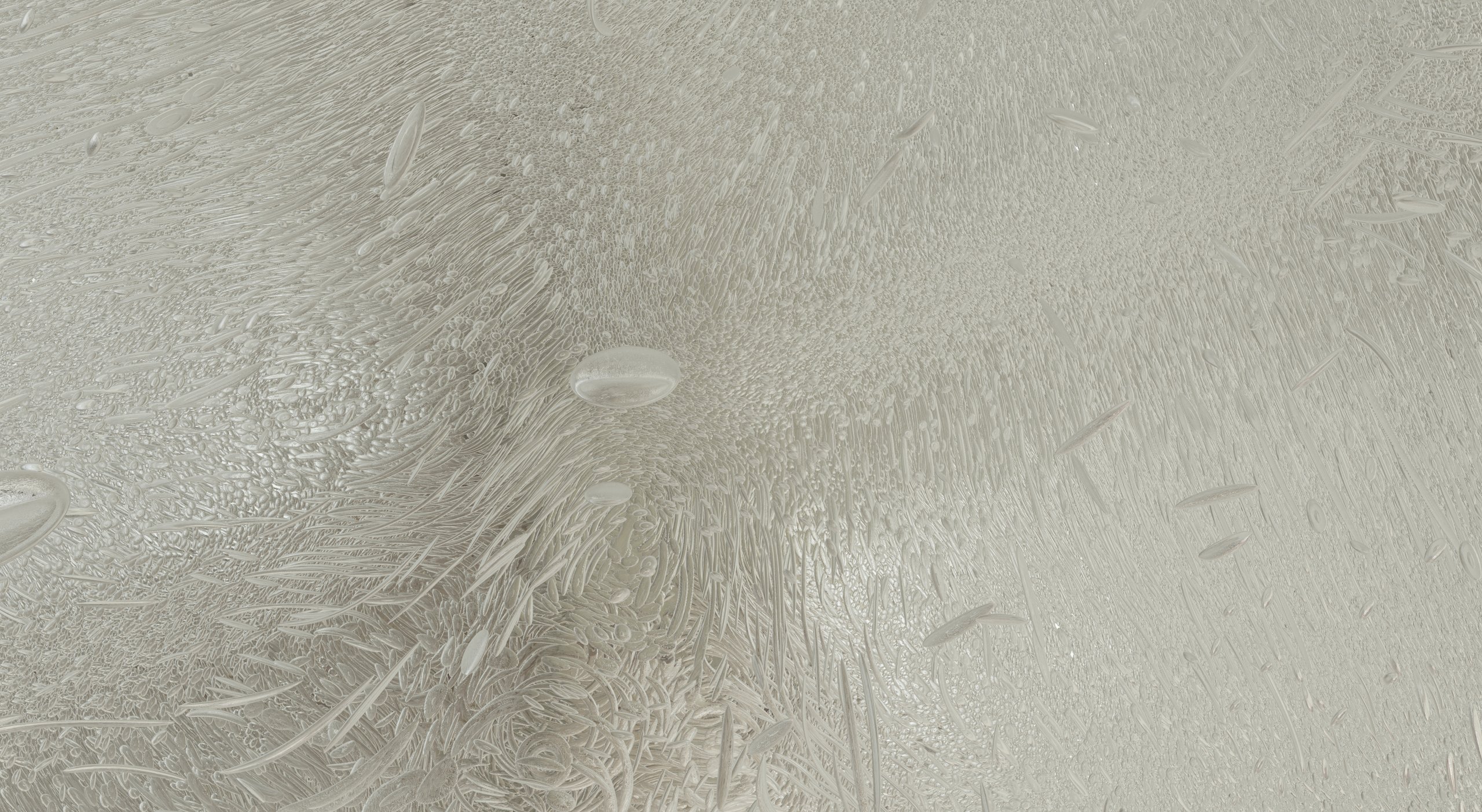} &
  \includegraphics[width=0.5\linewidth]{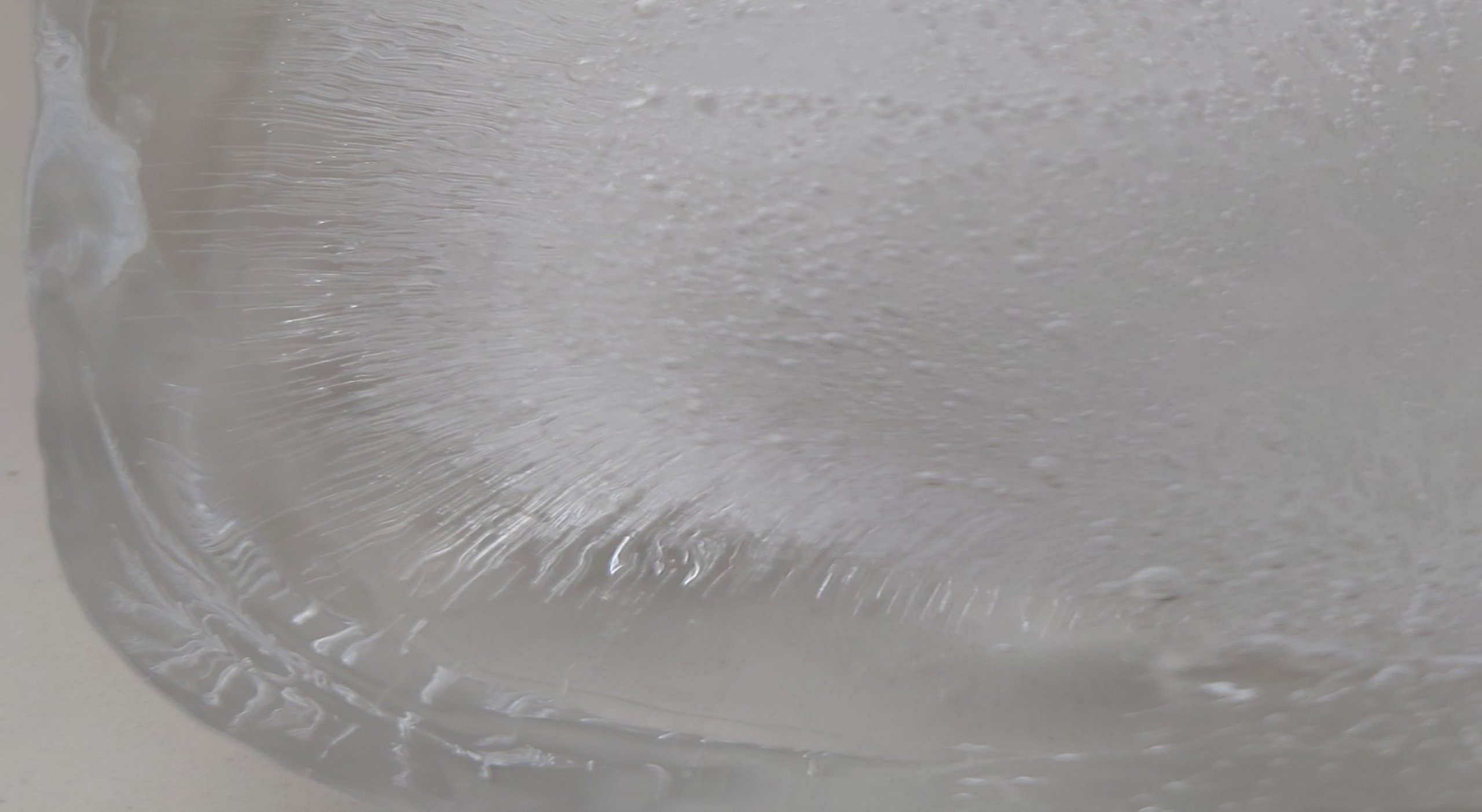} 
  \\[-0.5ex]
  \end{tabular}
\caption{Inspired by a photo of air bubbles in ice (right), we used our multi-phase particle cloud approach to model a similar material (left). Our model has ice as the host medium and contains air particles that vary in size and shape with the spatial location in the medium, transitioning from spherical to non-spherical.}
\label{fig:ice_multiphase}
\end{figure}

The frequency of function $f$ can generate noise to particle surfaces and affect their sizes in different regions. Alongside $f$, we use grid cell width $w$ and set limits on particle size $s$ to control number density and size distributions. Constraints on $\bm{x}$ help balance regularity and randomness in particle positioning within the cloud. Transforming point $\bm{p}$ with the turbulent matrix generates multi-phase particle clouds (see Fig. ~\ref{fig:ice_multiphase}) and particles with noisy patterns. 


 Boolean operations can be effectively used for merging particle inclusions. By utilizing the minimum (or union) operator, we integrate SDFs with unique local variations, enabling the blending of different particle size and density distributions (see Fig.~\ref{fig:svbunny}). For blending various shaped particles, we sample a piecewise constant pdf using random instantiation and a method akin to Russian roulette, based on predetermined shape probabilities. We also employ a rejection control technique with a spatially varying function \( g(\bm{p}) \) between 0 and 1, which determines the likelihood of accepting a particle in each grid cell. In each cell, we utilize a Russian roulette method with a random number \( \xi_j \); if \( \xi_j < g(\bm{p}) \), the particle’s SDF is \( d_{\bm{q}}(\bm{p}_w) \); otherwise, it is set to \( \infty \). Particle correlation governs the clustering or regular distribution of particles and is managed through the rejection control on particle clouds. 

 To bridge the digital scene with a physical length scale, it's essential to define the physical length of a unit within the digital domain, which we denote as the scene scale~$\ell$. Consequently, the physical width of a grid cell is given by $\ell w$, resulting in a particle number density of $N = (\ell w)^{-3}$ when there is one particle per grid cell. For a specified grid cell width~$w$, the maximum radius of the bounding sphere for a particle, according to our method, is $r_{\max} = \frac{1}{2}\ell w$. Within this framework, we can choose the particle shape and scale~$s$ freely. To create a tailored particle size distribution, we produce the pseudo-random numbers required for our random instantiation (refer to Sec.~\ref{sec:pointdistrib}) and sample $s$ based on a probability density function (pdf). A range of sampling methods (including normal distribution sampling) can be found in the work of Shirley et al.~\cite{shirley2019sampling}. This process allows us to simulate materials with specific particle size distributions. Furthermore, employing offsetting (the function $f$ detailed in Sec.~\ref{sec:particulate}) or adjusting the distribution parameters based on the location of interest $\bm{p}$ results in heterogeneous materials, where particle density varies throughout the volume.

\subsection{Modeling Particle Agglomeration} \label{sec:pa}

\begin{figure}[ht]
    \centering
   \begin{tabular}{@{}c@{}c@{}}
    \includegraphics[width=0.488\linewidth]{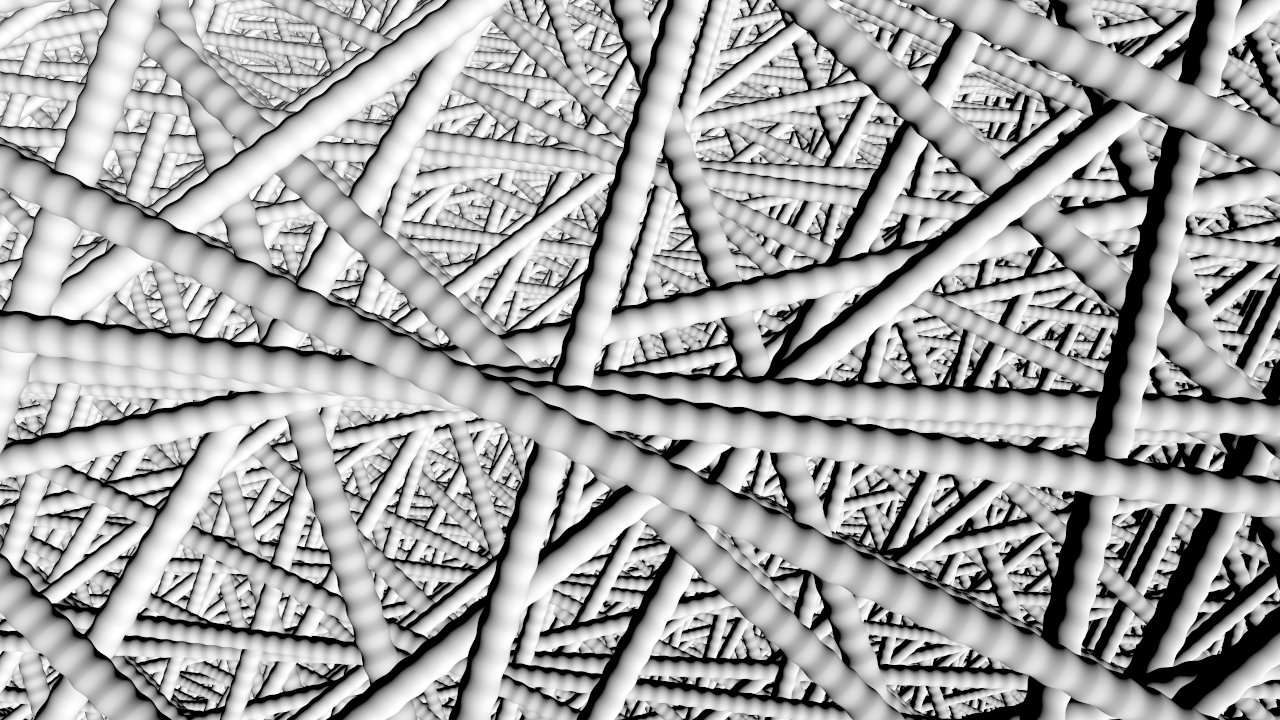}
    \includegraphics[width=0.5\linewidth]{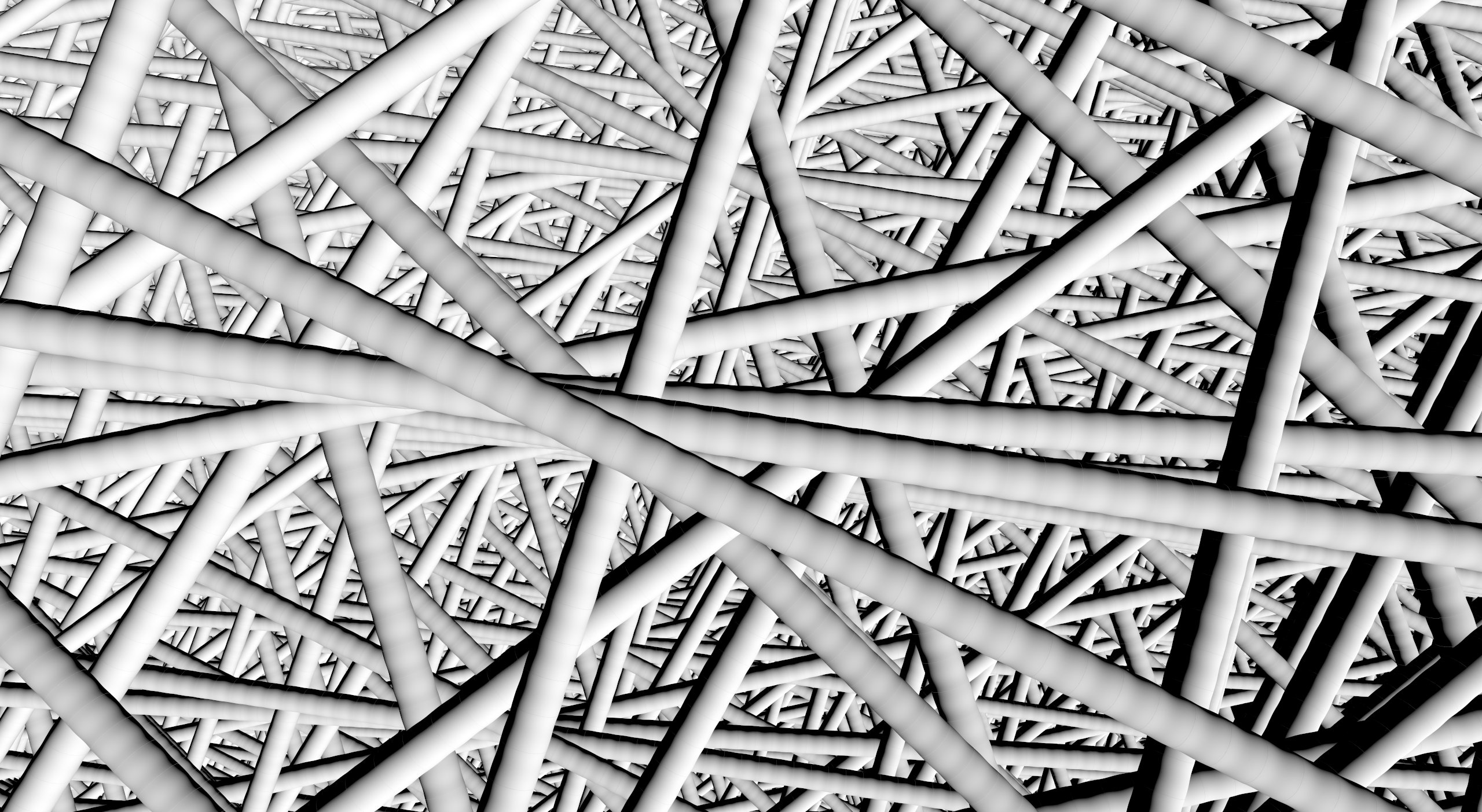}
    \end{tabular}
  \caption{ Left: Fibrous microstructure with diameter 1mm generated using particle agglomeration method, right: same microstructure as left but smooth it using the smooth min operator. We generate the snapshot of the processes of particle agglomeration (the clustering of particles) and crystal growth using polynomial functions on the regular lattice of particle clouds. }
  \label{fig:fibers_sitting}
\end{figure}

Particle agglomeration in suspended particulate media involves individual particles in a lattice growing and sticking together to form various geometric structures. We model this process using a cubic lattice of spherical particles, sampling clusters, and applying a union operator (the minimum in an SDF) for agglomeration, see Fig.~\ref{fig:fibers_sitting}. Particle size constraints limit this on the dual grid approach. To better model agglomeration, we examine the 27-neighborhood of each point for SDF evaluation, allowing particles to grow and form agglomerates. This method enables control over crystal growth mechanisms and the synthesis of different geometries, such as gels, fibers, laminar, and porous structures from the cubic lattice of spherical particles.

For the particle agglomeration, our starting point is a cubic regular lattice. For any point in space \(\bm{p}\), we find the index of the grid cell it occupies using \(\bm{q} = \lfloor\bm{p}/w\rfloor\), where \(w\) represents the width of the grid cell. This index can then be applied in conjunction with xor-shift bit scrambling~\cite{marsaglia2003xorshift} to generate a seed for PRNG, which may be useful for random instantiation or for controlling geometry rejection within a grid cell. We utilize the index \(\bm{q}\) to explore the 27-neighborhood of grid cells surrounding the point \(\bm{p}\). If a particle within cell \(\bm{q}\) has a radius of at most \(w\), it will not overlap with any cells beyond the 27-neighborhood.

With respect to the 27-neighbourhood of the point point \(\bm{p}\), we change the Eqs. \ref{eq:p-sdf} and \ref{eq:g-sdf}, in the following manner for generating the particle agglomeration. We can utilize any particle shape within a grid cell if its bounding sphere radius satisfies \( r \leq w \). By scaling the point of interest as \( \bm{p}_w = \bm{p}/w \) and defining \( \bm{c} \) as the relative center of the sphere within the grid cell, the SDF that describes a sphere located in the grid cell indexed by \( \bm{q} \) is given by.
\begin{equation}
d_{\bm{q}}(\bm{p}_w) = \|\bm{p}_w - \bm{q} - \bm{c}\| - \frac{r}{w} \,.
\label{eq:particlesdf}
\end{equation}

Letting $\bm{q}_i$ denote the index of one of the 27 nearest grid cell neighbors surrounding a point $\bm{p}$; we evaluate the SDF using the Eq.~\ref{eq:g-sdf} with the value of the $k$ is 27.

Given the restriction of $r \leq w$, particles that lie outside the 27-neighborhood cannot overlap with any point in the grid cell containing $\bm{p}$. Nevertheless, there may be particles beyond this neighborhood that are closer than what is represented by the geometry of these 27 grid cells. To control overlapping with any particle positioned in a grid cell outside the 27-neighborhood during sphere tracing, we limit $d_g$ to $w/2$.

This framework does not limit itself to using non-spherical particles as the basic geometric primitive. It is entirely feasible to utilize a variety of other primitives, such as cubes, tori, cylinders, ellipsoids, hexagonal prisms, thin hexagonal prisms, etc.  

\begin{figure}[ht]
    \centering
    \begin{tabular}{@{}c@{}c@{}}
 \includegraphics[width=0.5\linewidth]{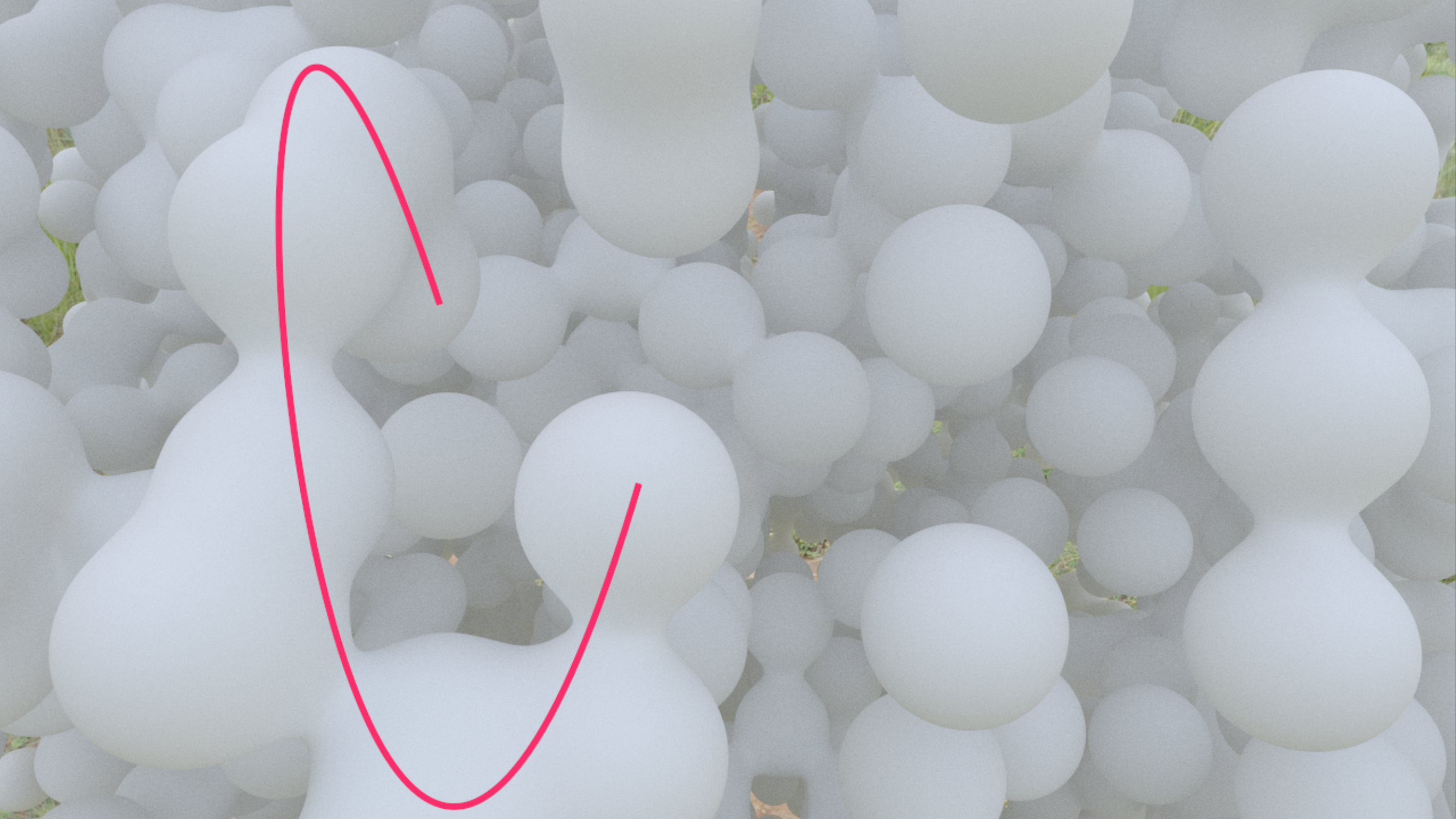} 
 \includegraphics[width=0.5\linewidth]{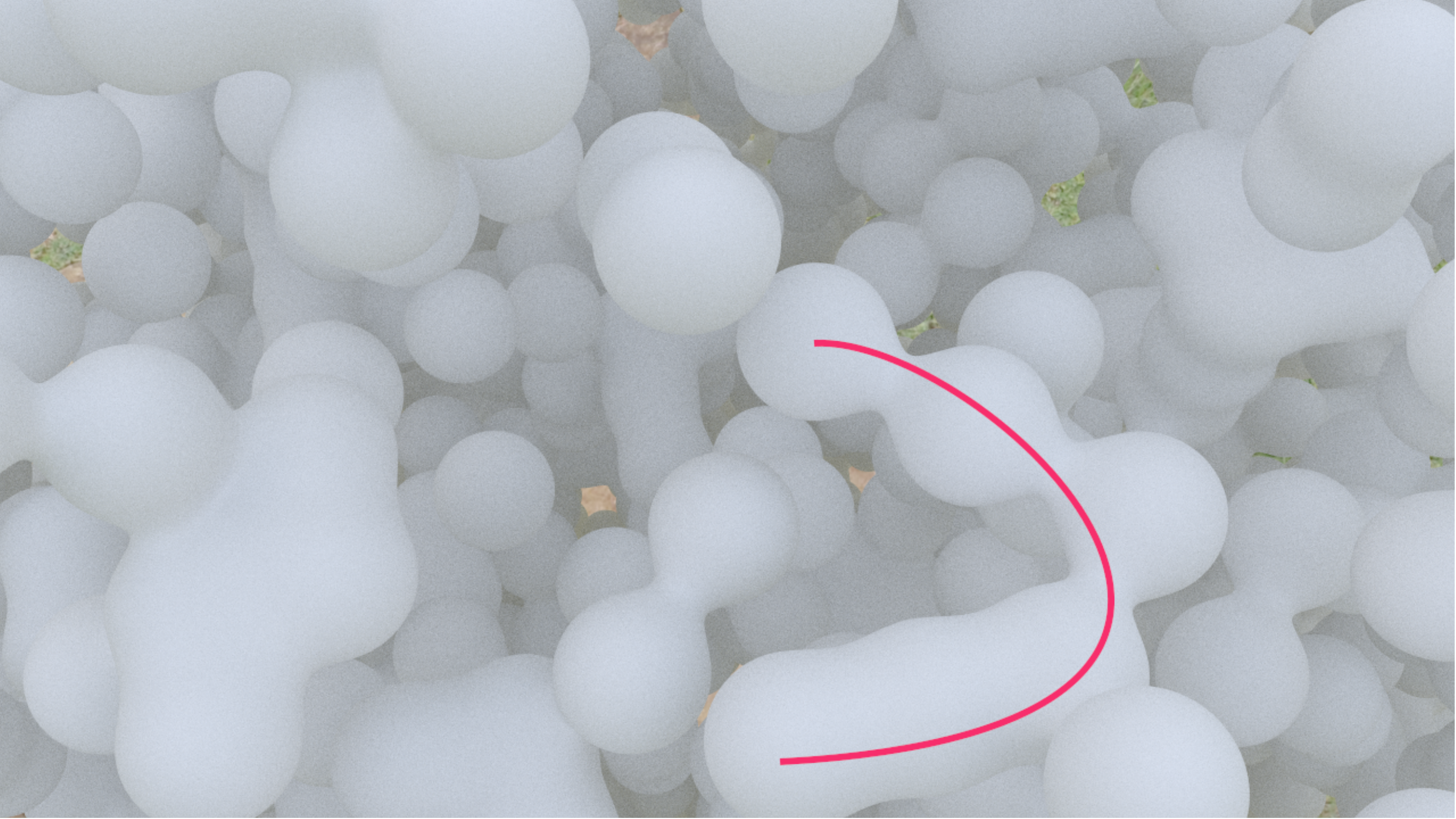} \\[-2ex]
 \end{tabular}
  \caption{Particle agglomeration structures generated using B{\'e}zier curves as polynomial functions on grid cell $\bm{q}$ in the 27-neighbourhood of the point $\bm{p}$. The shape of each agglomerate is different from the others. It is spatially varying throughout a large section of microgeometry. The particle arrangement mimics a snapshot of a gelation process.}
  \label{fig:bc}
\end{figure}

We synthesize meso- and microscopic materials using spherical particle clouds arranged in regular lattices. The particulate material SDF, \( d_g(\bm{p}, w) \), generates regularly distributed particles of size \( s \) per grid cell. By applying an integer-valued polynomial \( (P) \) or rational function \( (R) \) to the grid index, we perform rejection control, selectively removing particles. The remaining particles can form fibers, laminar structures, or porous lattices. Using set theory, we extract subsets of particles to create arbitrary microgeometries, such as gels shaped by Bézier curves in Fig.~\ref{fig:bc}. Additionally, jittering a regular cloud via polynomial-based rejection yields highly irregular distributions. Every micro- or mesoscopic structure is defined by a particle cloud, where each particle is either retained (1) or removed (0). For a volume of size \( u \times v \times w \), we have \( 2^u \times 2^v \times 2^w \) possible structures.

Suppose we let $D: \mathbb{Z}^3 \to \mathbb{B}$ be a function mapping a grid cell index $\bm{q}$ to a Boolean value. This function selects a subset of particles from the cubic regular lattice in such a way as to synthesize a microstructure. This is sufficiently general to define particulate, fibrous, laminar, porous, or mixed geometries. When the function returns true (1), the SDF of the particles is included in this part of the space. If $D$ returns false (0), we set the SDF to $w/2$. We define the non-Euclidean SDF for this type of material by
\begin{equation} \label{eq:sdf}
d(\bm{p}, w) = \begin{cases} 
d_g(\bm{p}, w)\,, & \text{for\ }  \bigwedge_{i=0}^{26}[D(\bm{q}_i) = 1], \\
w/2\,, & \text{otherwise},
\end{cases}
\end{equation}
where $\bm{q}_i$ are the grid cell indices of the 27-neighborhood around $\bm{p}$ and $[C]$ is an Iverson bracket returning $1$ if the condition $C$ is true and $0$ otherwise. We specify the grid cells containing geometry using
\begin{equation}
D(\bm{q}) = 
\begin{cases} 
1\,, & \text{if } \bigoplus_{i=1}^{k_1} \bigoplus_{j=1}^{k_2} [P_i(\bm{q}) \equiv c_{ij} m_{ij}\!\!\pmod{m_i}], \\
0\,, & \text{otherwise}.
\end{cases}
\label{eq:pf}
\end{equation}
where $P_i(\bm{q})$ are multivariate integer-valued polynomial functions, and $m_i$ with $i=1,\dots,k_1$ are positive integer moduli for the polynomials $P_i(\bm{q})$. We can also use inequalities instead of congruency in the Iverson bracket.

The symbol \(\bigoplus\) represents either the reduction operator version of logical OR ($\bigvee$) or logical AND ($\bigwedge$). The inner \(\bigoplus\) operator can be used to select different congruence classes of the same polynomial, while the outer \(\bigoplus\) operator can be used to select congruence classes across the polynomials.  
The binary coefficients \(c_{ij}\) can be either -1 or 1 and are used to select a specific congruence class for the individual polynomials with respect to their modulo; if it is -1 then the specific congruence class is not included, and when it is 1 the congruence class can be included. The integers \(m_{ij}\) are constants such that \(0 \leq m_{ij} < m_i\), while \(m_i\) is the modulus for the \(i\)-th polynomial,  and the integers \( k_1 \) and \( k_2 \) are selected such that \( k_1 \geq 1 \) and \( k_2 \leq m_i \). 
In some cases, a rational function could be used instead of a polynomial function in the above equation. See supplemental material for more details about the polynomials used to generate gels based on Bézier curves shown in Fig.~\ref{fig:bc}.

%

\subsection{Modeling Granular Media with Particle Piling}\label{sec:pp}

After modeling particle suspensions and particle agglomerations, we focus on particle piling, where particles are settled in a contact position and can not merge or intersect. We use the same regular lattice of spherical particles used for the particle agglomeration, which is basically based 27-neighborhood of the point $\bm{p}$. Like traditional finite element methods, we don't need to find the collisions between particles(or grains). We consider rock piles and sand as interesting case studies for piling structures. We place spherical particles in a cubic regular lattice with the maximum possible size, and we do grid deformation or particle deformation to generate piling structures. Here, no union and intersection operations on particle clouds are allowed because every grid cell already contains a particle. Only one rotation operation with respect to the coordinates systems x, y, and z is allowed.

The SDF Eq. \ref{eq:g-sdf} for the particle material can be changed in the following manner for the generation of particle piling
\begin{equation}
d_g(\bm{p}, w) = f(\bm{p}) + \min_{i=0,\dots,26} d_{\bm{q}_i}\left(\frac{1}{w}((\bm{R_{ \theta}}\bm{p}) + \bm{P}(\bm{p})\bm{N}(\bm{p}))\right) \,, \label{eq:pp-sdf}
\end{equation}

In the above Eq., the term $\bm{R_{ \theta}}$ refers to the rotation matrix with angle  $\theta = P$, where we consider polynomial in $\bm{p}$ as the rotation angle. The term $\bm{P}(\bm{p})\bm{N}(\bm{p}))$ generates the grid deformation using a product of the polynomial function $ P(p)$ and the noise function $N(p)$. For instance, sparse convolution noise \cite{frisvad2007fast} is used as $N(p)$ and $\theta$ as the gyroid function in $\bm{p}$ to generate structures similar to the rock piles in Fig.\ref{fig:svbunny}. The term $ P(p)$ in the above equation controls the degree of convexity to the concavity of the grains in the same particle cloud, where the smooth transition between convexity to concavity happens in the same cloud.

\subsection{Implicit Periodic Functions} \label{sec:pip}
Continuous functions can be used to avoid grid discretizations and improve performance, as they do not require the evaluation of geometry in neighboring grid cells. One such family of functions is trigonometric functions. The periodic functions, sine and cosine, are particularly useful in this context, as every microstructure can be represented by combinations of these functions with varying frequencies and amplitudes. We exploit the periodic behavior of these functions, which offers flexibility in producing various microstructures through point transformations applied to them. This is directly related to Fourier transforms.


The general formula for a sine-cosine function defined by a summation over \( u \) trigonometric functions, each of these functions again the product of \( v \) trigonometric functions is
\begin{equation}
SC(\bm{p}) = \sum_{i=1}^{u} \prod_{j=1}^{v} T_{ij}(\bm{p})^{k_{ij}} + w \,,
\label{eq:tf}
\end{equation}
where \( w \) represents the width of the microstructure and \( k_{ij} \geq 0 \) are the powers associated with each trigonometric term \( T_{ij}(\bm{p}) \) defined by

\[
    T_{ij}(\bm{p}) \in \left\{ A_k \cdot f(\omega_k p_d + \phi_k) \mid f \in \{\sin, \cos\}, \, p_d \in \{p_x, p_y, p_z\} \right\}.
\]

where \( A_k \) is the amplitude, \( \omega_k \) is the frequency, and \( \phi_k \) is the phase shift of the trigonometric function.  

One collection of such functions that we use in our system is the family of Triply Periodic Minimal Surfaces (TPMS); for instance, the Gyroid, Diamond, and Primitive functions~\cite{han2018overview}, see below for equations of these structures.

Many microstructures can be defined using these trigonometric functions:
Let us consider a $3\times3$ matrix $\bm{T}$, which performs affine transformations such as scaling, rotation, and translation. Sometimes, one or more of these operations are combined in the $3\times3$ matrix using octaves. Each entry in this matrix is generated from the summation of multiple sine and cosine waves with varying amplitudes and frequencies, and typically, each value is normalized between $-1$ and $1$.

\begin{figure}[ht]
    \centering
    \includegraphics[width=\linewidth]{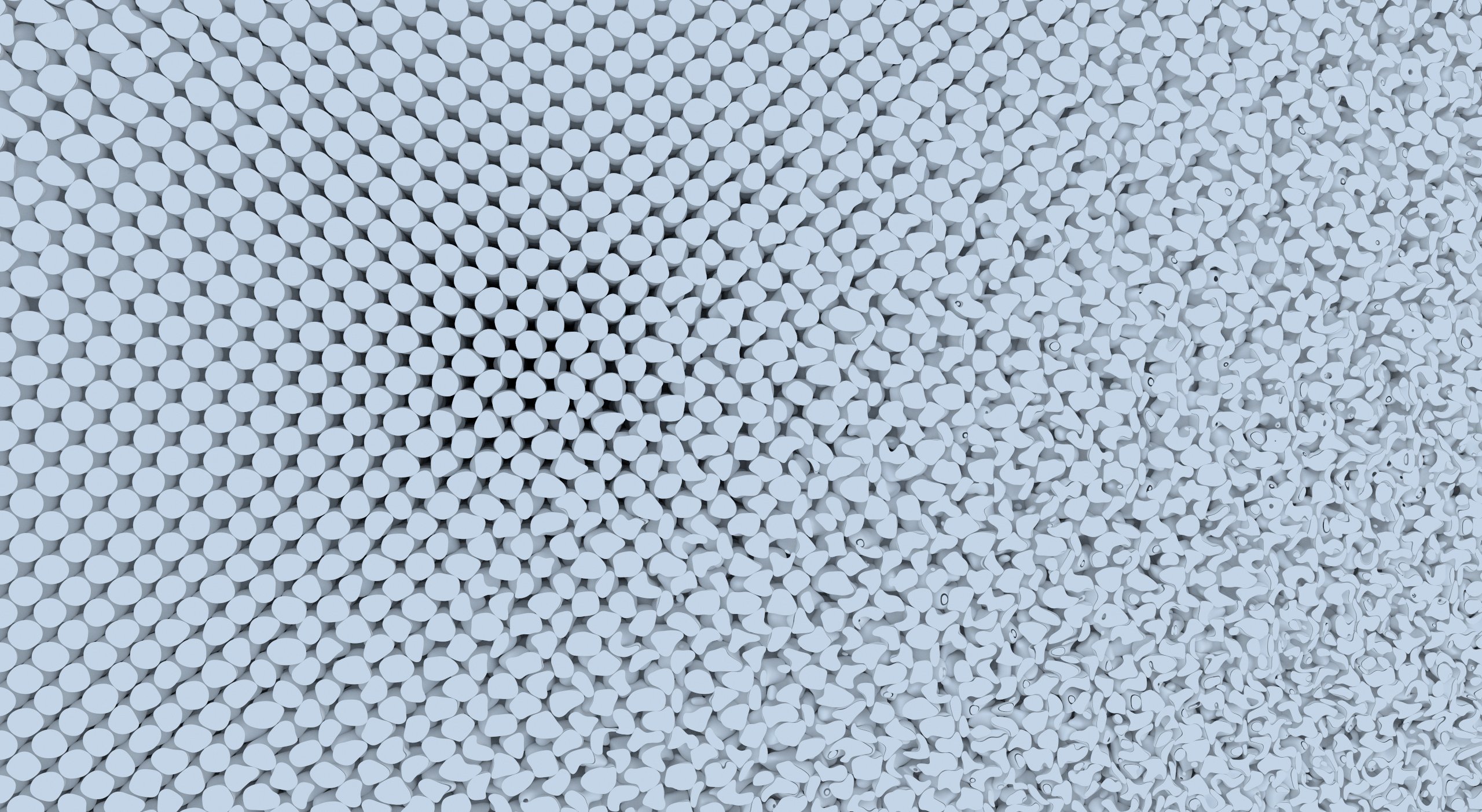}   
  \caption{We can control convexity to the concavity of the grains in the same cloud with smooth transitions using the polynomial function and the noise functions, where in the above result, we use $ P(p) = p_x+p_y+p_z$ and Perlin noise as $N(p)$ in the Eq.\ref{eq:pp-sdf}.}
  \label{fig:rockpiles-convex_concave}
\end{figure}
\section{Multiscale Geometry Rendering} \label{sec:rendering}

We use conventional ray tracing methods to render the multiscale structures generated by our framework based on implicit functions $d_g$. Unlike statistical scattering-based volumetric methods, we instatiate the mesoscale and microstructures on the fly during rendering, using surface shaders along with volume absorption~\cite{pharr2023physically} to simulate volumetric light transport. This approach not only enables high-fidelity visualization at multiple scales, but also allows users to flexibly study how custom shapes, spatial variation, anisotropy, and structural correlation influence the final material appearance.

To find the intersection of the surface on different scales, we follow Hart's sphere tracking approach~\cite{hart1996sphere} in the form of adaptive sphere tracking with different SDF and shaders for the different scales. The normal vector used by the surface shading at a surface intersection point is computed with the gradient of the SDF using central differences. In addition, we divide $d_g$ by the greatest magnitude of the gradient to ensure that the functions represent signed distance despite potentially being transformed to a non-Euclidean domain. We are adapting the distance bound by the values of the affine transformation matrix. We apply a Russian roulette technique based on absorption at the particle surfaces or within volumes to terminate paths probabilistically. A maximum trace depth is also set to prevent indefinite ray tracing, starting with a low value leading to a darker appearance due to early path termination gradually increasing until brightness stabilizes.
\section{Sphere Tracing Optimization} \label{sec:asl}
\begin{figure}[ht]
    \centering
    \includegraphics[width=\linewidth]{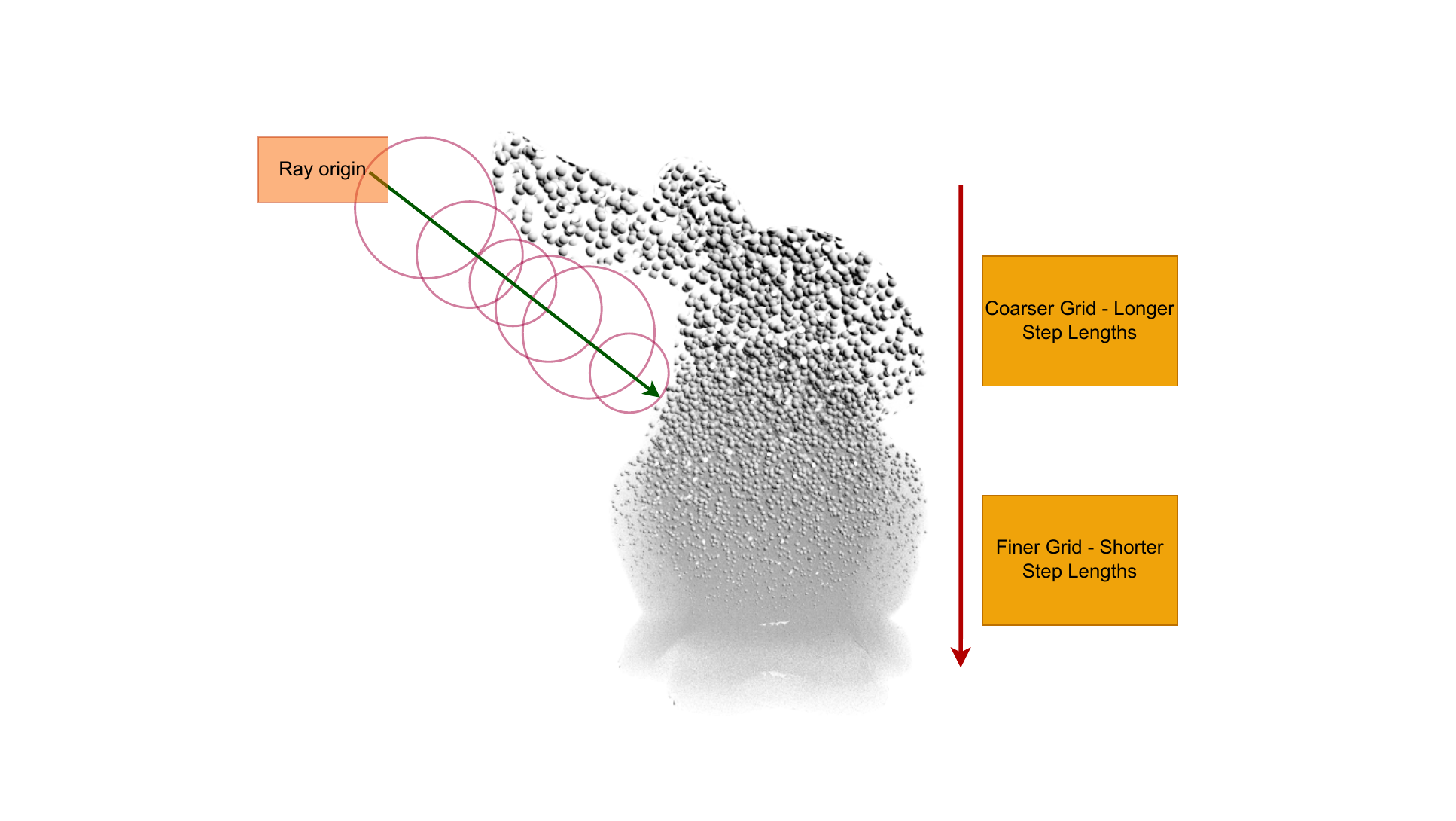}   
  \caption{We enhance the performance of the sphere tracing algorithm through adaptive step lengths by tracking grid scale variations using SDF gradients when the polynomial checking condition \(D(p)\) is true. This approach eliminates the need to compute gradients at every step, as they are calculated only for certain steps based on polynomial checks, allowing us to sample gradients from different regions selectively. By employing various types of polynomials, we can effectively monitor grid scale variations throughout the volume. This is particularly beneficial since different grid scales require different step lengths, so finer grids necessitate shorter steps, while coarser grids can use longer ones for improved efficiency.}
  \label{fig:st-gridscales}
\end{figure}

We optimize the performance of the sphere tracing for multiscale geometry using adaptive step lengths. Multiscale geometry features spatially varying grid scales, where coarser grids require larger steps and finer grids require shorter steps, see Fig.~\ref{fig:st-gridscales}. Fixed step-length sphere tracing is inefficient as it must use small steps throughout to accommodate all scales, reducing performance. We adapt step lengths based on SDF gradients to address this and dynamically identify coarse and fine regions without precomputed grid scale data. Larger gradient values indicate finer grids and vice versa. Since computing gradients at every step is costly, we optimize by evaluating them every N steps, significantly improving performance over fixed-step methods. Further, we optimize by computing gradients for adaptive step lengths at a select few points, $\bm{p}$, using Eq.~\ref{eq:pf}. These points are randomly chosen based on a polynomial condition similar to that used in particle agglomeration. By evaluating gradients at these locations, we can assess grid-scale variations and effectively capture adaptive step lengths in finer and coarser-scale grids to enhance performance. See the supplemental material for more details on the optimization of sphere tracing for the multi-scale grid variations.

\section{Multiscale Geometry Reconstruction}
\label{sec:optimization}

In this section, we focus on reconstructing procedural microstructures generated by our framework. We explore two modalities: SDF, potentially extracted from noisy MicroCT scans, and RGB images, potentially obtained from Transmission Electron Microscopy (TEM) or Scanning Electron Microscopy (SEM) samples. For the first modality, we apply direct parameter fitting to find the parameters of our procedural microstructures with observed data. For the second, we employ analysis-by-synthesis, iteratively refining reconstructions by comparing simulated and real electron microscopy images. 

\subsection{Parameter Fitting}
\label{parameter_fitting_approach}

This approach formulates microstructure reconstruction as a model-fitting task, where the goal is to optimize the parameters of a procedural SDF to match the microstructure's SDF ground truth accurately. Given a set of spatial coordinates $\Omega = \{\mathbf{x} \mid \mathbf{x} \in \mathbb{R}^3\}$, a ground-truth SDF $\hat{S}:~\mathbb{R}^3 \rightarrow \mathbb{R}$ describing the surface at the point $x$, a bounded parameter space $\Phi = [a_1, b_1] \times [a_2, b_2] \times \cdots \times [a_n, b_n]$ with $n$ parameters, and a loss function $\mathcal{L}: \mathbb{R} \times \mathbb{R} \rightarrow \mathbb{R}$, the problem is defined as minimizing the discrepancy between the fitted function $f$ and the ground truth $\hat{S}$ across the domain: 
\begin{equation}
\min_{\boldsymbol{\phi} \in \Phi} \int_{\Omega} \mathcal{L}(f(\mathbf{x}, \boldsymbol{\phi}), \hat{S}(\mathbf{x})) d\mathbf{x}.
\label{eq:parameter_fitting}
\end{equation}
Procedural SDFs allow efficient extrapolation of the microstructure beyond the optimization domain without modifying the fitted function. In addition, our compact parametrized microstructures are described with a relatively low number of parameters. We propose a loss function to evaluate the discrepancies in the frequency domain using the discrete Fourier transform (DFT) to consider the periodic behavior common in many microstructures. The loss function compares the amplitude and phase spectra of the Fourier coefficients, using the logarithm of the mean squared error (MSE) as an error metric. This metric is closely related to the spectral density function commonly used in microstructure analysis~\cite{bostanabad2018computational}. 

\subsection{Analysis by Synthesis}
\label{analysis_by_synthesis_approach}

The analysis-by-synthesis approach optimizes the SDF by iteratively rendering it using a non-differentiable pipeline and comparing the results in image space. Given a camera intrinsics $\mathbf{K} \in \mathbb{R}^{3\times 3}$ taken from a set of intrinsic matrices $\mathcal{K}$, the camera extrinsics $\mathbf{E} = [\mathbf{R} \mid \mathbf{t}] \in \mathbb{R}^{3\times 4}$  taken from a set of extrinsic matrices $\mathcal{E}$, a labeling function $\hat{S}:~\mathbb{R}^3 \rightarrow \mathbb{R}$, a bounded parameter space $\Phi = [a_1, b_1] \times [a_2, b_2] \times \cdots \times [a_n, b_n]$ with $n$ parameters, a signed distance fitting function $f: \mathbb{R}^3 \times \Phi \rightarrow \mathbb{R}$, a loss function $\mathcal{L}: \mathbb{R}^{W\times H \times 3} \times \mathbb{R}^{W\times H \times 3} \rightarrow \mathbb{R}$, with $W, H$ being width and the height respectively, a set of functions $\mathcal{A}$, and finally the rendering function $\mathcal{R}:~\mathcal{A}\times \mathcal{K} \times \mathcal{E} \rightarrow \mathbb{R}^{W\times H \times 3}$, we can express the optimization problem as follows:
\begin{equation}
    \min_{\boldsymbol{\phi} \in \Phi} \mathcal{L}\left( \mathcal{R}\left( f\left( \cdot, \boldsymbol{\phi}\right), \mathbf{K}, \mathbf{E}\right), \mathcal{R} ( \hat{S}\left( \cdot \right), \mathbf{K}, \mathbf{E} ) \right).
\label{analysis_by_synthesis_eq}
\end{equation}
$\mathcal{R}$ renders the microstructure bounded by a unit sphere from the given view with the camera parameters using the sphere-tracing algorithm~\cite{hart1996sphere} in the CUDA-based framework OptiX \cite{parker2010optix}. For the loss function, we used the two-dimensional DFT to transform the image signal into the frequency domain. By focusing on the amplitudes of the Fourier coefficients, we capture the key periodic features of the image, which are crucial for the effective global optimization of microstructures. While a multi-view optimization is possible, we opted for a single-view approach to minimize computational overhead.

\subsection{Optimization Algorithms}

Selecting a suitable optimization algorithm is critical for solving unconstraint global optimization problems. The loss used in parameter fitting is differentiable, allowing us to use gradient-based optimization algorithms as the operations involved are differentiable, including the DFT. In this paper, we consider two first-orded methods taking advantage of the gradient information: Basin-Hoppin(BH)~\cite{wales2003energy} and Simplicial Homology Global Optimization (SHGO)\cite{endres2018simplicial} algorithms. The BH algorithm is particularly effective for problems with funnel-like loss functions. Its key hyperparameters include the step size $s$ and the temperature $T$, which define the bounds of the uniformly distributed random perturbations and the acceptance probability for the candidate function, respectively. SHGO global optimization algorithm approximates locally convex subdomains using homology groups. The main hyperparameters for SHGO include the number of iterations  $l$ and the number of samples $m$ for building the complex. The Fourier losses functions are differentiable, but certain procedural SDFs and the rendering function $\mathcal{R}$ are non-differentiable, leading to non-differentiability of the final composite function. We consider two gradient-free optimization algorithms: Covariance Matrix Adaptation Evolutionary Strategy (CMA-ES) algorithm~\cite{hansen2019pycma} and modified Powell method~\cite{press1988numerical, powell1964efficient}. The CMA-ES algorithm is highly effective, particularly for rugged, noisy, and non-differentiable problems. Its key hyperparameters are the population size $\mathcal{P}$ and the initial $\sigma_0$ of the covariance matrix, typically chosen depending on the complexity and dimensional of the problem. The modified Powell method \cite{press1988numerical, powell1964efficient} is a local optimizer without hyperparameters that can be effective when a reasonably close initialization is provided.\\
\section{ Results}

We implemented our framework using NVIDIA OptiX (v.7)~\cite{parker2010optix}, and most of the experiments were executed on a NVIDIA RTX 4090 GPU.

\subsection{Multiscale Geometry and Appearance}

\begin{figure*}[tb]
 \centering
  \begin{tabular}{@{}c@{}c@{}c@{}}
  
  \begin{tikzpicture}
    \node[anchor=south west,inner sep=0] at (0,0) {\includegraphics[width=0.33\linewidth]{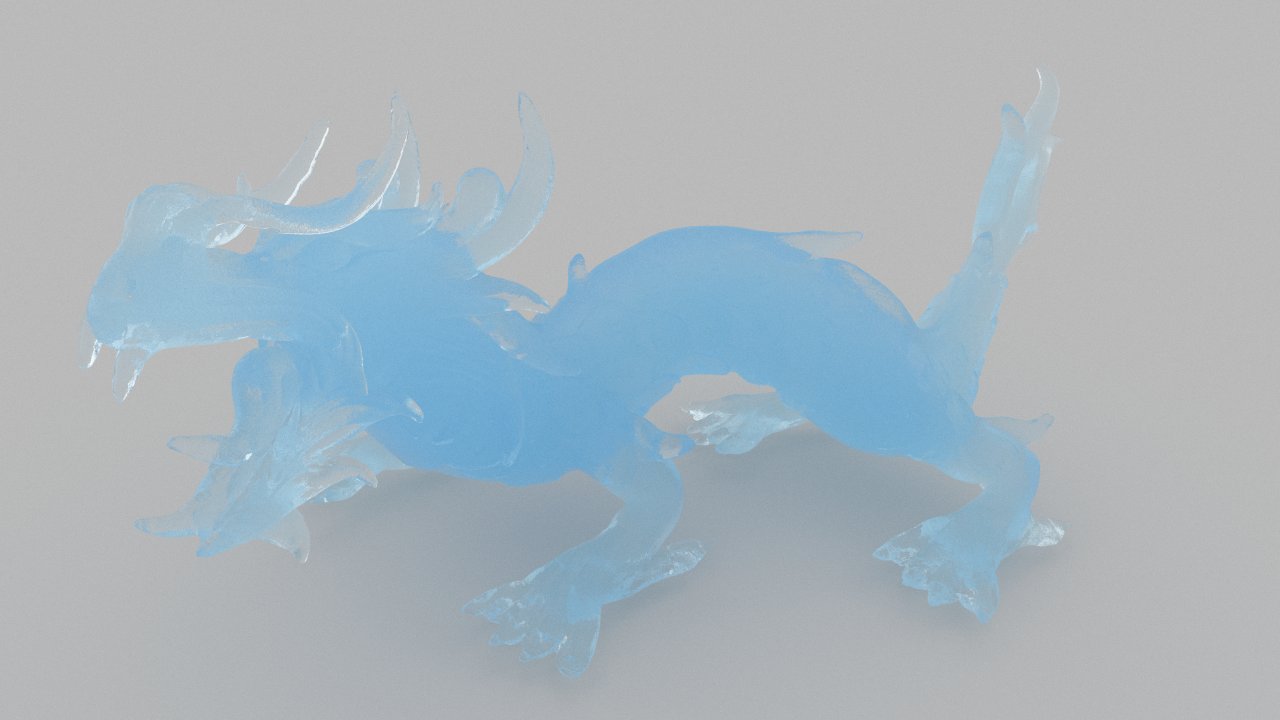}};
    \node[anchor=north east,inner sep=0] at (1.0,0.8) {\includegraphics[width=0.0512\linewidth]{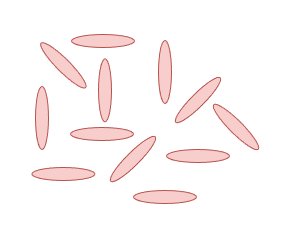}};
  \end{tikzpicture} &

  \begin{tikzpicture}
    \node[anchor=south west,inner sep=0] at (0,0) {\includegraphics[width=0.33\linewidth]{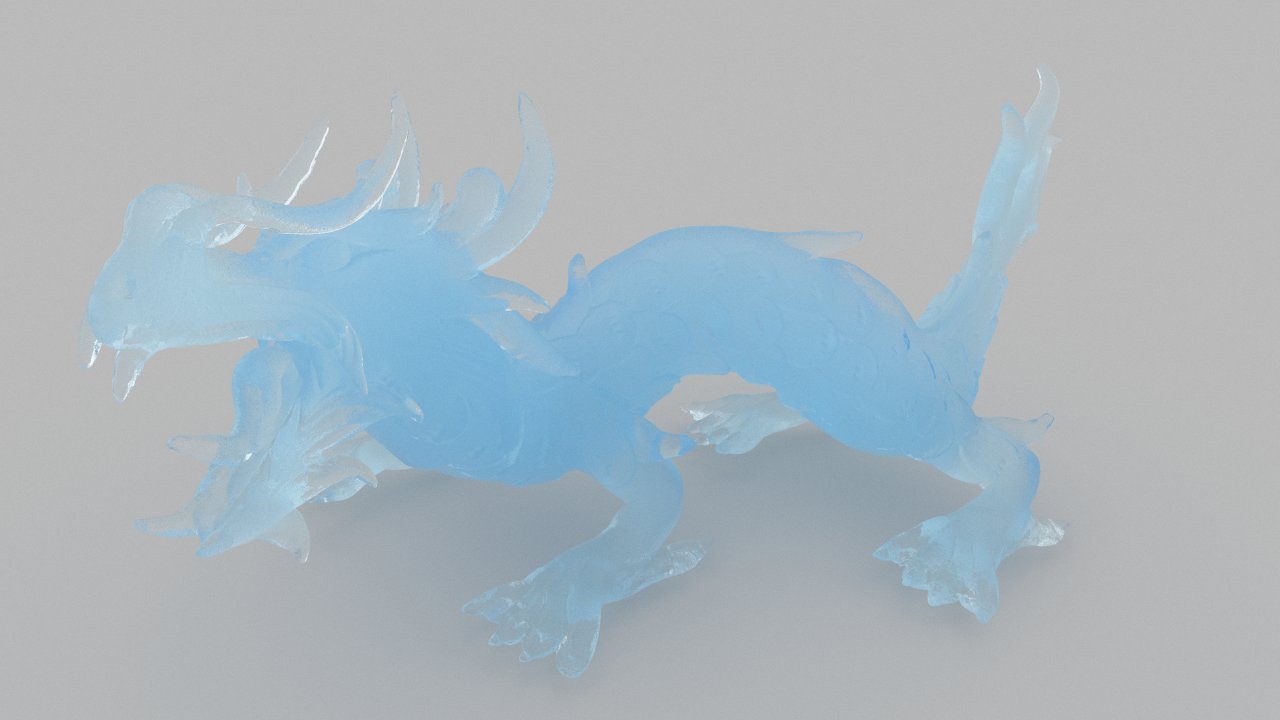}};
    \node[anchor=north east,inner sep=0] at (1.0,0.8) {\includegraphics[width=0.0512\linewidth]{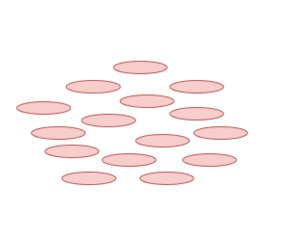}};
  \end{tikzpicture} &

  \begin{tikzpicture}
    \node[anchor=south west,inner sep=0] at (0,0) {\includegraphics[width=0.33\linewidth]{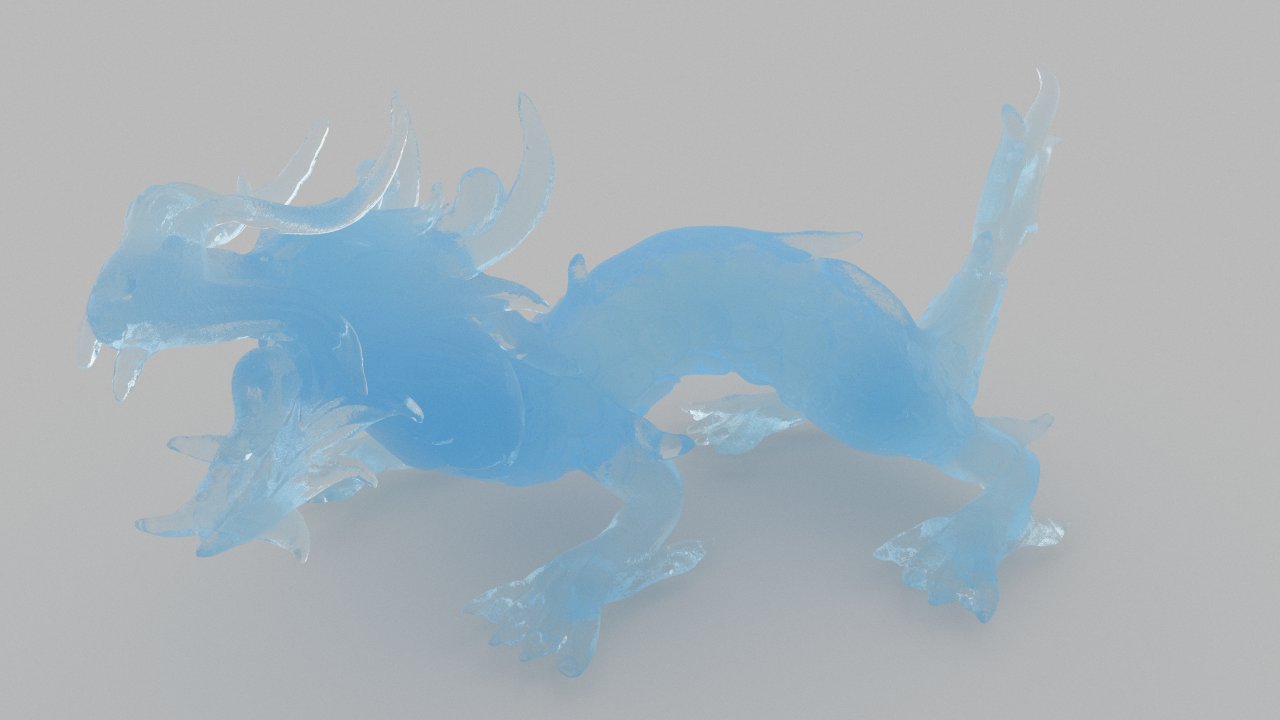}};
    \node[anchor=north east,inner sep=0] at (1.0,0.8) {\includegraphics[width=0.0512\linewidth]{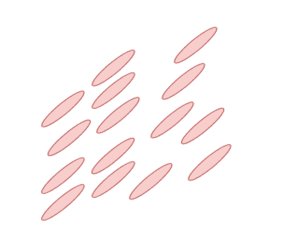}};
  \end{tikzpicture} \\[-1ex] 
   \begin{tikzpicture}
            \node[anchor=south west, inner sep=0] (image) at (0,0) {\includegraphics[width=.33\linewidth]{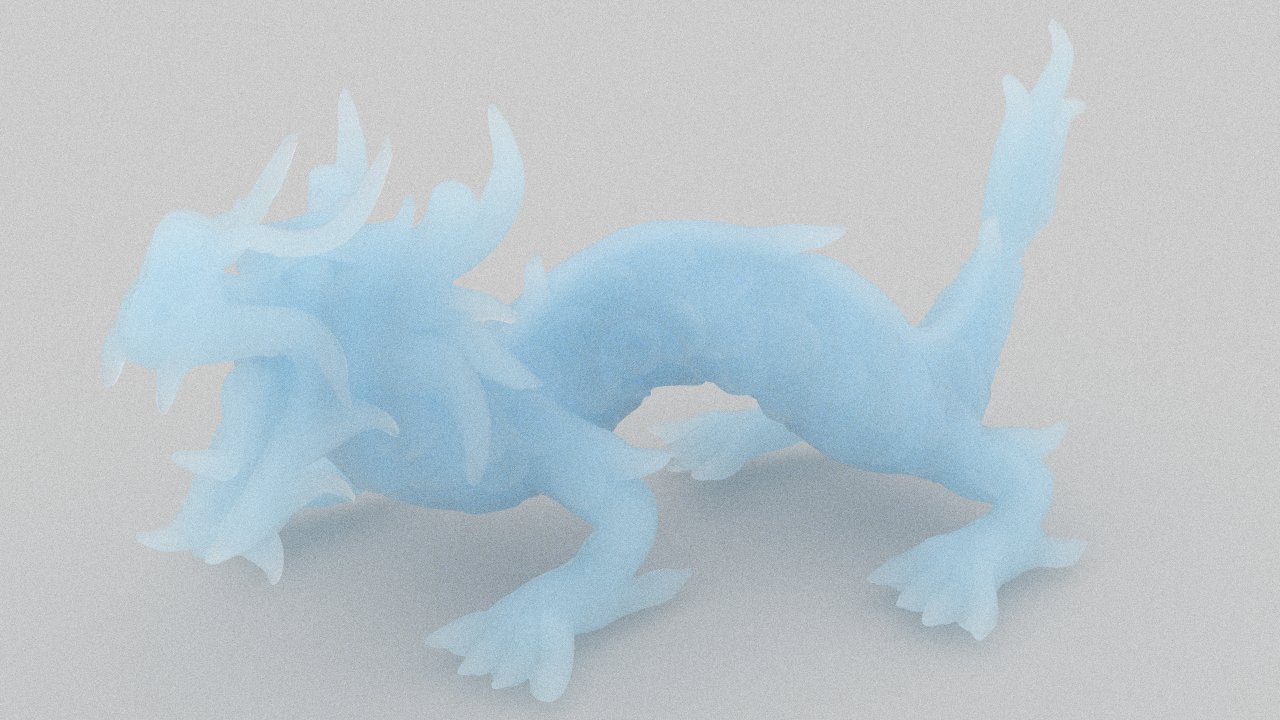}};
            \node[anchor=north east,inner sep=0] at (1.0,0.8) {\includegraphics[width=0.050\linewidth]{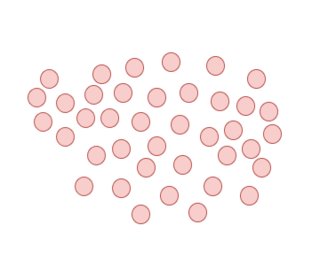}};
        \end{tikzpicture} &

        \begin{tikzpicture}
            \node[anchor=south west, inner sep=0] (image) at (0,0) {\includegraphics[width=.33\linewidth]{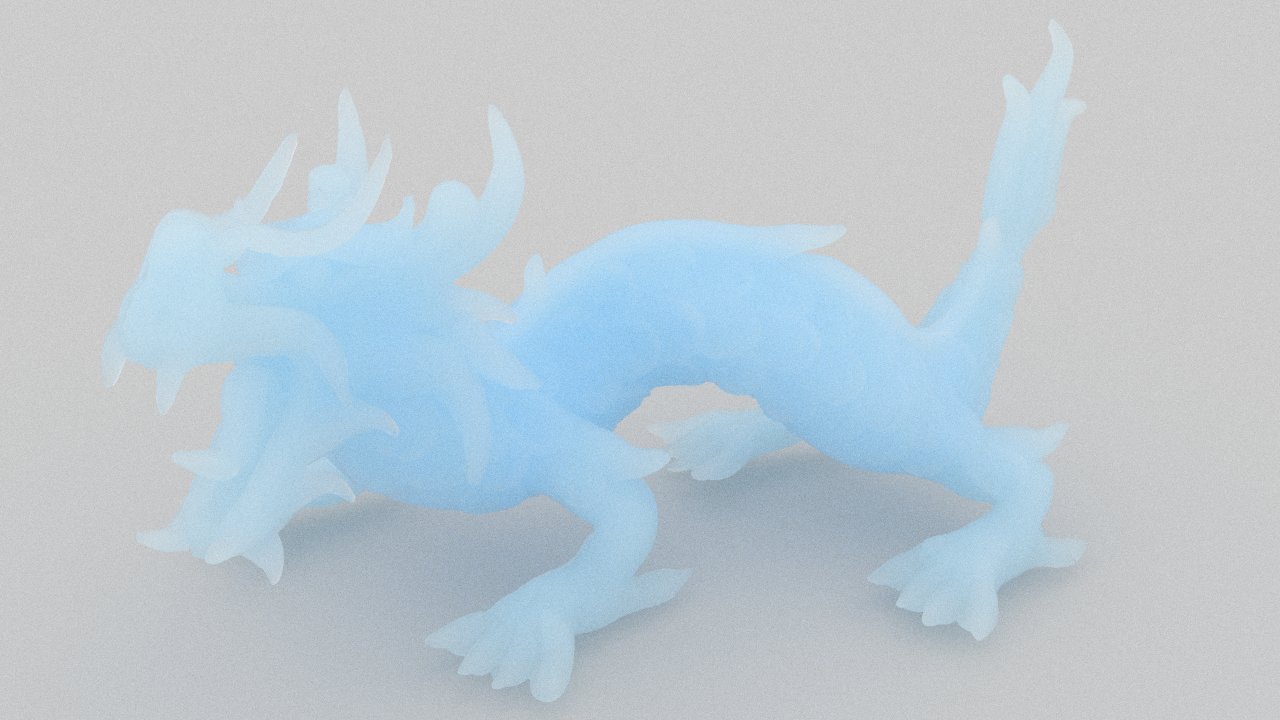}};
            \node[anchor=north east,inner sep=0] at (1.0,0.8) {\includegraphics[width=0.050\linewidth]{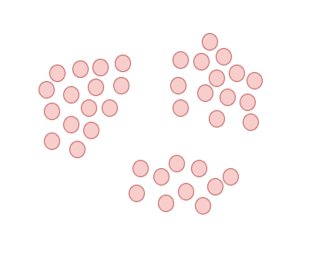}};
        \end{tikzpicture} &

        \begin{tikzpicture}
            \node[anchor=south west, inner sep=0] (image) at (0,0) {\includegraphics[width=.33\linewidth]{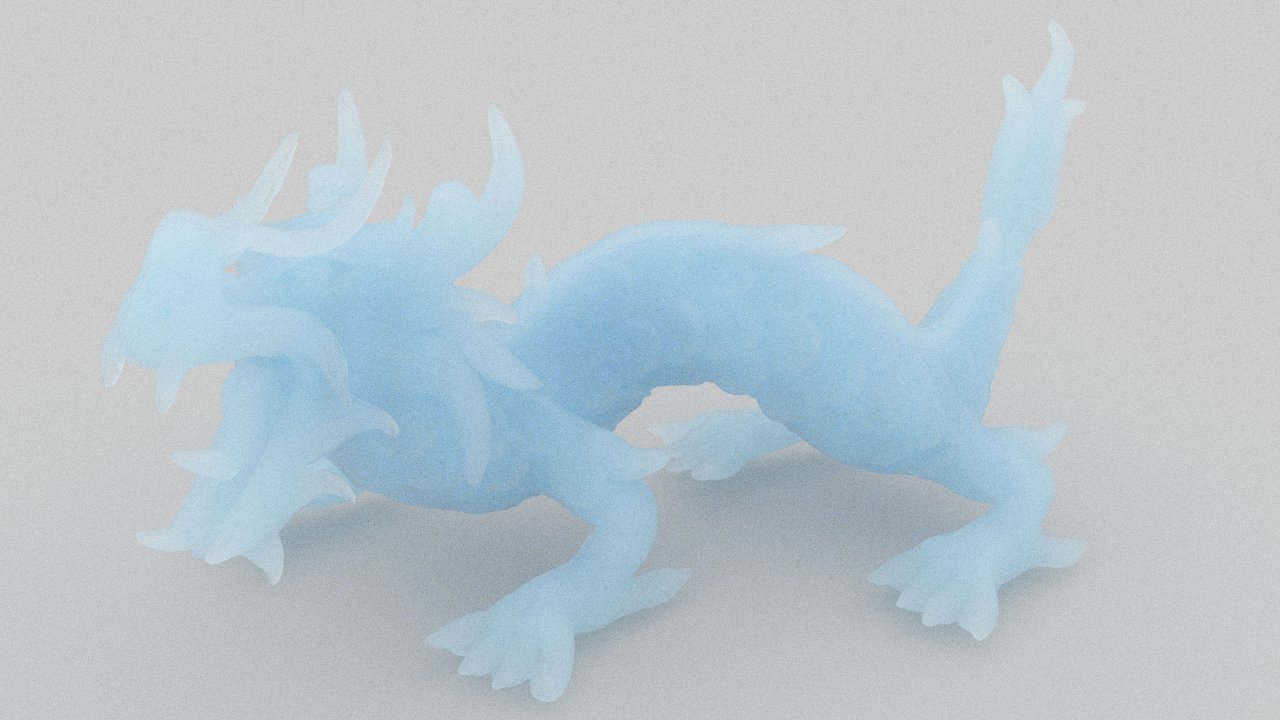}};
            \node[anchor=north east,inner sep=0] at (1.0,0.8) {\includegraphics[width=0.050\linewidth]{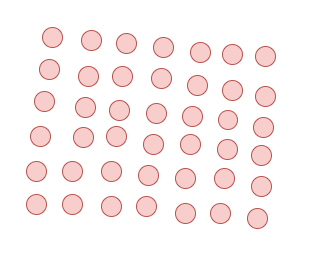}};
        \end{tikzpicture} \\[-1ex]


  \end{tabular}
\caption{Top row: Macroscopic material appearance changes due to aligned ellipsoidal particles, leading to material anisotropy. From left to right, the materials containing particle distributions are isotropic (random orientation) and anisotropic in the $x$ and $z$ directions. Light transport varies with material anisotropy, with consistent particle number density across cases. Bottom row: The spatial correlation of particles influences macroscopic visual appearance. On the left, a random distribution of air bubbles in ice shows no spatial correlation. In the middle, a positive spatial correlation results in clustered particles. On the right, a negative spatial correlation yields a regular distribution. The particle number density is unchanged in all cases. This study builds on Jarabo et al.~\cite{jarabo2018radiative} and uses fully defined particle geometry for rendering. }
\label{fig:anisotopic_ice}
\end{figure*}

Different types of implicit functions are explored to generate microstructures on the fly without the need for pre-computation. The performance evaluation of fibrous materials using three different methods is presented in Table~\ref{tab:comparison_eta}. The results suggest that periodic functions can synthesize microstructures quickly and that particle agglomeration is slower because it requires evaluating the polynomial for each discretized grid cell. Table~\ref{tab:step_length_performance} presents performance details for modeling multiscale geometry using fixed, bijection, and adaptive step lengths in sphere tracing.
\begin{figure}
    \centering
     \begin{tabular}{@{}c@{}c@{}c@{}}
    \includegraphics[width=0.5\linewidth]{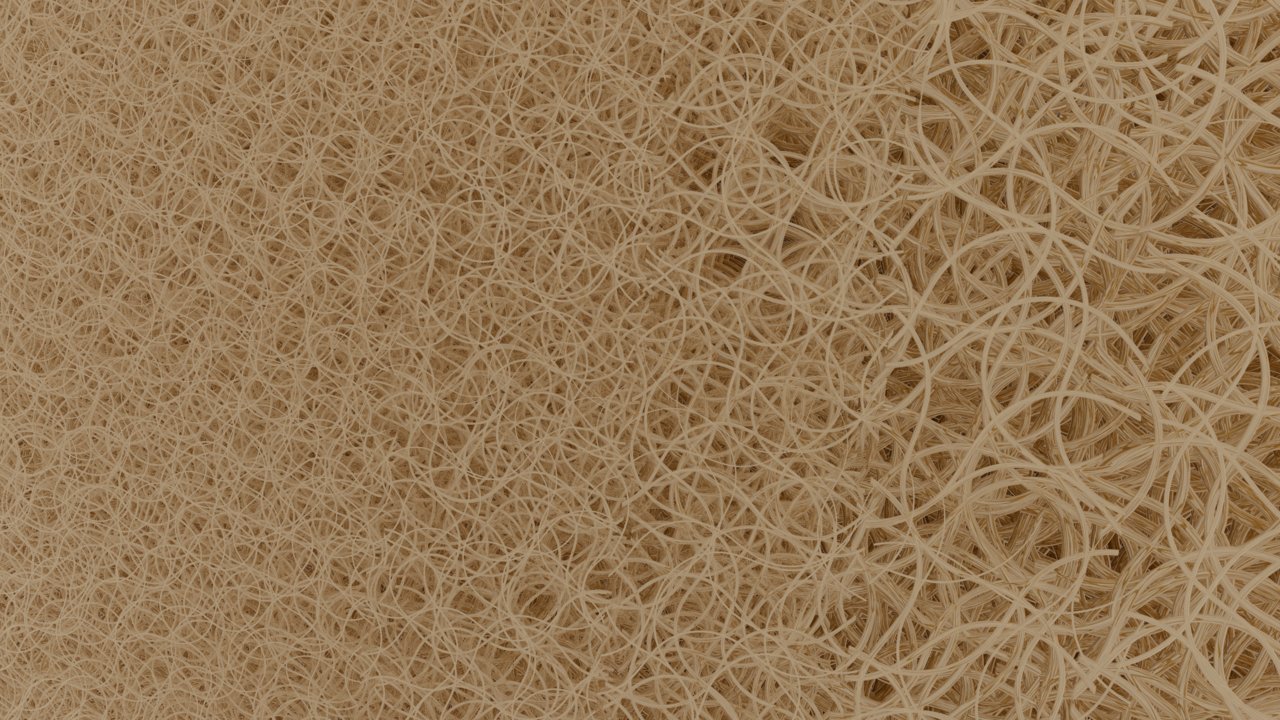}
    \includegraphics[width=0.5\linewidth]{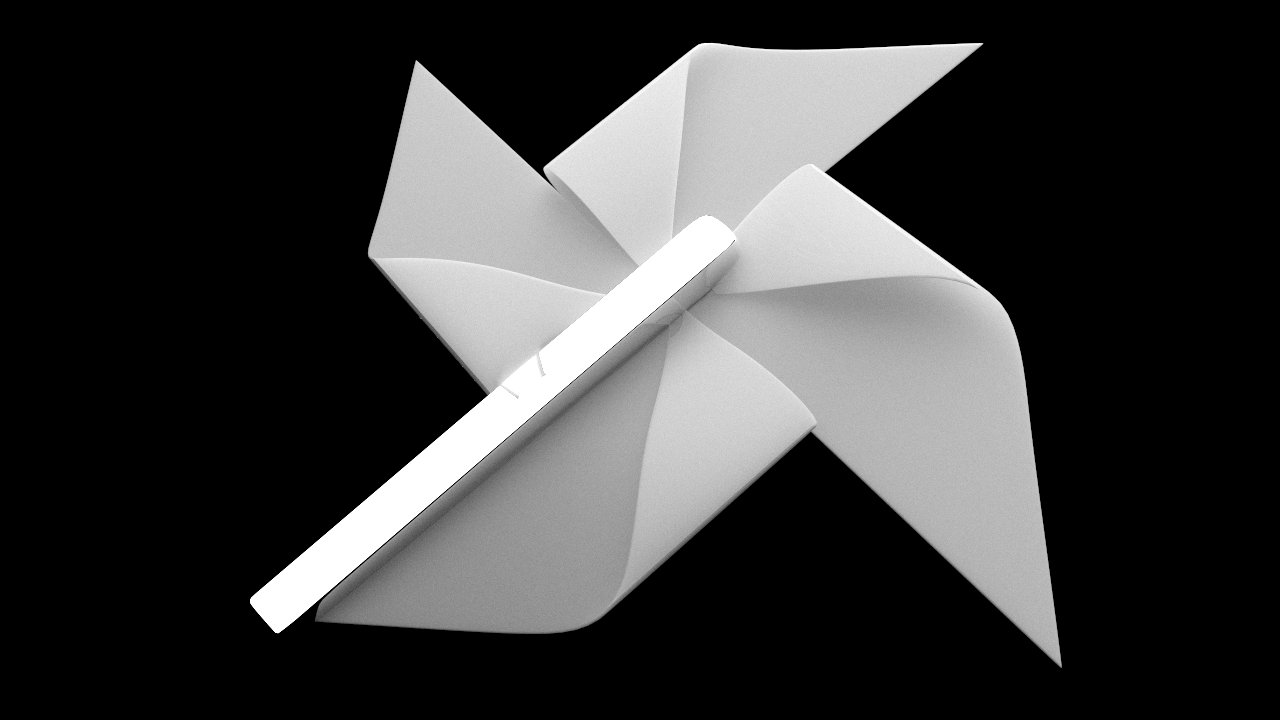}

    \end{tabular}
    
  \caption{Appearance of paper: Chaotic fibers (with gold material) (left) are generated using the gyroid function. Even on the left one, we see the smooth transition between multiple scales. This demonstrates an application where a paper spinner (right) is rendered using fibrous microgeometry with cellulose material. The fibrous microgeometry exists in parallel layers, similar to the real paper microstructure. Paper spinner OBJ file is available at \href{https://www.turbosquid.com/Search/Index.cfm?keyword=paper+spinner&media_typeid=2}{https://www.turbosquid.com/} }
  \label{fig:fibers-gyroid}
\end{figure}

\begin{figure*}
    \centering
    \begin{tabular}{@{}c@{}c@{}c@{}c@{}}
    \includegraphics[width=0.25\linewidth]{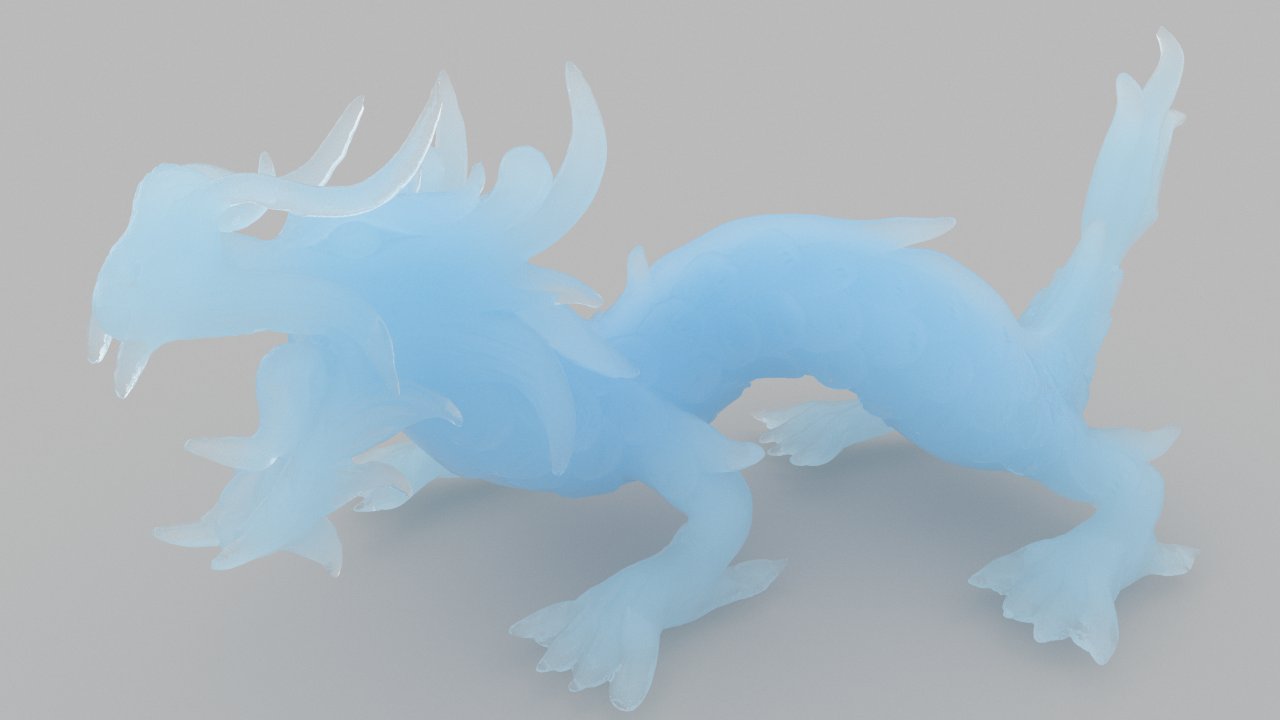}&
    \includegraphics[width=0.25\linewidth]{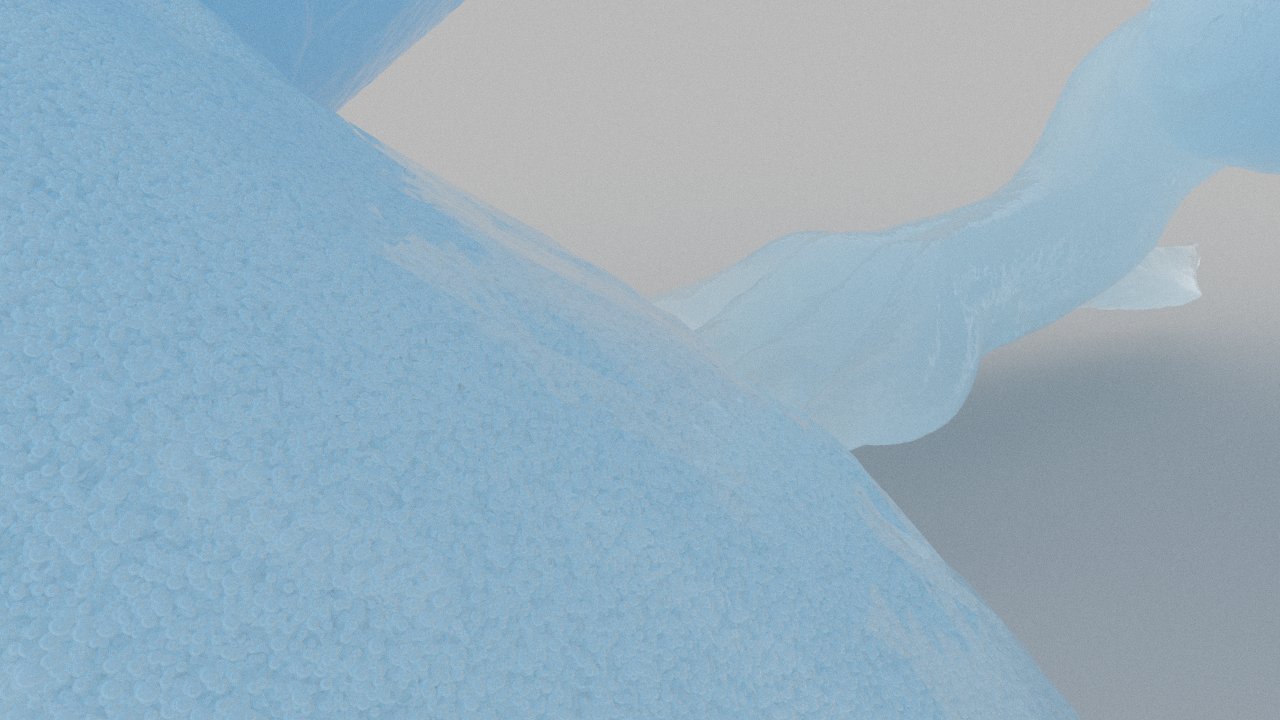}&
    \includegraphics[width=0.25\linewidth]{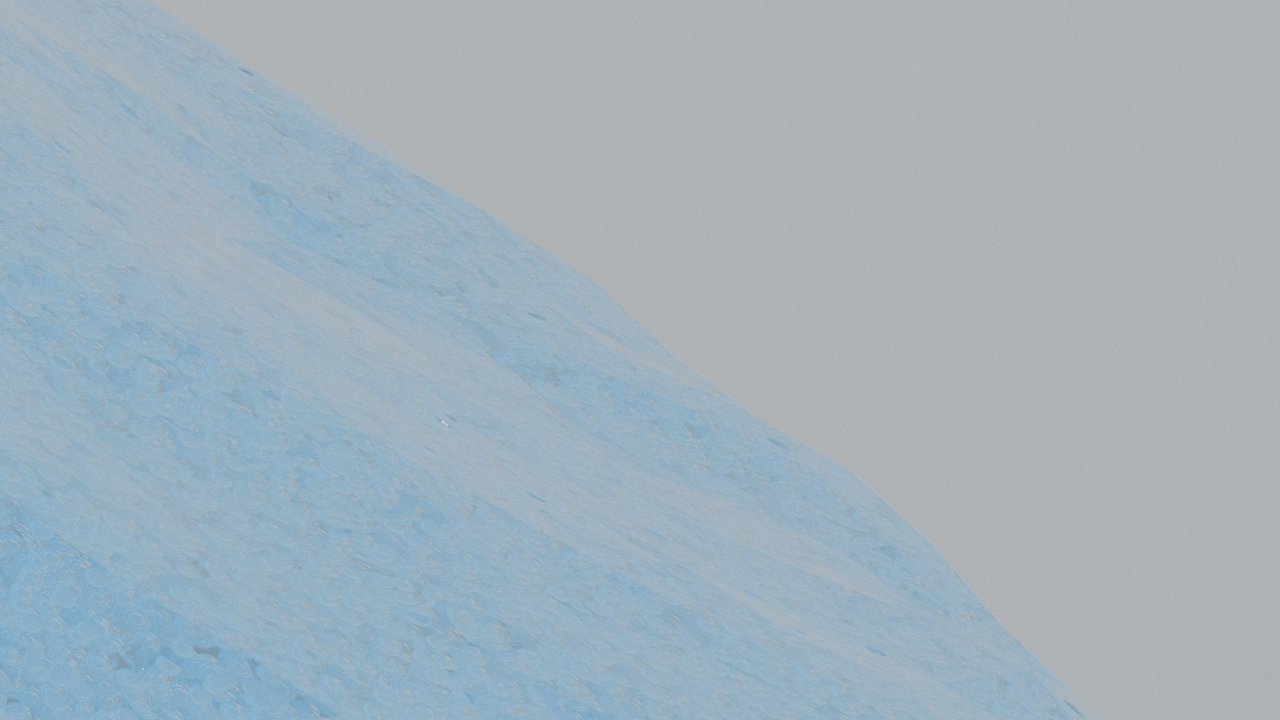}&
    \includegraphics[width=0.25\linewidth]{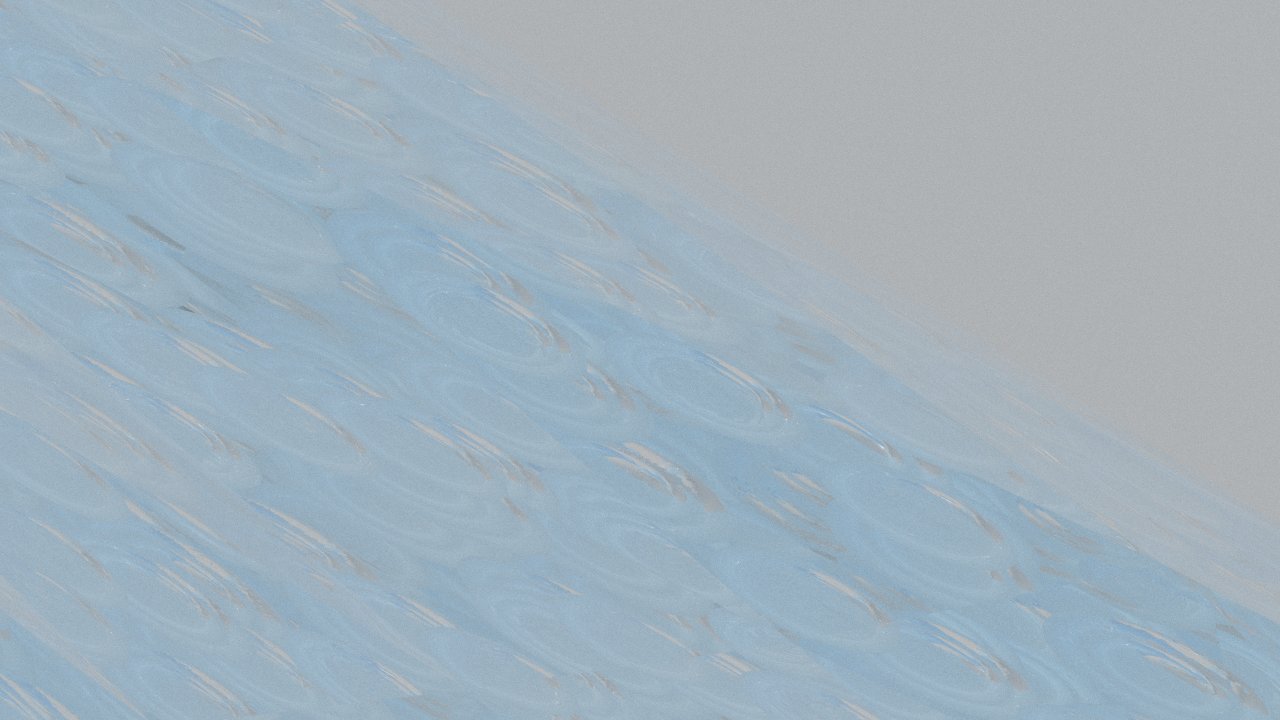} \\ [-0.9ex]
    \includegraphics[width=0.25\linewidth]{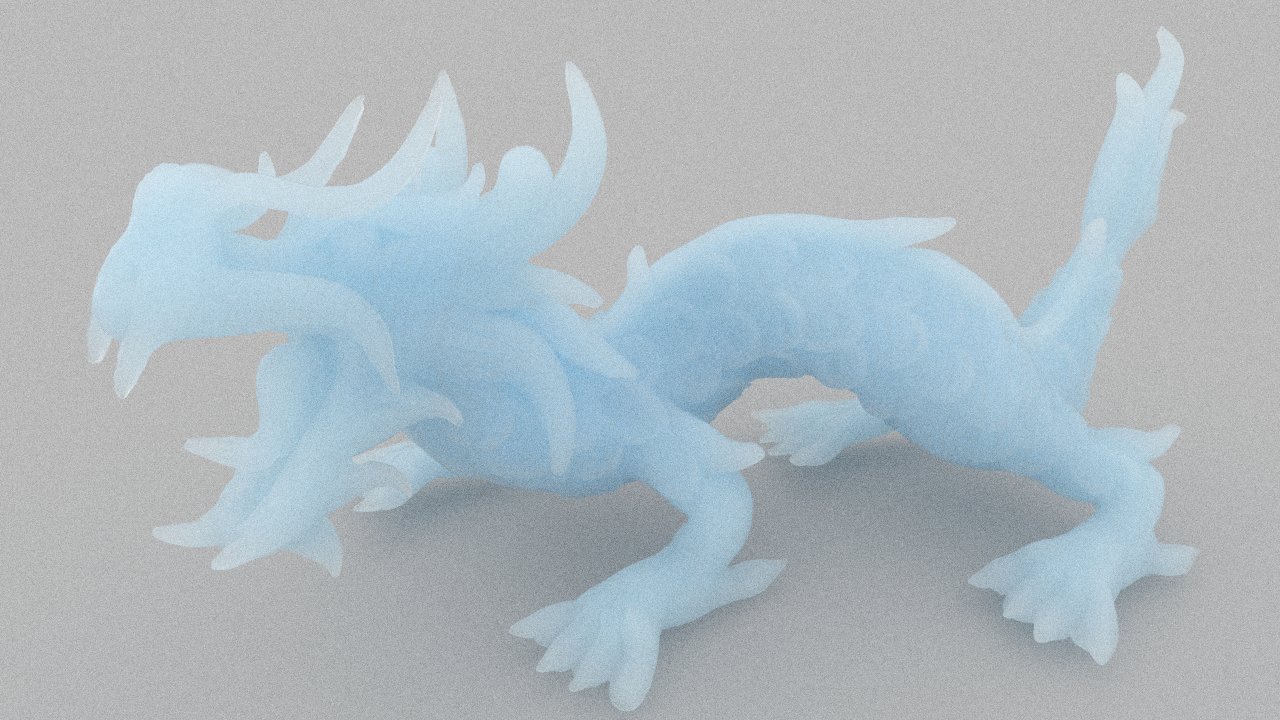}&
    \includegraphics[width=0.25\linewidth]{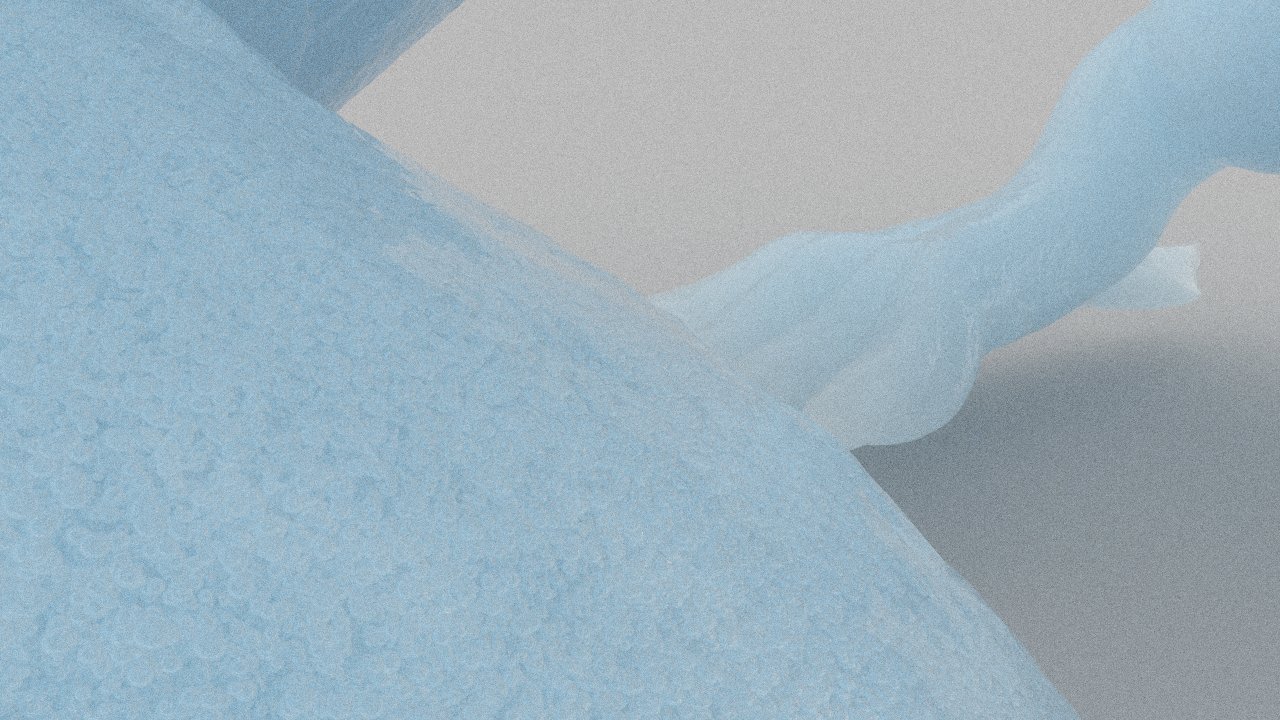}&
    \includegraphics[width=0.25\linewidth]{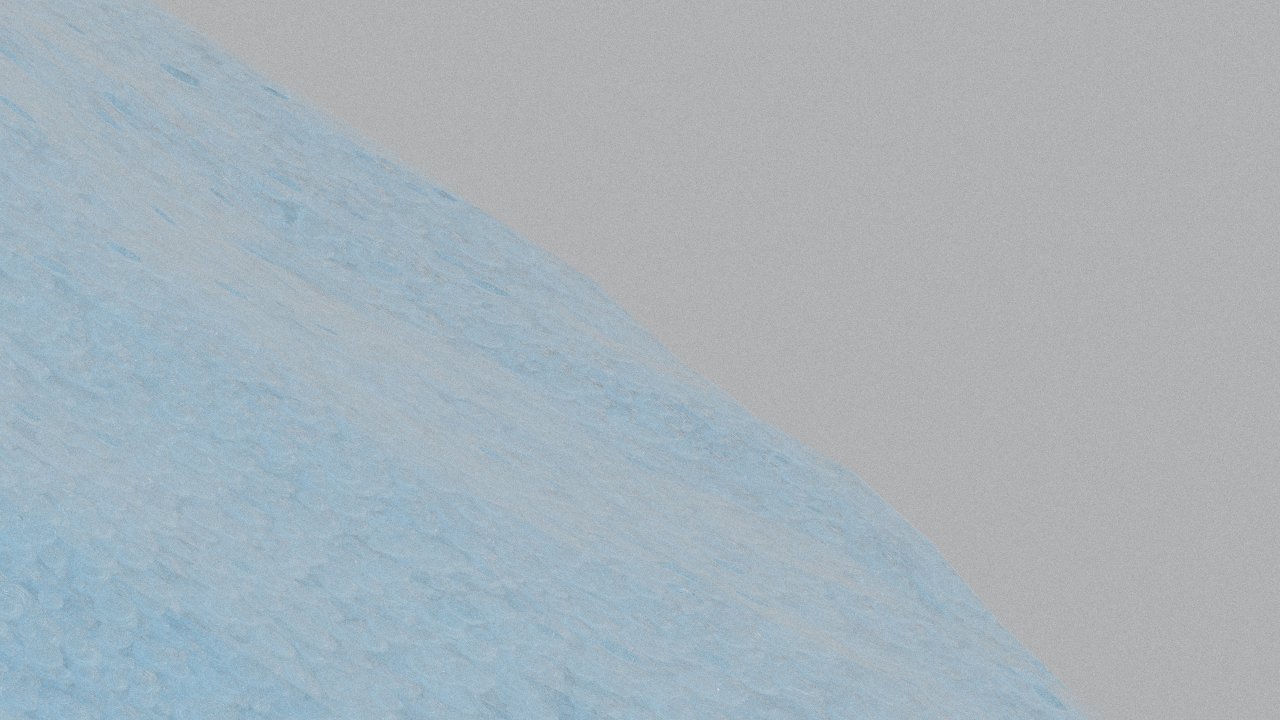}&
    \includegraphics[width=0.25\linewidth]{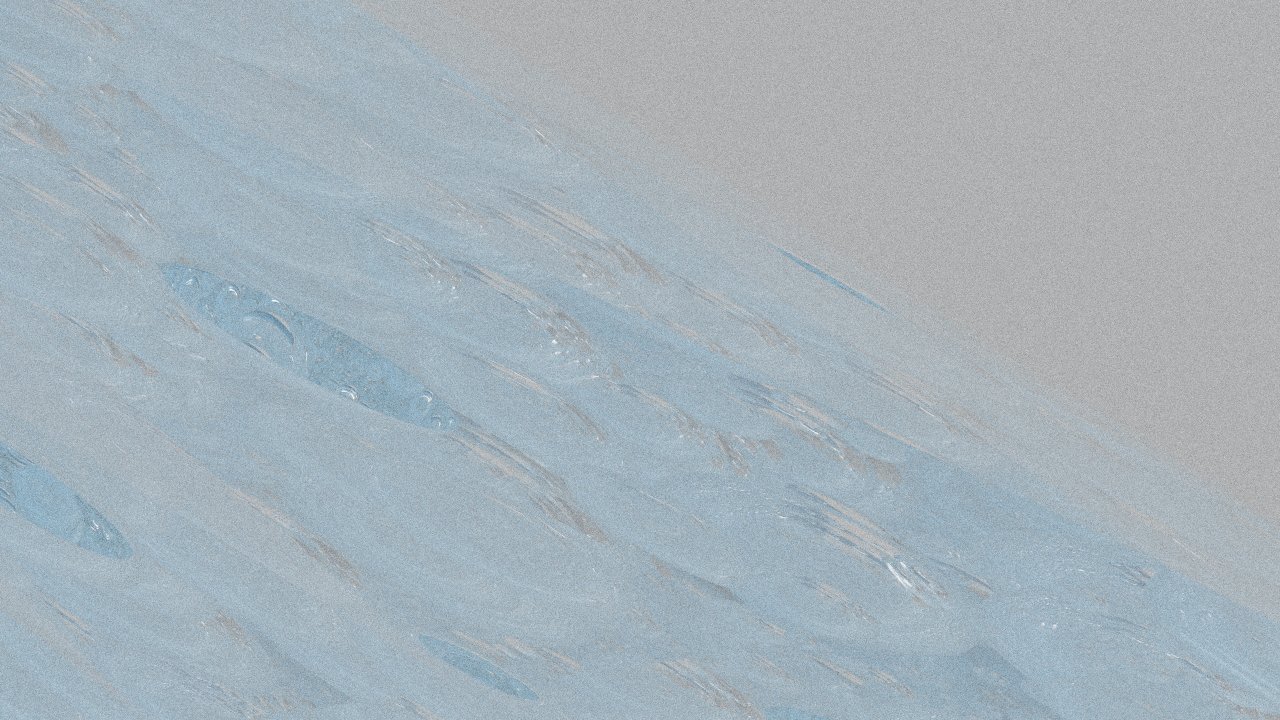} \\
    \end{tabular}
  \caption{The first row shows renderings of particulate material with a particle number density of $10^6$ air particles per cubic meter embedded in ice. The particles have a radius of 1 mm. This results in an air bubble volume fraction of 0.00419. The particulate material in the second row has air bubbles embedded in ice with a volume fraction of 0.0419, and the size of the air particles is normally distributed with a mean of 1 mm and a standard deviation of 0.1 mm. It shows that our method effectively generates both mesoscopic surface and volume rendering effects simultaneously, whereas other
methods are typically limited to only one of them. Each row shows the zoom-in images of the material, showing that we can move the camera close to the material to capture surface-level glittering effects. }
  \label{fig:particle-dist}
\end{figure*}

Some subsurface scattering effects require an additional effort in order to be rendered using volume path tracing. The radiative transfer framework needs an update to capture material anisotropy~\cite{jakob2010radiative}, spatial correlation of particles~\cite{jarabo2018radiative}, or near-surface particles~\cite{mudaliar2013multiple}. With our representation of the individual particles, we can first represent such complications in a geometric optics approximation. Fig.~\ref{fig:anisotopic_ice} shows the presentation of material anisotropy and its influence on the macroscopic appearance of the material rial using ellipsoidal particles oriented in specific directions. Fig.~\ref{fig:anisotopic_ice} presents an example of the correlation between particles with appearance effects similar to those observed by Jarabo et al.~\cite{jarabo2018radiative}. See Fig.~\ref{fig:particle-dist} for the particle distributions based on fixed and normal distributions on particle sizes. Since our method provides direct control of the particles in a material, we can more easily control and render the spatial correlation between particles, as well as many other properties of the particle distribution in a particulate material. In addition, we can see the particles in a close-up. If real ice is observed at a short distance, we can inspect the particles and use our method to mimic their shapes. An example is provided in Fig.~\ref{fig:ice_multiphase}. Inspired by photographed air bubbles in ice, we use the multiphase particle cloud concept to model a digital version of similar air particles in ice that exhibit spatial variations in shape, size, and density. See the supplemental material for more details about the modeling of multiphase particles and spatial correlations in particles.



On-the-fly granular media modeling is definitely a great asset to the procedural modeling toolbox. In this paper, we present rock piles as an example, where we control various parameters of rock piles, such as concavity and multiscale transitions between grains. Fig.~\ref{fig:rockpiles-convex_concave} presents rock piles that demonstrate a continuous transition from convex to concave grains within the same cloud. 

Implicit periodic functions combined with affine transformations can be useful for synthesizing various geometric structures, such as particles, fibers, laminar structures, and porous materials, with faster evaluation. Fig.~\ref{fig:fibers-gyroid} shows that the fibrous geometry is modeled using the gyroid pattern, and the appearance of the paper is molded using fibrous microgeometry. The fibers in the paper are arranged in parallel layers, similar to the microgeometry of the real paper. See the supplemental material for details of the correspondence and mapping between discrete grids and implicit periodic functions with Taylor series expansion. We refer the reader to the supplemental for more details.


\begin{table}
\centering
\resizebox{0.48\textwidth}{!}{
\begin{tabular}{|c|c c|c c|c c|}
\hline
$\null$ &\multicolumn{2}{c|}{Periodic functions} & \multicolumn{2}{c|}{Direct primitive} & \multicolumn{2}{c|}{Particle agglomeration} \\ \hline
  $Grid\ noise\ scale$  & RT (ms) & FPS & RT(ms) & FPS & RT (ms) & FPS \\ \hline
10.0  &  2.6 & 365.6   & 4.1 & 177.6 & 51.7 & 19.8   \\ \hline
50.0  & 2.6 & 364.2     & 5.9 & 165.4  & 94.9 & 10.7   \\ \hline
100.0 & 2.6 & 362.5    & 6.5 & 150.4  & 137.2 & 7.3   \\ \hline
200.0 & 2.6 & 360.4  & 7.1 & 138.1 & 157.9 & 6.2    \\ \hline
\end{tabular}
}
\caption{The performance of the different methods listed in the table, where the volume with fibrous microstructure is generated using three methods: periodic functions, direct cylindrical primitive, and particle agglomeration methods. It shows that microgeometry synthesis using periodic functions is fast because it requires no evaluation in the 27-neighborhood grid cells of the point $\bm{p}$, while other methods need that. Interestingly, the evaluation time is constant for the periodic functions method, even when increasing the amount of microgeometry required to generate the volume.}
\label{tab:comparison_eta}
\end{table}

\begin{table}[h]
    \centering
    \resizebox{0.24\textwidth}{!}{
    \begin{tabular}{|c|c|c|}
        \hline
        \textbf{Step Length} & \textbf{RT (ms)} & \textbf{FPS} \\ \hline
        Fixed & 9.3 & 105.7 \\ \hline
        Bijection & 9.0 & 108.5 \\ \hline
        Fully Adaptive & 44.1 & 22.6 \\ \hline
        Adaptive (N 10) & 11.5 & 85.8 \\ \hline
        Adaptive (N 100) & 8.1 & 121.1 \\ \hline
        Adaptive (N 500) & 8.1 & 121.5 \\ \hline
        Bijection + Adaptive (N 500) & 8.0 & 122.0 \\ \hline
        Adaptive ($D(\bm{p}) = 1$) & 7.8 & 124.5 \\ \hline
    \end{tabular}
    }
    \caption{We compare sphere tracing step lengths for multiscale particulate media, as shown in Fig.~\ref{fig:svbunny}. Macroscopic scales require longer steps, microscopic scales require shorter steps, and mesoscopic scales require intermediate steps for optimal performance. We evaluate fixed, bijection-based, fully adaptive, and adaptive stepping with gradient checks every N steps (10, 100, 500), adaptive stepping with bijection, and adaptive stepping with gradient checks for step lengths based on polynomials. Fully adaptive stepping is less efficient, while increasing the N value in adaptive step-length methods improves performance. Adaptive stepping with polynomial checks outperforms fixed and bijection-based step lengths, enhancing rendering efficiency by 16\%. The experiment used 1,500 steps for Lambertian particles. }
    \label{tab:step_length_performance}
\end{table}

\subsection{Light Path Analysis}

\begin{figure}[ht]
    \centering
    \includegraphics[width=\linewidth]{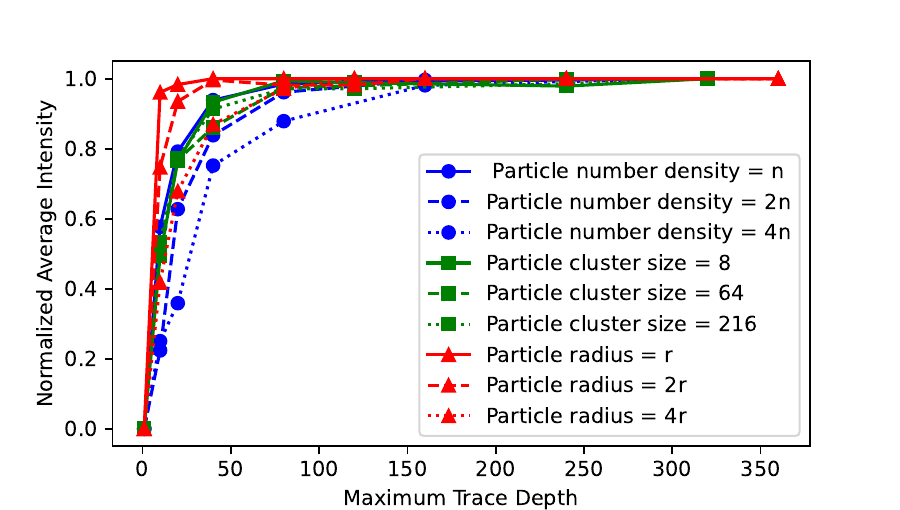}
    \caption{Since we return black when the maximum trace depth is reached, any loss of average intensity in the rendered image indicates an error due to a small maximum depth. These plots show how the normalized average intensity converges as the trace depth increases. We test this on semi-infinite particle clouds under varying conditions: Blue curves indicate particle number densities, green curves denote cluster sizes, and red curves reflect particle radii. The vertical axis is normalized for clear trend comparison. All experiments use air bubbles in ice as the material.}
    \label{fig:clustersizeintensity}
\end{figure}

\begin{figure}[ht]
 \includegraphics[width=\linewidth]{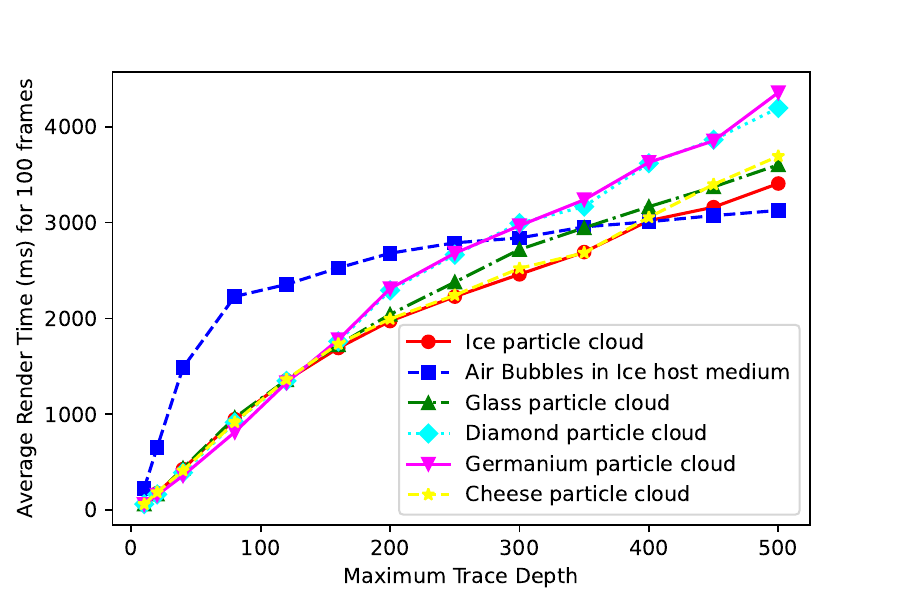}
    \caption{Rendering time plot for different materials against different maximum trace depths, where all the experiments are with 500 samples per pixel, and the rendering times are shown for the 100 frames for each material. It shows that an increase in trace depth leads to a logarithmic increase in rendering time for most of the materials. Here the density of the particles per cubic meter is $10^6$, and the radius is 1\,mm.}
  \label{fig:tracedepth-render-time}
\end{figure}

Due to the very large number of surfaces in our particulate media, the maximum trace depth, that is, the number of surface bounces along a path that the renderer allows, is an important parameter. The plots in Fig.~\ref{fig:clustersizeintensity} illustrate that our renderings converge for different parameters like size and number density around a maximum trace depth of 200. A larger maximum trace depth comes with an impact on rendering performance. In Fig.~\ref{fig:tracedepth-render-time}, we see that an increase in the maximum trace depth for most of the materials increases the rendering time, but the trend is more logarithmic in nature than linear.

\subsection{Reconstruction Results}

We evaluated the performance of the optimization approaches and algorithms introduced in Section \ref{sec:optimization} on several synthetic microstructures created with our framework as a proof of concept. As we create the synthetic dataset, we have access to the ground truth SDF, the RGB image, and the optimal parameters. For the parameter fitting, we sample an oriented point cloud by multiplying the 3D points with a rotation matrix, adding a small perturbation using a Gaussian noise, and evaluating the given SDF. To generate the RGB image for the analysis-by-synthesis approach, we render the microstructure using the $\mathcal{R}\Bigl(\hat{S}(\cdot), \mathbf{K}, \mathbf{E} \Bigr)$ with a fixed $\mathbf{K}, \mathbf{E}$ later used for the optimization. For more details on the synthetic dataset, the optimization procedure, and additional results, we refer the reader to the supplemental material.\\ 



\begin{figure}[ht]
    \begin{adjustbox}{max width=\linewidth, max height=\textheight}
    \setlength{\tabcolsep}{0.5pt}
    \begin{tabular}{c c c c c c c}
        & \tiny Reference & \tiny SHGO-SDF & \tiny BH-SDF & \tiny CMA-ES-SDF & \tiny CMA-ES-S & \tiny Powell-S \\ 
         \multirow{2}{*}{{\rotatebox{90}{ \tiny Gyroid [1D]}}} & \includegraphics[width=0.08\textwidth]{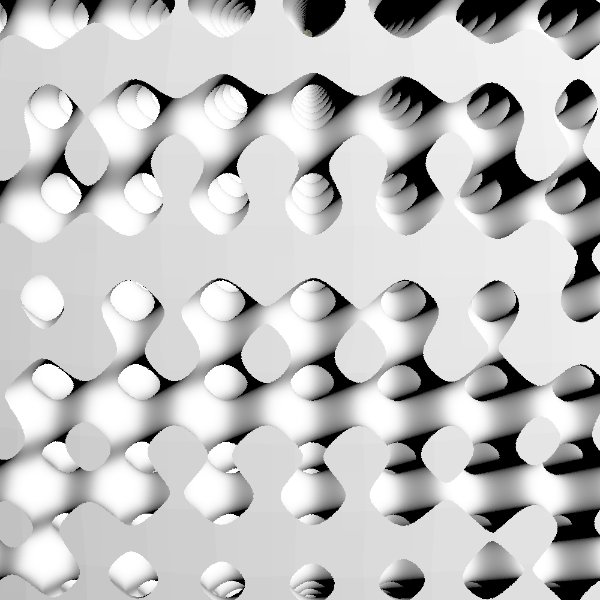} & 
        \includegraphics[width=0.08\textwidth]{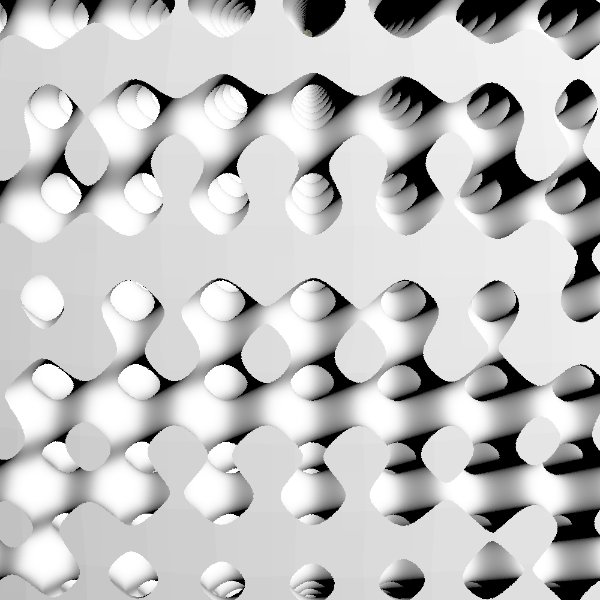} &
        \includegraphics[width=0.08\textwidth]{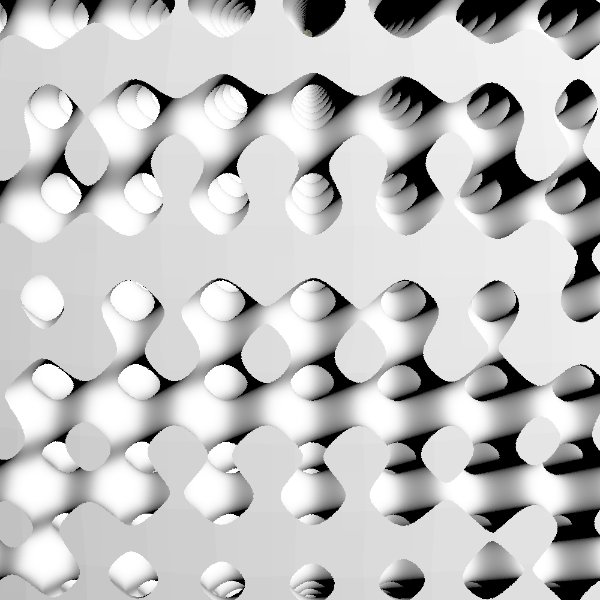} &
        \includegraphics[width=0.08\textwidth]{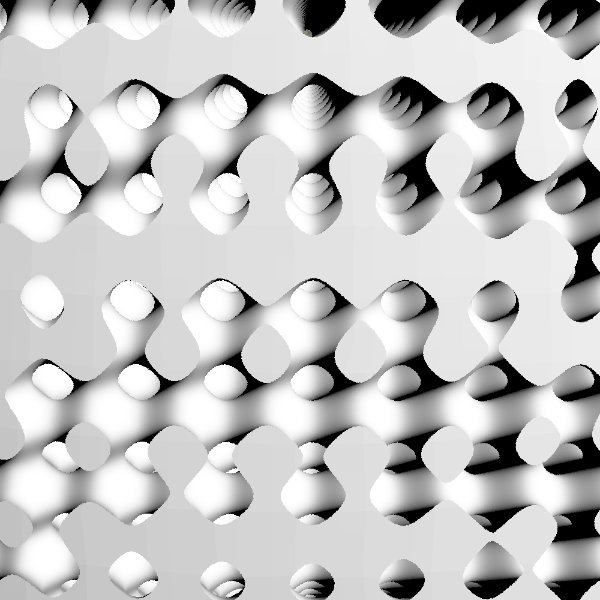} &
        \includegraphics[width=0.08\textwidth]{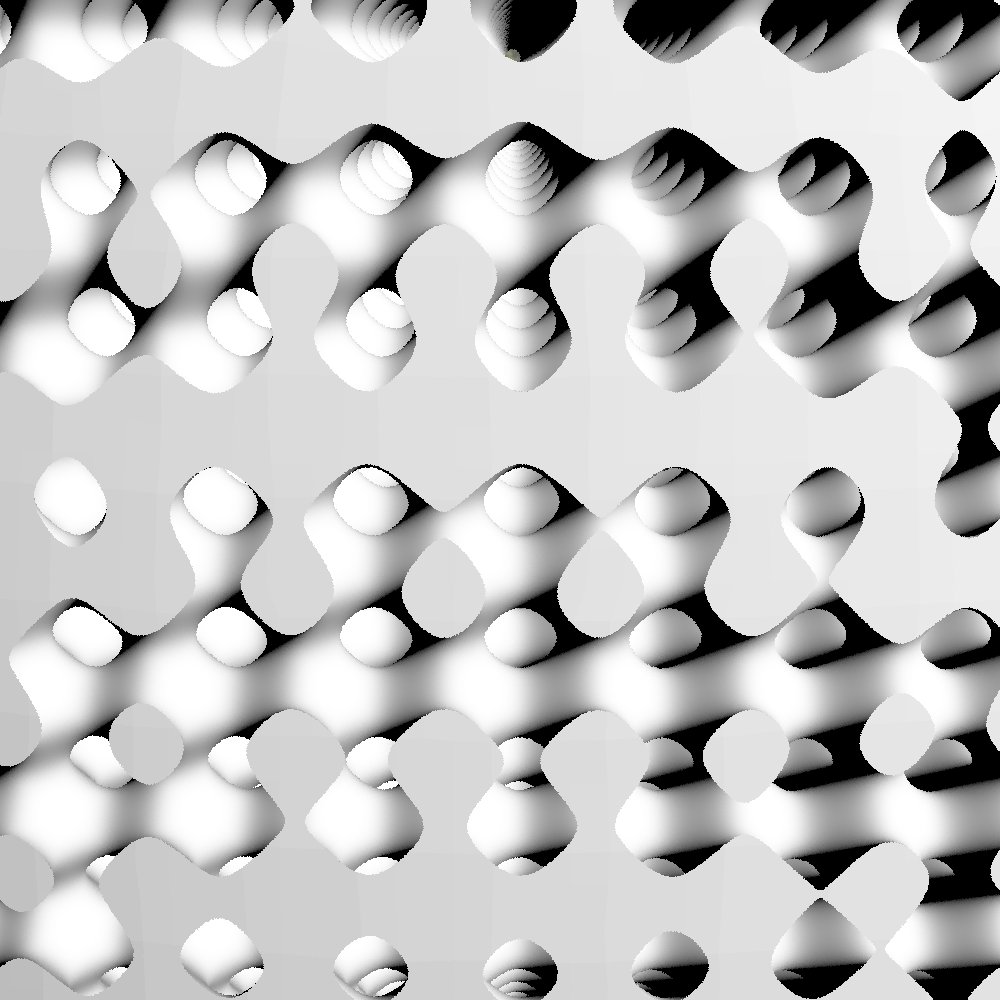} &
        \includegraphics[width=0.08\textwidth]{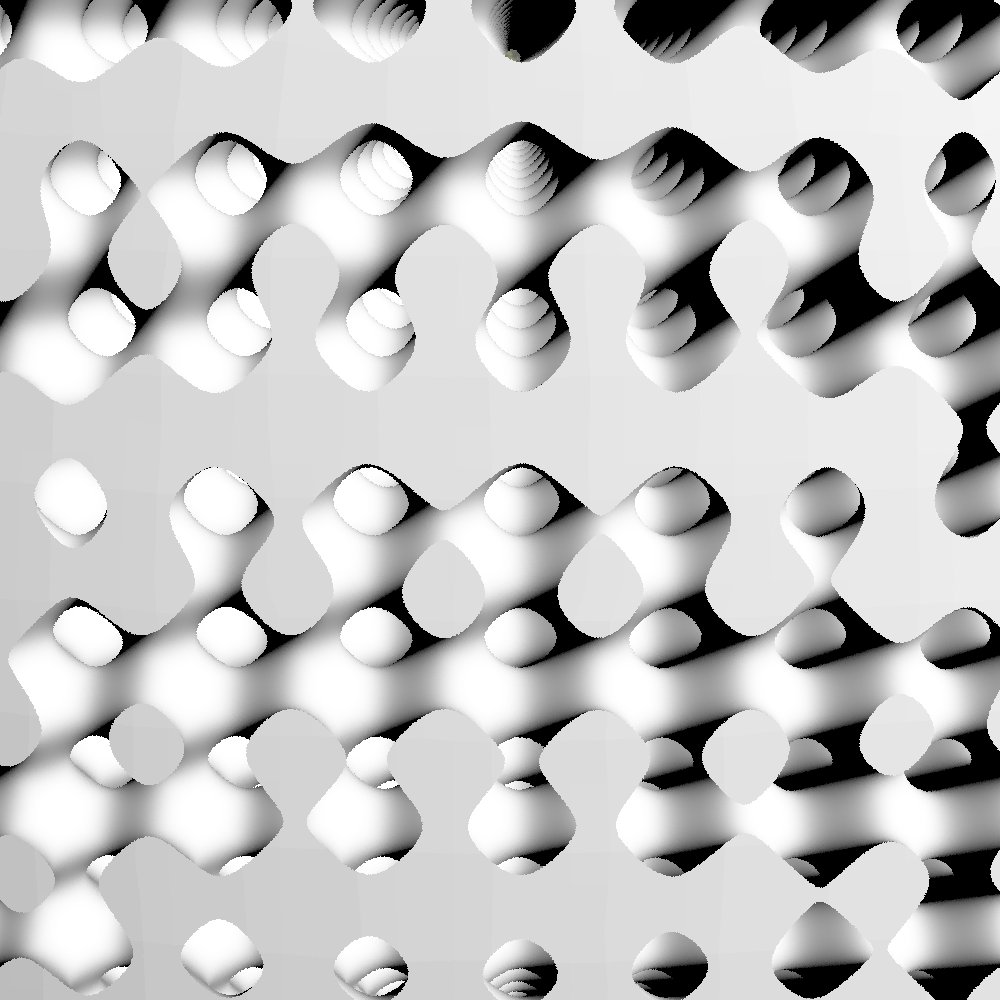} 
        \\ 
        & \includegraphics[width=0.08\textwidth]{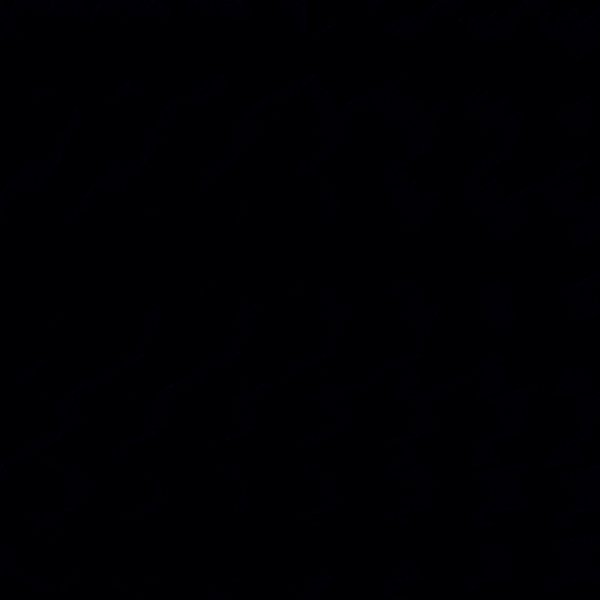} & 
        \includegraphics[width=0.08\textwidth]{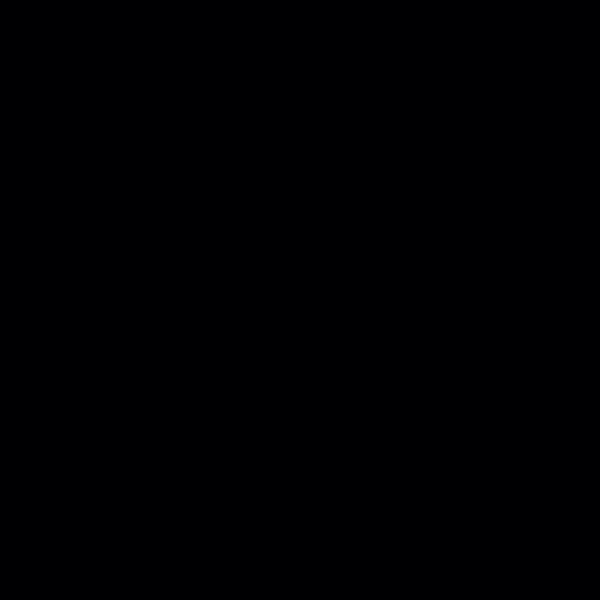} &
        \includegraphics[width=0.08\textwidth]{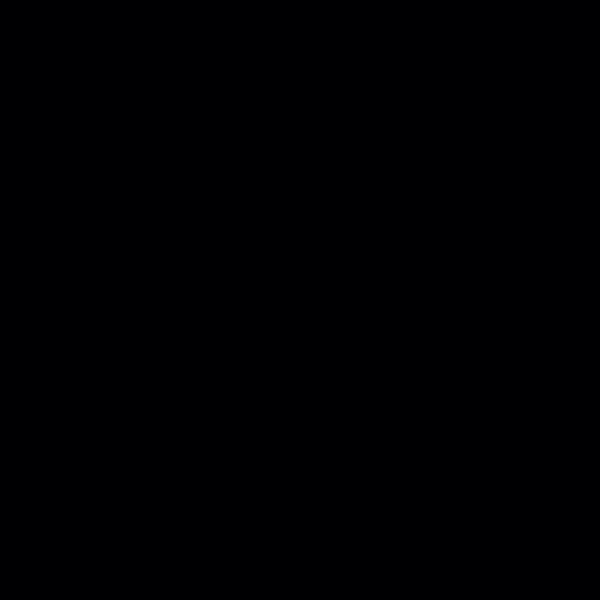} &
        \includegraphics[width=0.08\textwidth]{figures/optimization_results/errors_flip/CMA-ES_0.jpg} &
        \includegraphics[width=0.08\textwidth]{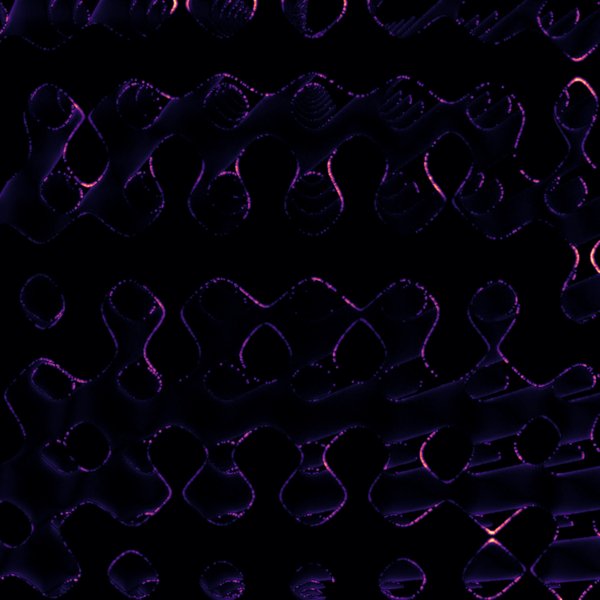} &
        \includegraphics[width=0.08\textwidth]{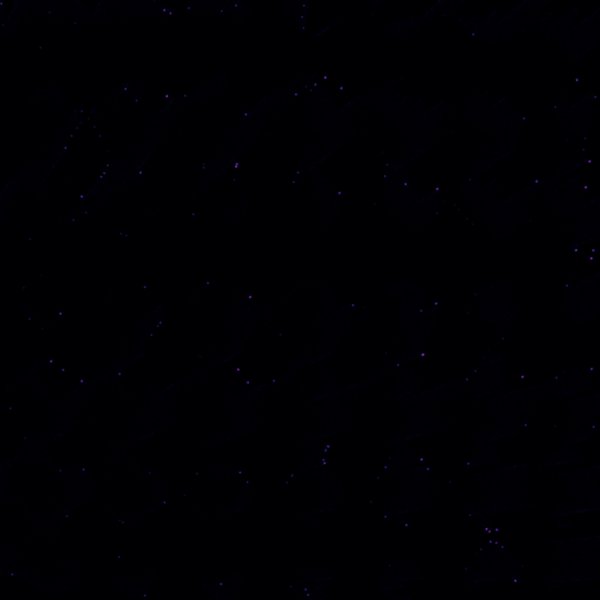} \\ 
         \multirow{2}{*}{{\rotatebox{90}{ \tiny Fibers [2D]}}} & \includegraphics[width=0.08\textwidth]{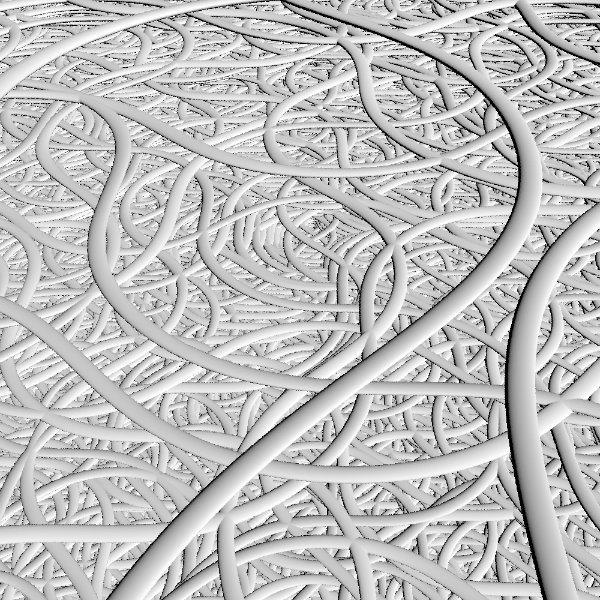} & 
        \includegraphics[width=0.08\textwidth]{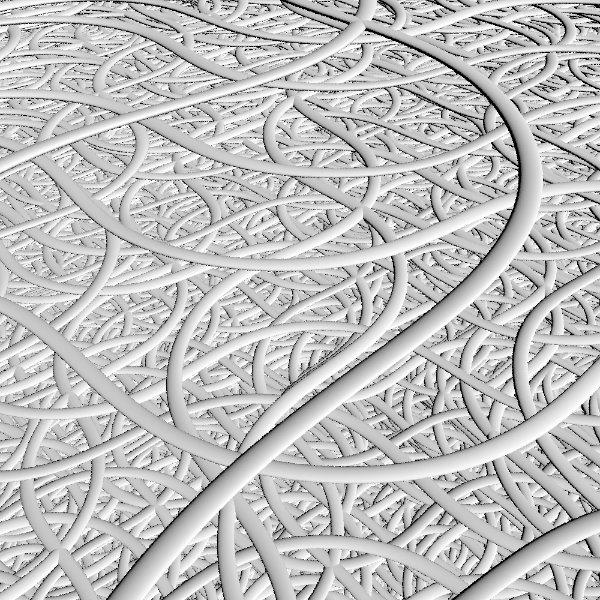} &
        \includegraphics[width=0.08\textwidth]{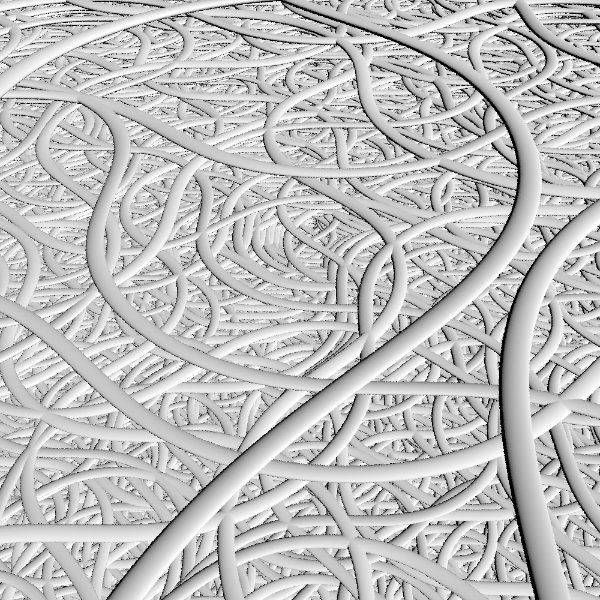} &
        \includegraphics[width=0.08\textwidth]{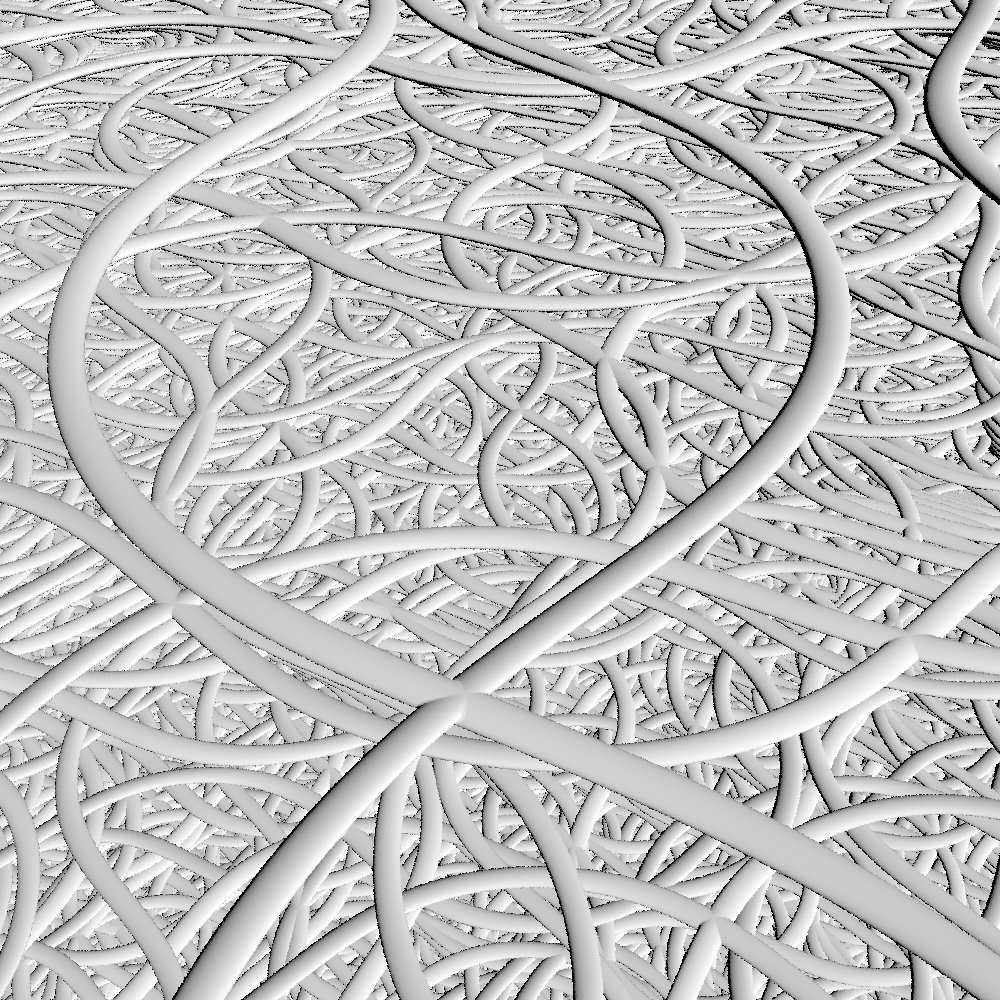} &
        \includegraphics[width=0.08\textwidth]{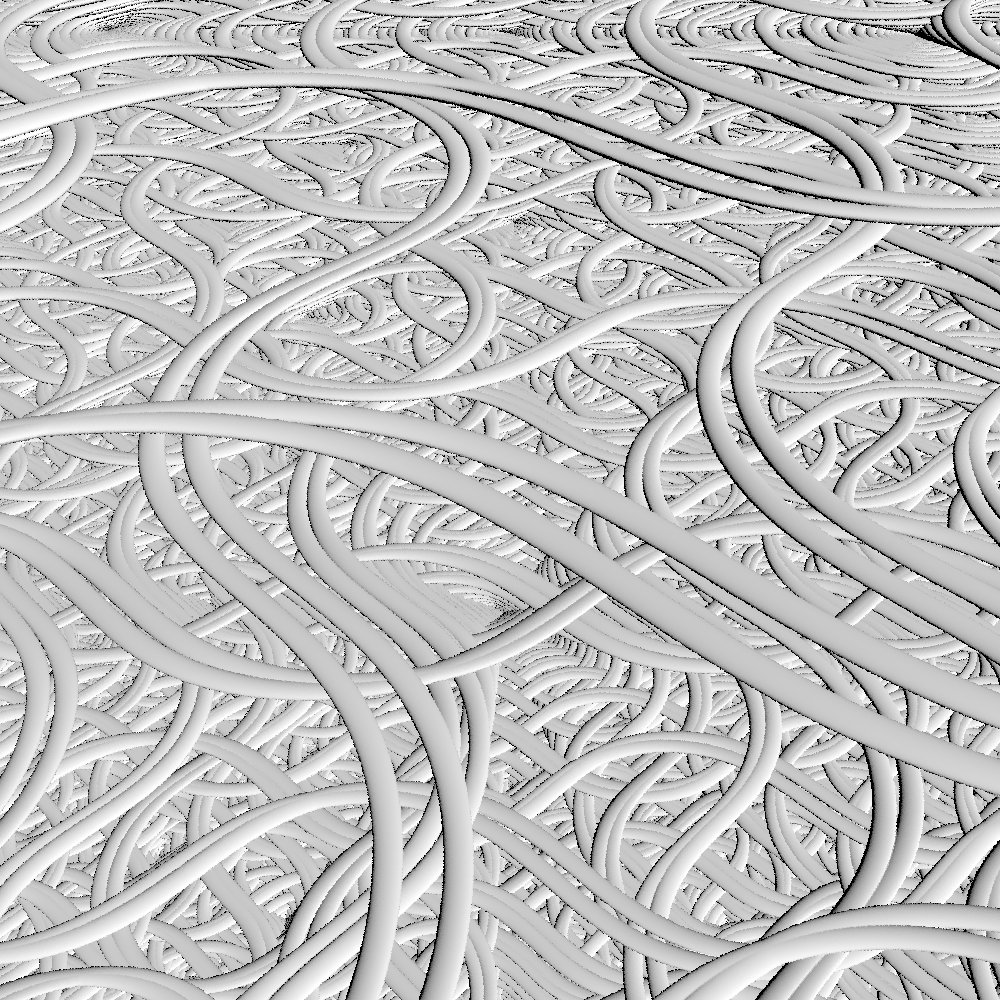} &
        \includegraphics[width=0.08\textwidth]{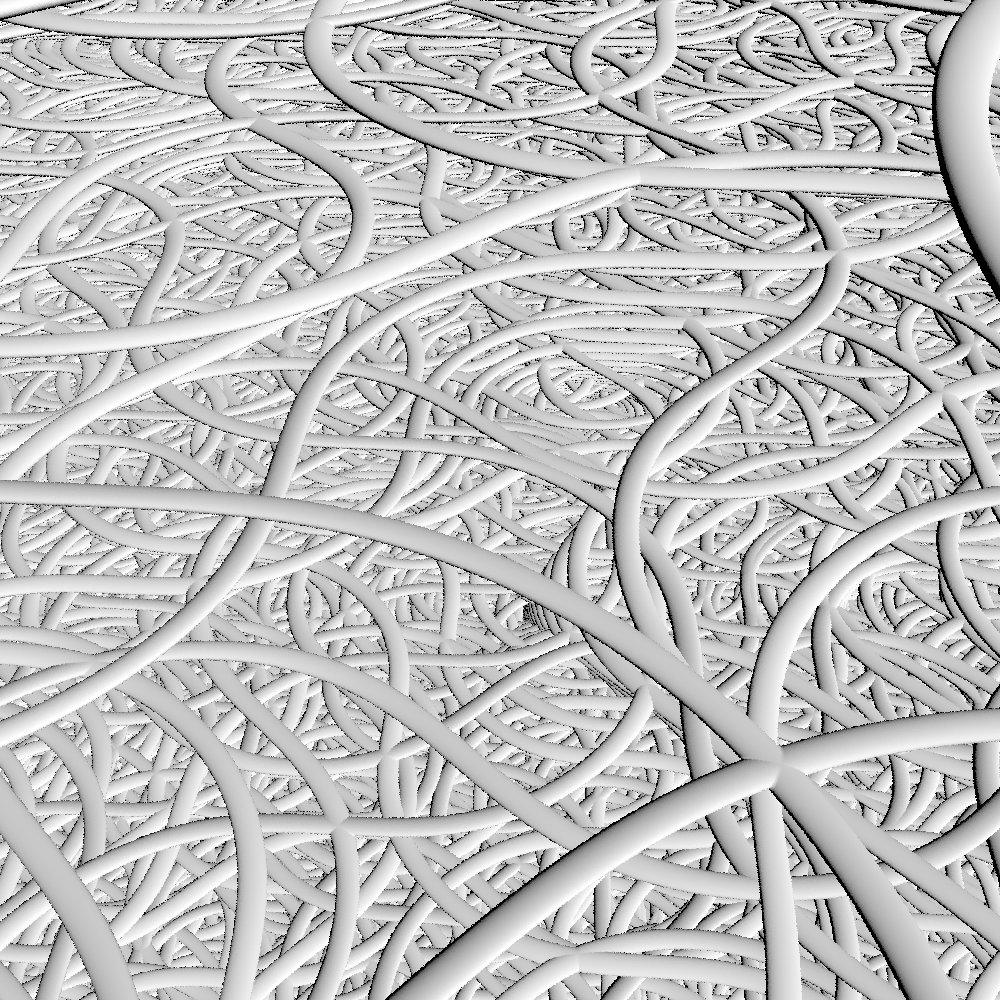} 
        \\ 
        & \includegraphics[width=0.08\textwidth]{figures/optimization_results/errors_flip/CMA-ES_0.jpg} & 
        \includegraphics[width=0.08\textwidth]{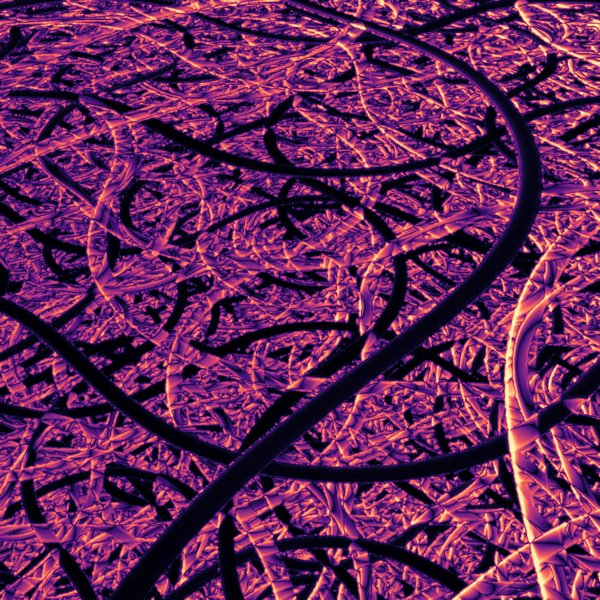} &
        \includegraphics[width=0.08\textwidth]{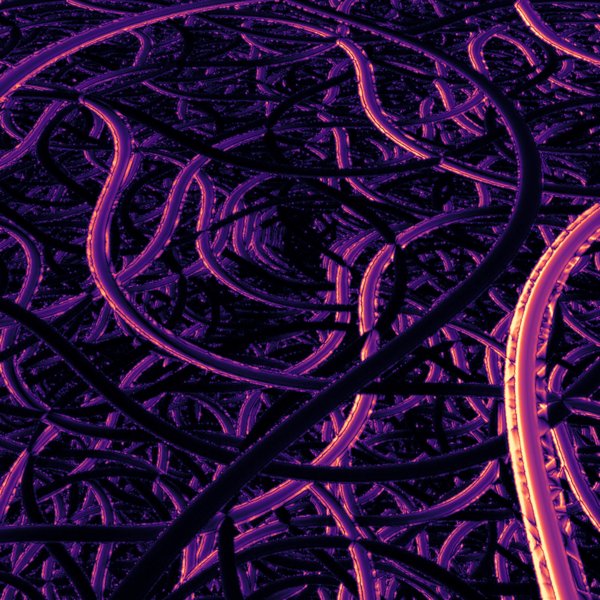} &
        \includegraphics[width=0.08\textwidth]{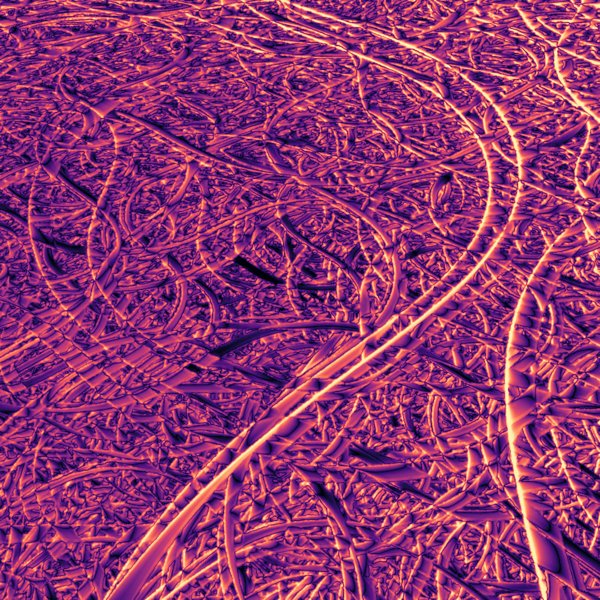} &
        \includegraphics[width=0.08\textwidth]{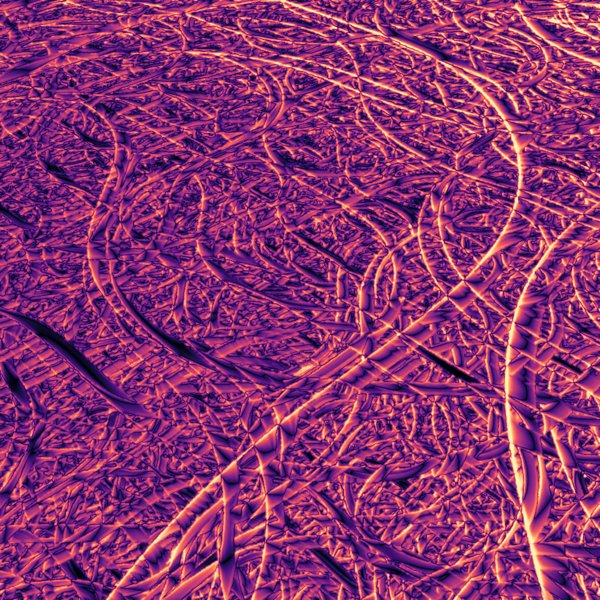} &
        \includegraphics[width=0.08\textwidth]{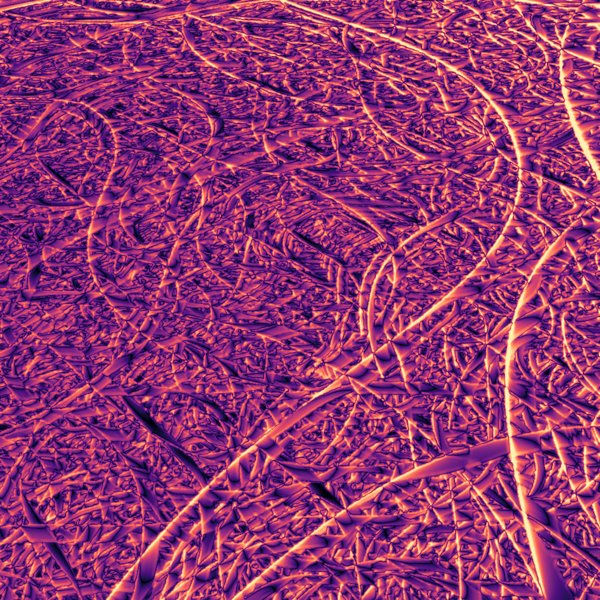} \\
        \multirow{2}{*}{{\rotatebox{90}{ \tiny Spheres [2D]}}} & \includegraphics[width=0.08\textwidth]{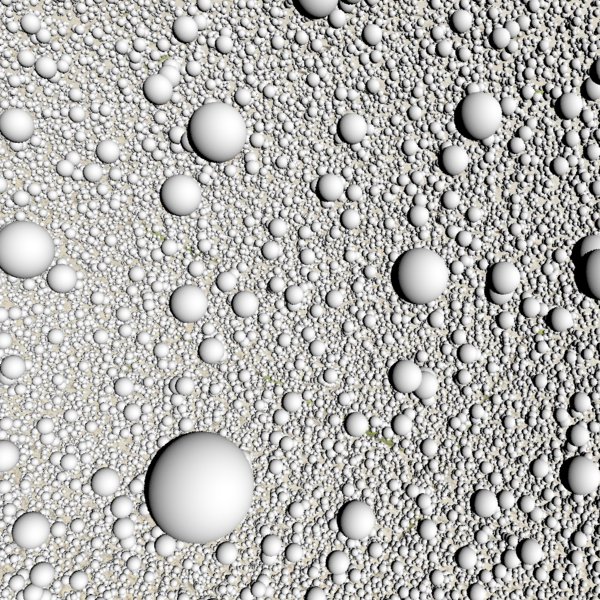} & 
        \includegraphics[width=0.08\textwidth]{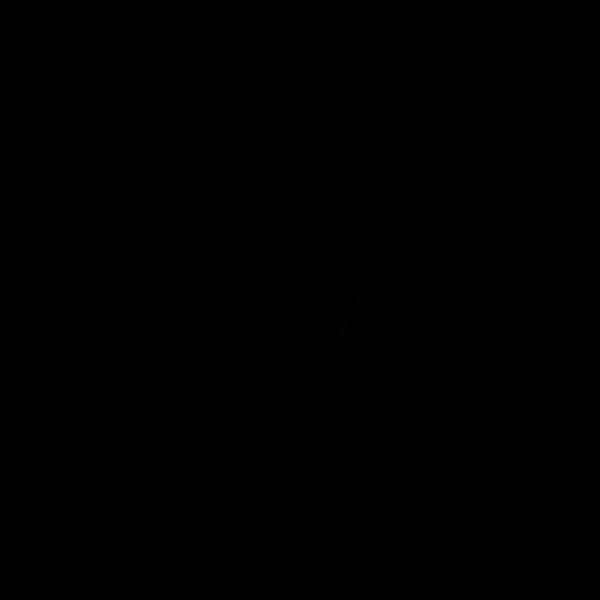} &
        \includegraphics[width=0.08\textwidth]{figures/optimization_results/not_applicable_rescaled.jpg} &
        \includegraphics[width=0.08\textwidth]{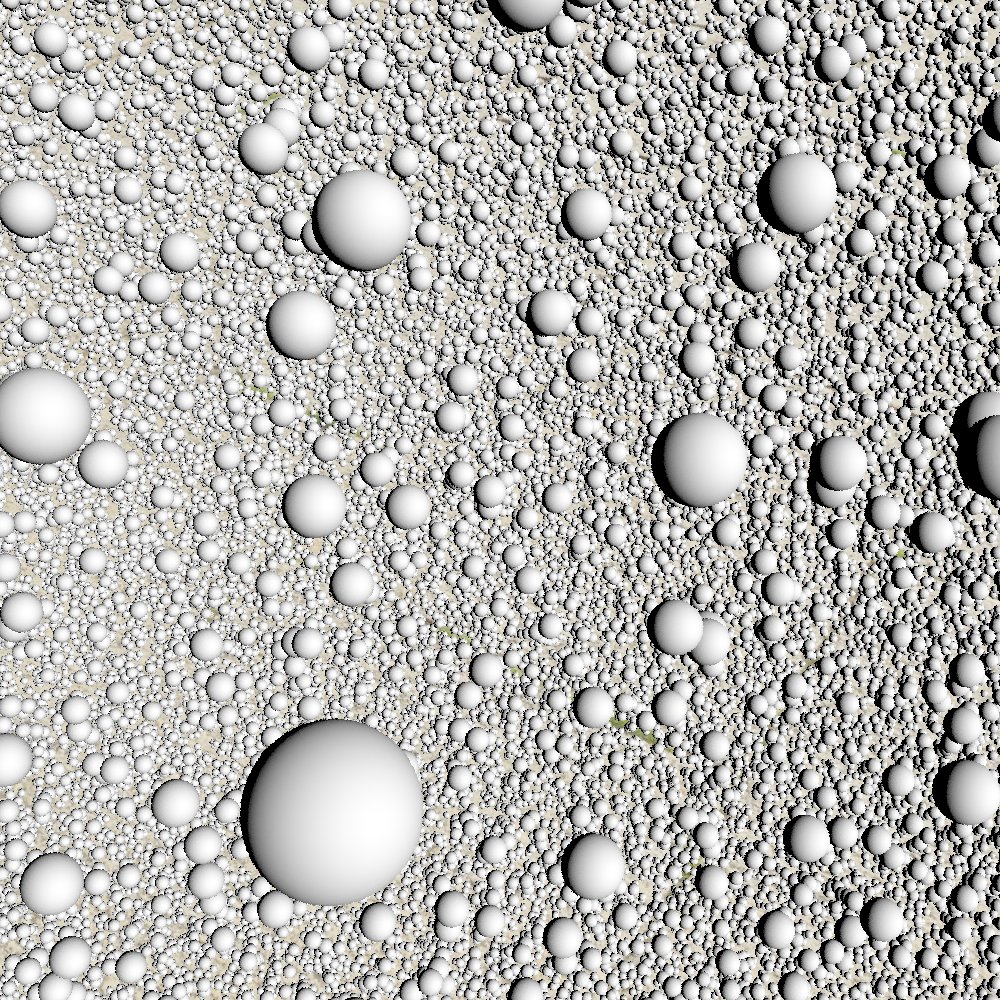} &
        \includegraphics[width=0.08\textwidth]{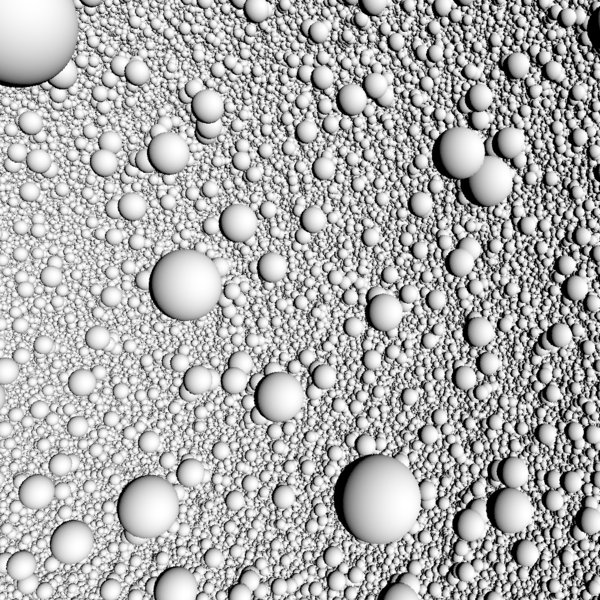} &
        \includegraphics[width=0.08\textwidth]{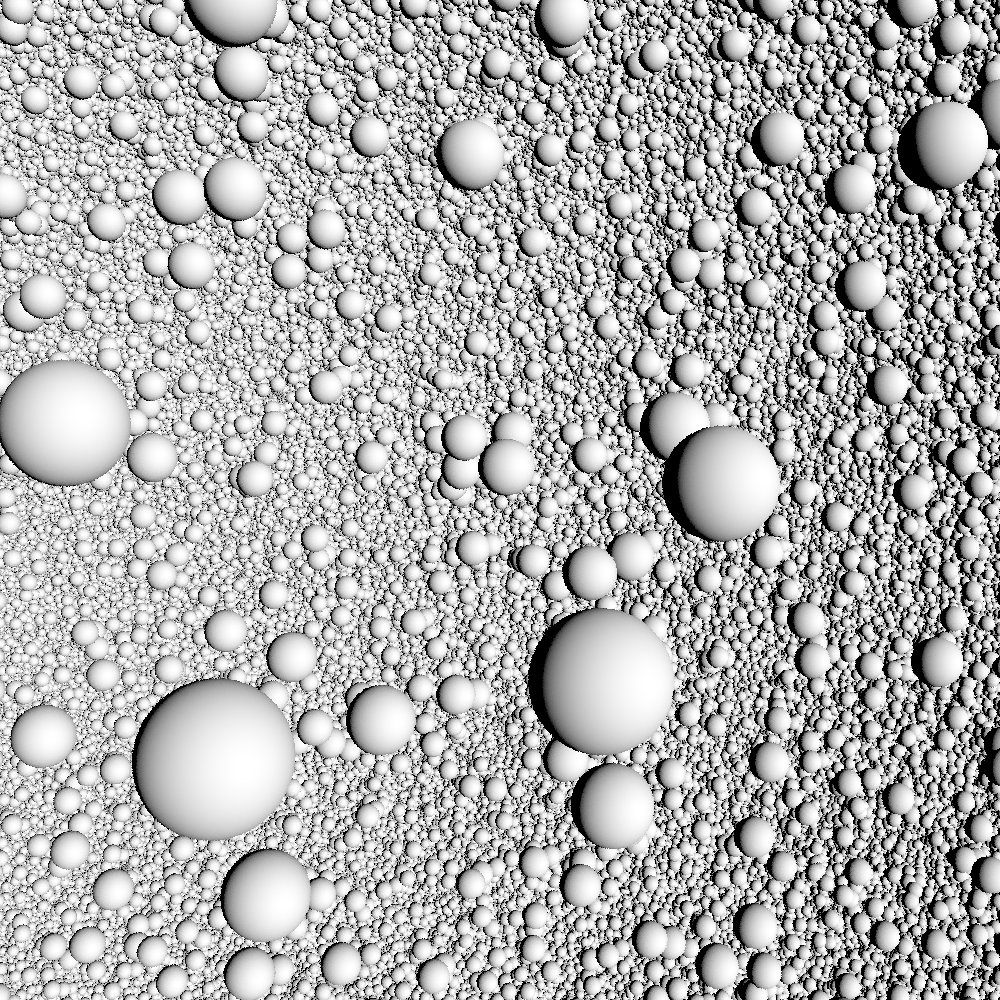} 
        \\ 
        & \includegraphics[width=0.08\textwidth]{figures/optimization_results/errors_flip/CMA-ES_0.jpg} & 
        \includegraphics[width=0.08\textwidth]{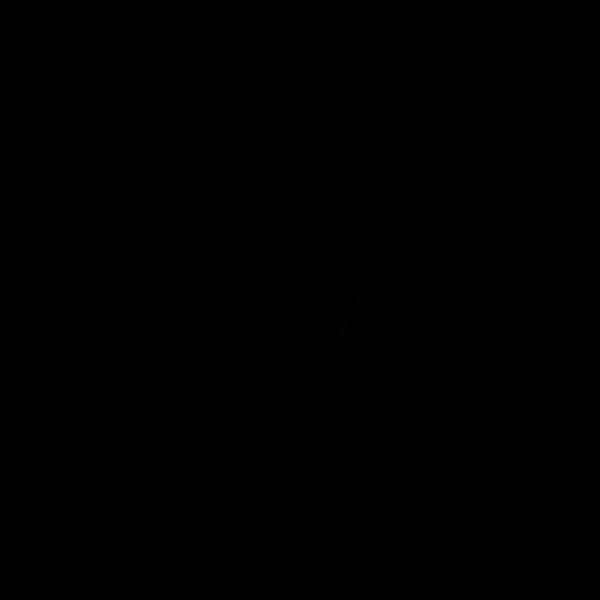} &
        \includegraphics[width=0.08\textwidth]{figures/optimization_results/errors_flip/not_applicable_rescaled.jpg} &
        \includegraphics[width=0.08\textwidth]{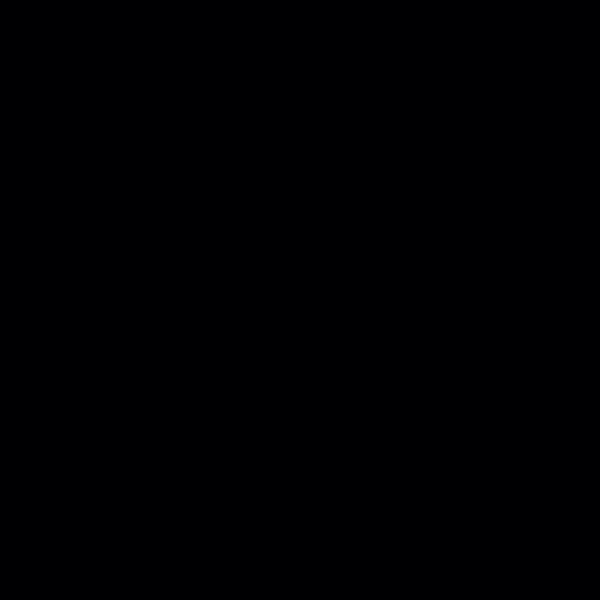} &
        \includegraphics[width=0.08\textwidth]{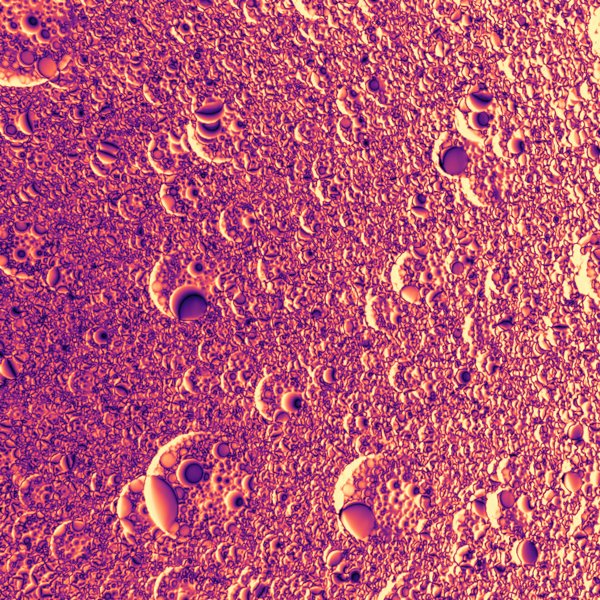} &
        \includegraphics[width=0.08\textwidth]{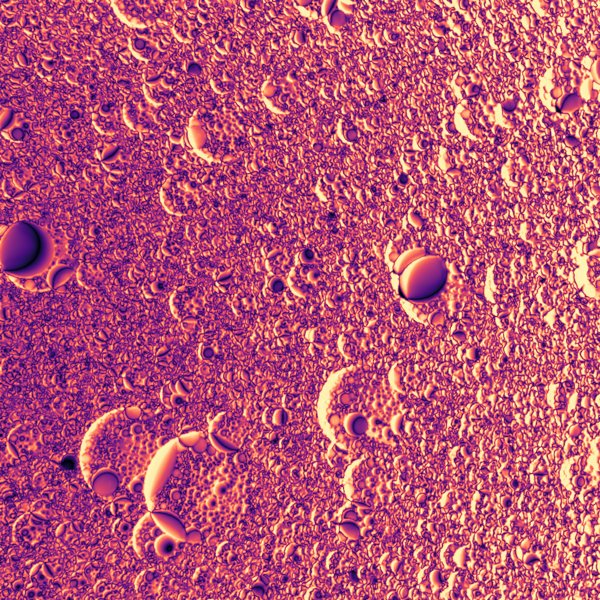}
        \\
         \multirow{2}{*}{{\rotatebox{90}{ \tiny Gyroid [5D]}}} & \includegraphics[width=0.08\textwidth]{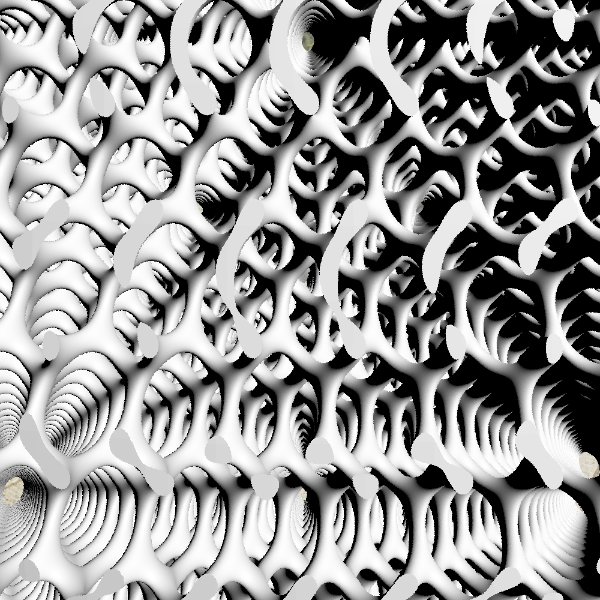} & 
        \includegraphics[width=0.08\textwidth]{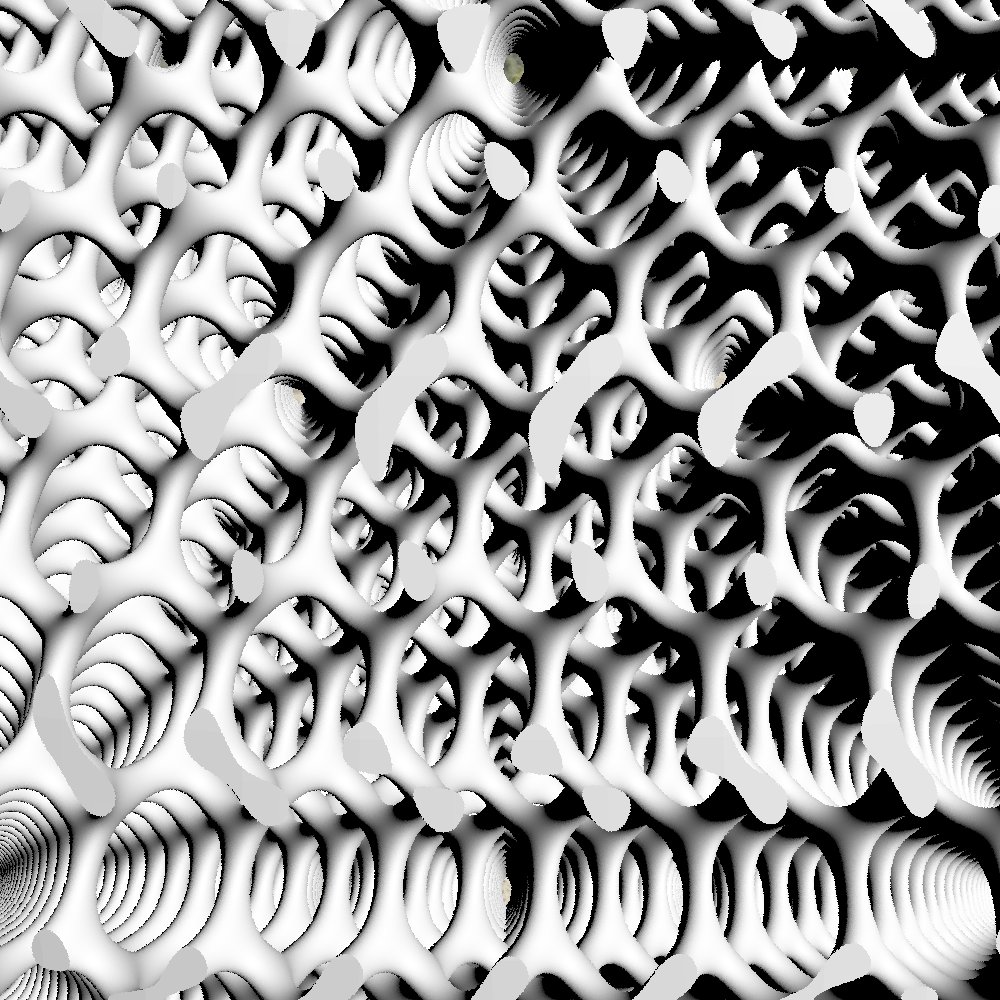} &
        \includegraphics[width=0.08\textwidth]{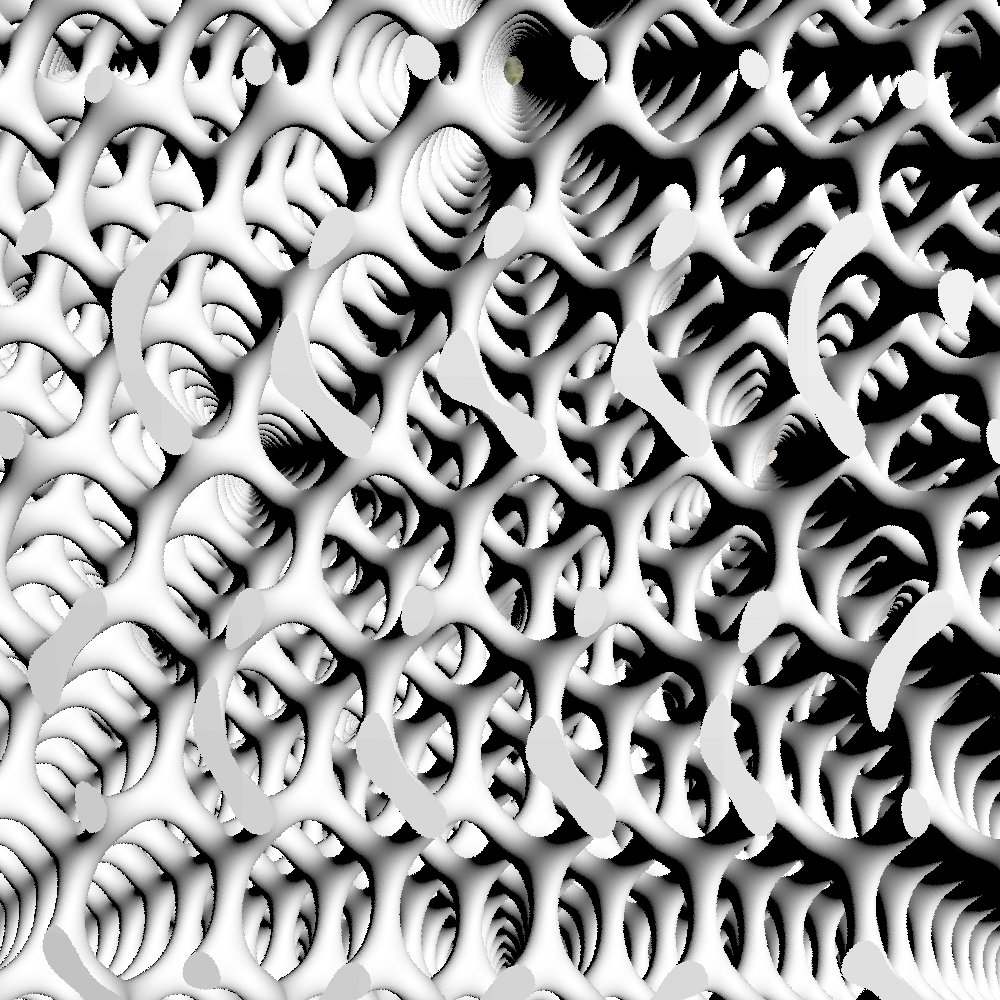} &
        \includegraphics[width=0.08\textwidth]{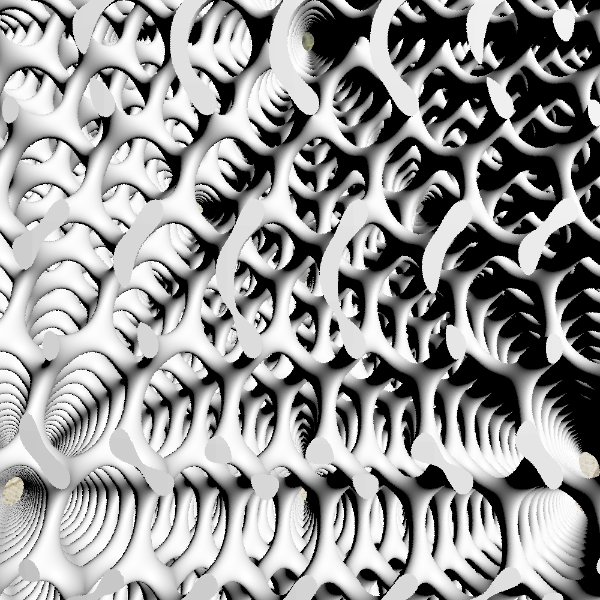} &
        \includegraphics[width=0.08\textwidth]{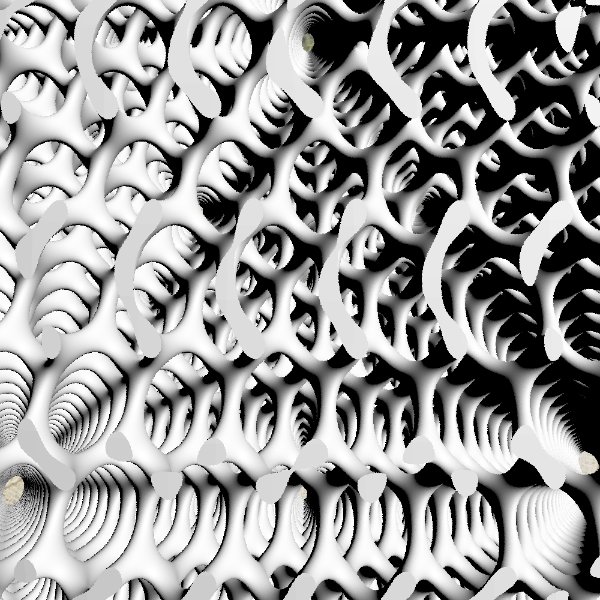} &
        \includegraphics[width=0.08\textwidth]{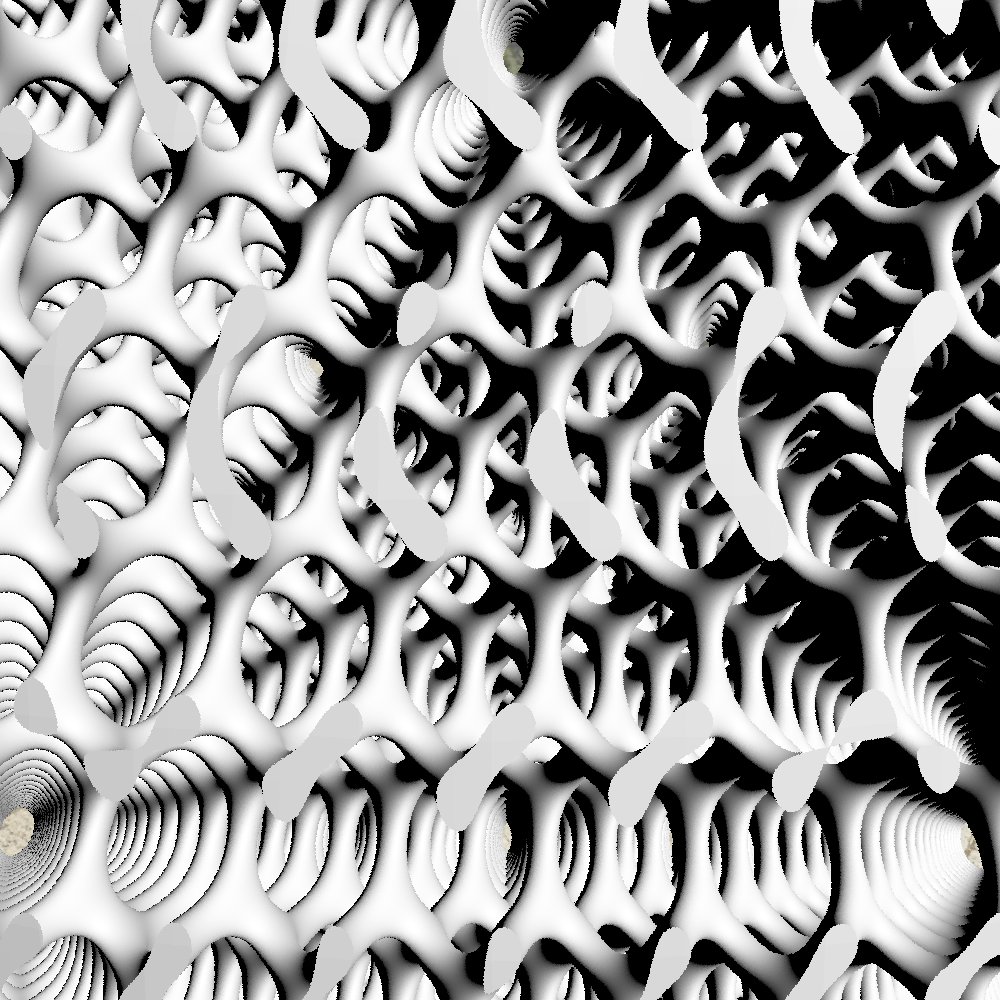} 
        \\ 
        & \includegraphics[width=0.08\textwidth]{figures/optimization_results/errors_flip/CMA-ES_0.jpg} & 
        \includegraphics[width=0.08\textwidth]{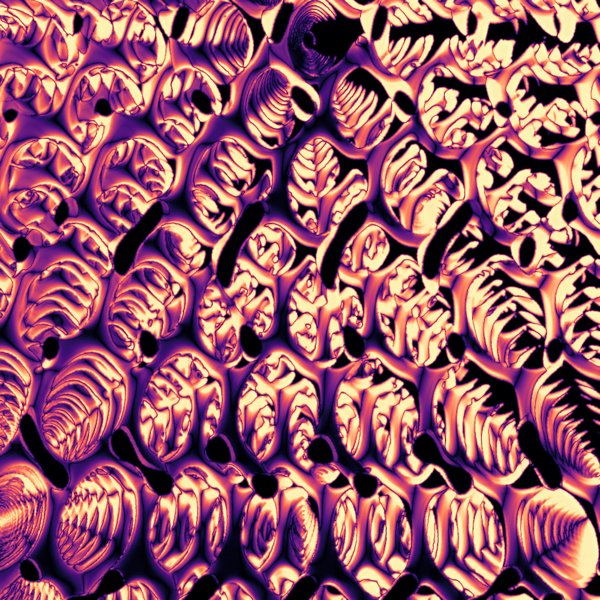} &
        \includegraphics[width=0.08\textwidth]{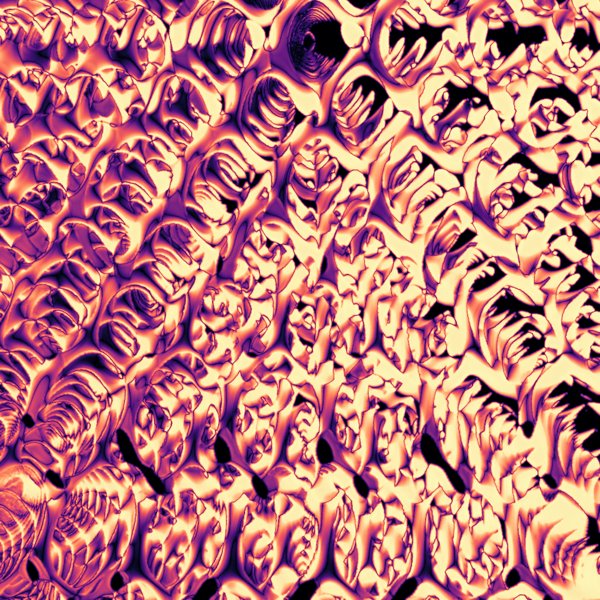} &
        \includegraphics[width=0.08\textwidth]{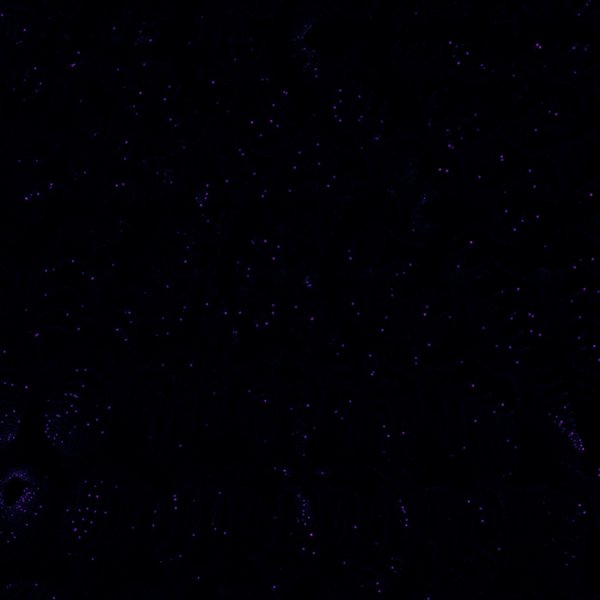} &
        \includegraphics[width=0.08\textwidth]{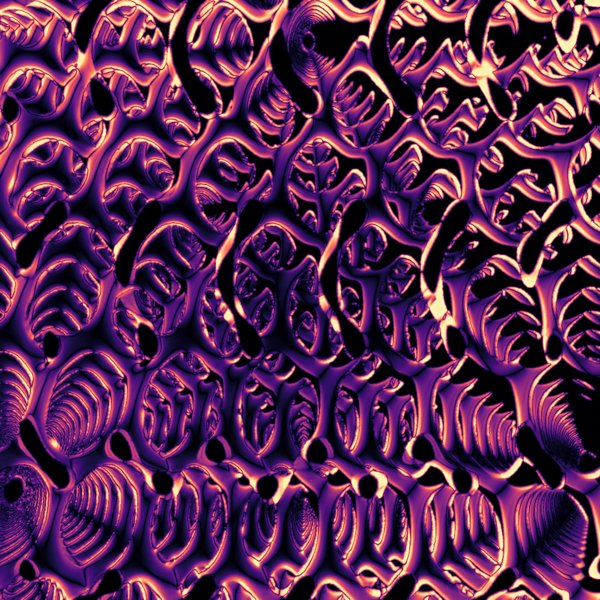} &
        \includegraphics[width=0.08\textwidth]{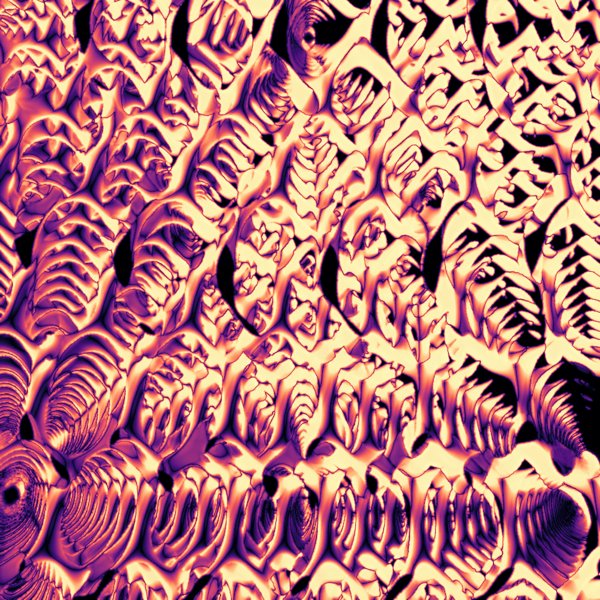}
    \end{tabular}
    \end{adjustbox}
    \caption{Reconstruction results for different optimization algorithms (columns) and synthetic procedural microstructures (rows). We show a comparison of rendered ground truth exemplars and the rendered approximations~(\textit{top row per microstructure}) with the measured \FLIP (\textit{bottom row per microstructure}).}
    \label{fig:reconstruction_renders}
\end{figure}

In Table \ref{optimization_table_results}, we report the performance of different algorithms for different microstructures. In Fig. \ref{fig:reconstruction_renders}, we show the corresponding renders. We set a maximum time limit for the optimization of approximately 20 minutes, as results typically do not improve significantly beyond this point.  The parameters $\boldsymbol{\phi}$ obtained per microstructure and optimization algorithm can be found in the supplemental material. The results show that no single optimization algorithm consistently outperforms the others across all microstructures. SHGO-SDF and BH-SDF perform well for small-variable problems but struggle with higher-dimensional cases (e.g., Gyroid [5D], Porous [28D]) and cannot be applied to certain discrete problems (e.g., Spheres [2D]). CMA-ES-SDF demonstrates solid performance across different microstructures, including discrete ones like Spheres [2D]. As expected, parameter fitting methods outperform analysis-by-synthesis in both accuracy and efficiency, given the larger complexity of the latter. Interestingly, the CMA-ES-S resembles Porous [28D] more than the other methods, highlighting its potential for high-dimensional reconstructions.\\

\definecolor{darkorange}{RGB}{255,140,0}  
\begin{table}[ht]
\begin{adjustbox}{max width=\textwidth, max height=0.17\textheight, center}
\begin{tabular}{l r | c c c c c} 
Microstructure & Metrics & SHGO-SDF & BH-SDF & CMA-ES-SDF & CMA-ES-S & Powell-S\\
\toprule
    \multirow{4}{*}{Gyroid [1D]} & Val. Error$\downarrow$ & \cellcolor{orange!25}$\mathbf{0.00}$ & \cellcolor{orange!25}$\mathbf{0.00}$ & \cellcolor{orange!25}$\mathbf{0.00}$ & $10^{-4}$ & \cellcolor{yellow!25}$10^{-6}$\\ 
& LPIPS$\downarrow$ & \cellcolor{orange!25}\textbf{0.00} & \cellcolor{orange!25}\textbf{0.00}& \cellcolor{orange!25}\textbf{0.00} & \cellcolor{yellow!25}0.02 & \cellcolor{orange!25}\textbf{0.00}\\ 
& \FLIP$\downarrow$ &  \cellcolor{yellow!25}$3\times10^{-7}$  & \cellcolor{orange!25}$\mathbf{1\times10^{-7}}$ &  $1\times10^{-3}$  &  $2\times10^{-3}$ &  $2\times10^{-3}$\\ 
& Time$\downarrow$ & \cellcolor{orange!25}\textbf{2.85s}  & \cellcolor{yellow!25}16.67s & \cellcolor{orange!25}\textbf{2.23s}  & 20m & 7m\\ 
\\
\multirow{4}{*}{Fibers [2D]} & Val. Error$\downarrow$ & \cellcolor{yellow!25}0.34 & \cellcolor{orange!25}$\mathbf{1\times10^{-3}}$ & 0.65 & 1.00 & 1.00\\ 
& LPIPS$\downarrow$ & \cellcolor{yellow!25}0.41 & \cellcolor{orange!25}\textbf{0.13} & 0.54 & 0.56 & 0.56   \\ 
& \FLIP$\downarrow$ &  \cellcolor{yellow!25}0.3  & \cellcolor{orange!25}\textbf{0.14} & 0.42  &  0.40 &  0.43 \\ 
& Time$\downarrow$ & 13m  & 20m & \cellcolor{yellow!25}6m  & 20m & \cellcolor{orange!25}\textbf{5m}\\ 
\\
\multirow{4}{*}{Gyroid [3D]} & Val. Error$\downarrow$ & \cellcolor{orange!25}$\mathbf{1\times10^{-3}}$ & $6\times10^{-3}$ & 1.02& 0.15 & \cellcolor{yellow!25}$2\times10^{-3}$\\ 
& LPIPS$\downarrow$ & \cellcolor{orange!25}\textbf{0.01} & 0.04 & 0.65 & 0.35 & \cellcolor{yellow!25}0.02\\ 
& \FLIP$\downarrow$ &  \cellcolor{orange!25}\textbf{0.01}  & 0.05 &  0.61 &  0.42 &  \cellcolor{yellow!25}0.03 \\ 
& Time$\downarrow$ & \cellcolor{orange!25}\textbf{22.7s} & 20m & \cellcolor{yellow!25}7m & 14m & 20m\\ 
\\
\multirow{4}{*}{Spheres [2D]} & Val. Error$\downarrow$ & \sout{N/A} & \sout{N/A} & \cellcolor{orange!25}$\mathbf{1\times10^{-5}}$ & 1.01 & 1.01\\ 
& LPIPS$\downarrow$ & \sout{N/A} & \sout{N/A} & \cellcolor{orange!25}\textbf{0.00} & 0.53 & 0.53\\ 
& \FLIP$\downarrow$ &  \sout{N/A}& \sout{N/A} & \cellcolor{orange!25}$\mathbf{3\times10^{-7}}$  &  0.55 & 0.55 \\ 
& time$\downarrow$ &\sout{N/A} & \sout{N/A} & \cellcolor{orange!25}\textbf{4m}  & 20m & 5m\\ 
\\
\multirow{4}{*}{Gyroid [5D]} & Val. Error$\downarrow$ & 0.66 & 0.65 & \cellcolor{orange!25}\textbf{$\mathbf{3\times10^{-4}}$} & \cellcolor{yellow!25}0.03 & 0.44 \\ 
& LPIPS$\downarrow$ & 0.42 & 0.57 & \cellcolor{orange!25}\textbf{0.00} & \cellcolor{yellow!25}0.21 & 0.54 \\ 
& \FLIP$\downarrow$ &  0.42  & 0.62 &  \cellcolor{orange!25}\textbf{$\mathbf{6\times10^{-3}}$}  &  \cellcolor{yellow!25}0.28 &  0.60\\ 
& Time$\downarrow$ & \cellcolor{orange!25}\textbf{2m} & 15m & \cellcolor{yellow!25}11m  & 20m & 15m \\ 
\\
\multirow{4}{*}{Porous [28D]} & Val. Error$\downarrow$ & \sout{N/A} & \cellcolor{orange!25}\textbf{1.00} & \cellcolor{orange!25}\textbf{1.00} & \cellcolor{orange!25}\textbf{1.00} & \cellcolor{orange!25}\textbf{1.00}\\ 
& LPIPS$\downarrow$ & \sout{N/A} & \cellcolor{yellow!25}0.64 & \cellcolor{yellow!25}0.64 & \cellcolor{yellow!25}0.64 & \cellcolor{orange!25}\textbf{0.61}\\ 
& \FLIP$\downarrow$ &  \sout{N/A} & \cellcolor{orange!25}\textbf{0.51} & \cellcolor{yellow!25}0.52 & \cellcolor{yellow!25}0.52 & 0.59\\ 
& Time$\downarrow$ &  \sout{N/A} & \cellcolor{orange!25}\textbf{4m} & \cellcolor{yellow!25}8m & 20m & 20m\\ 
\end{tabular}
\end{adjustbox}
\caption{Reconstruction performance for different optimization algorithms (columns) and synthetic procedural microstructures (rows). We report for each microstructure the \textbf{a)} Validaton Error, \textbf{b)} LPIPS Perceptual loss, \textbf{c)} \FLIP error, \textbf{d)} Time of the optimization. We consider a numerical~0 to be a number below $1\times10^{-7}$. The \sout{N/A} implies that the optimization method could not be applied for the corresponding microstructure. The best results are
highlighted in orange, while the second-best results are marked in yellow. }
\label{optimization_table_results}
\end{table}

\section{Discussion and Future work}

We identify four categories for future research directions: geometry, rendering, reconstruction, and simulation. Given the flexibility of our multiscale geometry framework, exploring its applications in fabrication presents an exciting avenue for further investigation.

\paragraph*{Geometry}

Our procedural framework is a promising step toward creating digital multiscale materials, but it has limitations. While we emphasize its expressivity in generating diverse composite materials with powerful capabilities such as correlation, spatially varying properties, agglomeration, and intersect-free generation, extracting triangular meshes and locally manipulating structures remains challenging. Another general limitation of procedural modeling is the reconstruction from exemplars, which we try to mitigate with our reconstruction algorithms. 

\paragraph*{Rendering}
In this paper, we assume ray topics for the rendering algorithm. One potential avenue for future research is to enhance the current framework by incorporating wave optics-based simulations to capture interference and diffraction effects. Additionally, including polarization and fluorescence could provide more depth to the light transport simulation. A virtual gonioreflectometer could be useful for extracting key statistical distributions, such as the BSDF for surface and the volume phase functions, enabling integration with traditional rendering pipelines. 

\paragraph*{Reconstruction}
In this paper, we explore optimization strategies to reconstruct structures from direct parameter fitting, image exemplars, or incomplete volumetric information as a proof of concept with synthetic data. For future work, it would be interesting to reconstruct real microstructures using SEM/TEM images or microCT volumes. Reconstruction from SEM/TEM images would require the adaptation of the synthesis module to at least roughly mimic the appearance under an electronic microscope or, ideally, a suitable Monte Carlo simulator like Nebula. One possibility to handle the inability of our current reconstruction approaches to handle high-dimensional problems is to apply a surrogate loss~\cite{fischer2024zerograds} or explore neural network-based optimization techniques tailored for procedural models. The supplemental material includes preliminary reconstruction results using an MLP with positional encoding and the SIREN architecture~\cite{sitzmann2020siren}.

\paragraph*{Simulation}
In this work, we show how our multiscale geometry modeling framework can be leveraged to design the final appearance of materials. Additionally, it enables the fast generation of convex and concave piles without explicit simulation, such as rigid body simulation. A promising direction for future research is extending this framework to capture dynamic phenomena, such as particle flow effects for fluid motion simulation. A key challenge lies in performing these simulations without converting our representation into triangular meshes, which remains an open problem for further exploration.
\section{Conclusions}
We developed a framework for modeling multiscale materials using implicit functions, which can be useful for modeling materials with various properties. In this work, we only focused on modeling the materials' appearance as the case study for the modeled geometry. We started with particulate materials with various particle SDFs on discrete grids, such as particle suspensions, particle agglomerations, and particle piling, and we modeled various types of microgeometry. Further, we explored how periodic implicit functions are used for generating various types of microgeometry, and we also gave the mathematical relationship between SDFs in discrete grids and implicit periodic functions. One of the main central aspects is that polynomial functions are core to modeling geometry and even to optimizing the sphere tracing algorithm. Finally, we optimized SDFs for volume reconstruction from image and signed distance field exemplars using first-order and gradient-free optimization. 

\section*{Acknowledgements}

This project is part of the PRIME ITN initiative and has received funding from the European Union’s Horizon 2020 research and innovation program under the Marie Skłodowska-Curie grant agreement No. 956585. We would like to thank Andreas Bærentzen and Jeppe Revall Frisvad for their valuable feedback and guidance during the development of this project. The authors are also grateful to Adolfo Muñoz and Adrian Jarabo for their feedback and proofreading of the submitted manuscript. The 3D paper spinner model is available at \href{https://www.turbosquid.com/Search/Index.cfm?keyword=paper+spinner&media_typeid=2}{https://www.turbosquid.com/}




\newpage
\bibliographystyle{eg-alpha-doi} 
\bibliography{main}      

\clearpage

\begin{center}
    {\Huge \textbf{Supplementary Material}}
\end{center}

This document covers additional details of the modeling of the multiscale geometry in Sec.~\ref{sec:mgm}, the optimization of the sphere tracing algorithm for visualization of the multiscale geometry in Sec.~\ref{sec:sto}, and the reconstruction of the multiscale geometry in Sec.~\ref{sec:mgr}.
\section{Multiscale geometry modeling}\label{sec:mgm}
This section covers additional details and analysis of multiscale geometry modeling.
\subsection{Multi-phase Particle Cloud}\label{sec:mpc}

We use spatial transformations of points to generate multiphase particles within a volume. The following transformation generates particles with various shapes, smoothly transitioning between each type of shape in the volume, where each shape is considered a different phase. This works as a frequency modulation of the signal, and it almost generates the Phasor noise~\cite{tricard2019procedural} effect. These transformed points are used to generate the particle cloud using particulate SDF.

We can create a particle cloud containing specific regions where some regions are isotropic and others are anisotropic, with smooth transitions between these regions, see Fig.~\ref{fig:ice_multiphase} in the main document. This concept is similar to positional encoding, where particular regions in the volume behave in a specific manner. To achieve this, the transformed point
\begin{equation}
 \bm{p'} = \bm{p}\bm{T} \,,\quad \textrm{with $\bm{T} = \bm{A}\bm{D}$}\,,
\end{equation}
is used as the input point for the particulate material SDF. 
This generates effects like phasor noise, where the signal's frequency depends on spatial coordinates. The matrices
\begin{equation}
\begin{array}{c@{\quad}c}  
  \bm{A} = \begin{bmatrix} 
    A_1 & A_2 & A_3 \\ 
    A_4 & A_5 & A_6 \\ 
    A_7 & A_8 & A_9 
  \end{bmatrix},
  &
  \bm{D} = \begin{bmatrix} 
    \sin(p_y) & \sin(p_z) & \sin(p_x) \\ 
    \sin(p_z) & \sin(p_x) & \sin(p_y) \\ 
    \sin(p_x) & \sin(p_y) & \sin(p_z) 
  \end{bmatrix},
\end{array}
\end{equation}
contribute to these phasor noise generation effects. The matrix $\bm{A}$ acts as an amplitude matrix, typically with values between 0 and 1. Essentially, we use the original point to generate a noise matrix or turbulence~\cite{perlin1989hypertexture} matrix ($\bm{T}$), similar to octaves. Variations in particle turbulence affect the material translucency.

\subsection{Particle Correlation} \label{sec:correlation}
Our approach also allows for the modeling of spatial correlations among particles, a concept that is applicable in volume path tracing without the need to create individual surfaces for the particles. When the positions of particle centers $\bm{x}$ are fixed within the grid cells rather than being randomly initialized (as discussed in the main document), the particles are arranged in a regular cubic lattice, which is referred to as exhibiting negative spatial correlation. Conversely, if $\bm{x}$ is completely randomized within the range of $[0,1]^3$ through random instantiation, we achieve a stratified random distribution of particle locations, often described as spatially uncorrelated particles. Between these two extremes, it is also possible to create a positive correlation, where particles tend to cluster together in groups.

The rejection sampling technique is an effective tool for controlling positive correlation. Some particles can be removed to allow existing particles to form clusters. The number of clusters and the number of particles for each cluster depend on the function $g(\bm{p})$ used for rejection sampling. As an example, the modular function can be used to generate a positive correlation among the particles. 

\begin{equation}
g(\bm{p}) = \left\{\begin{array}{l@{\,,\quad}l}1 & \prod_{i \in \{x, y, z\}} \left| \operatorname{mod} (|p_i|, n) \right| < \frac{n}{2} \\ 0 & \textrm{otherwise}\end{array}\right.
\end{equation}
generates positively correlated particles of 
size $(\frac{n}{2})^3$.

 Another way to generate organic particle clusters is by using multi-level particle clouds, where we first generate a particle cloud and then generate a particle cloud inside every single particle. The following is the SDF representing particle clusters of size $n$:
\begin{equation}
d_{\text{clusters}}(\bm{p}, w) = \max(d_g(\bm{p}, 2w/n), d_g(\bm{p}, w)) \,.
\end{equation}

\subsection{Particle Agglomeration}

Based on the following polynomial inspired by B{\'e}zier curves:
\begin{align}
B_x &= (1-t)^2 q_x + 2(1-t)(q_x + n_1) + t^2 (q_x - n_1) \\
B_y &= (1-t)^2 q_y + 2(1-t)(q_y + n_2) + t^2 (q_y - n_2) \,,
\end{align}
where $n_1$ and $n_2$ are integer constants and the value of $t$ is between 0 and 1, we construct the following polynomials for $D$:
\begin{align}
P_1(q) &= q_x + q_y + nB_x \\
P_2(q) &=  q_y + q_xq_y + nB_y \,,
\end{align}
where $n$ is an integer constant.
We use these for the condition 
\begin{equation}
\text{mod}(P_1(q), 31) = 7 \quad \lor \quad \text{mod}(P_2(q), 11) = 7 \,.
\end{equation}
in Eq. \ref{eq:pf} in the main document to generate the gelation-like structures shown in Fig. \ref{fig:bc} in the main document.

\subsection{Mesoscale Surface Patterns on the Macrosurface}

the following SDF can be used to generate the mesoscale surface patterns on the macroscopic object with thickness control.
\begin{equation}
d(\bm{p}, w) =
\begin{cases}
d_g(\bm{p}, w), & \text{if } \exists \, \bm{q}_k \text{ such that } l_i < P_i(\bm{q}_k) < r_i, \\
& \forall k = 0, \dots, 26, \\
\frac{w}{2}, & \text{otherwise}.
\end{cases}
\label{eq:distance}
\end{equation}
where,
\begin{equation}
d_g(\bm{p}, w) = f(\bm{p}) \ + \min_{\substack{k=0 \\ \exists \, \bm{q}_k, \ l_i < P_i(\bm{q}_k) < r_i}}^{26} d_{\bm{q}_k}\left(\frac{1}{w}(\bm{p} + \bm{h}(\bm{p}))\right) \,, \label{eq:pameso-sdf}
\end{equation}

In this condition, the terms \( l_i \) and \( r_i \) represent the minimum and maximum bounds of the corresponding polynomial values \(P_i(\bm{q_k})\) when the \( i \)-th polynomial is evaluated with the point \(\bm{q_k}\). The interval length \( (l_i,r_i) \) controls the width (or thickness) of the macroscopic object. In some cases, it may be necessary to use $R$ instead of $P$. These polynomials can be used to create the macro-shape of the object based on the roots of the polynomials; basically, the particle jittering follows these polynomial functions.  When the roots of these polynomials are true, the microgeometry on the surface can be generated using the SDF \( d_g(\bm{p}, w) \).

For instance, the following polynomial can be used to generate the macrosurface with microstructure.
\begin{align}
P(q) &= 2q_x + q_y^2 + q_z^2 - q_x q_y - q_y q_z - q_z q_x \,.
\end{align}

\subsection{Correspondence Between Periodic Implicit Functions and Discretized Grids}
\begin{figure}
    \centering
    \begin{tabular}{@{}c@{}}

    \includegraphics[width=0.5\linewidth]{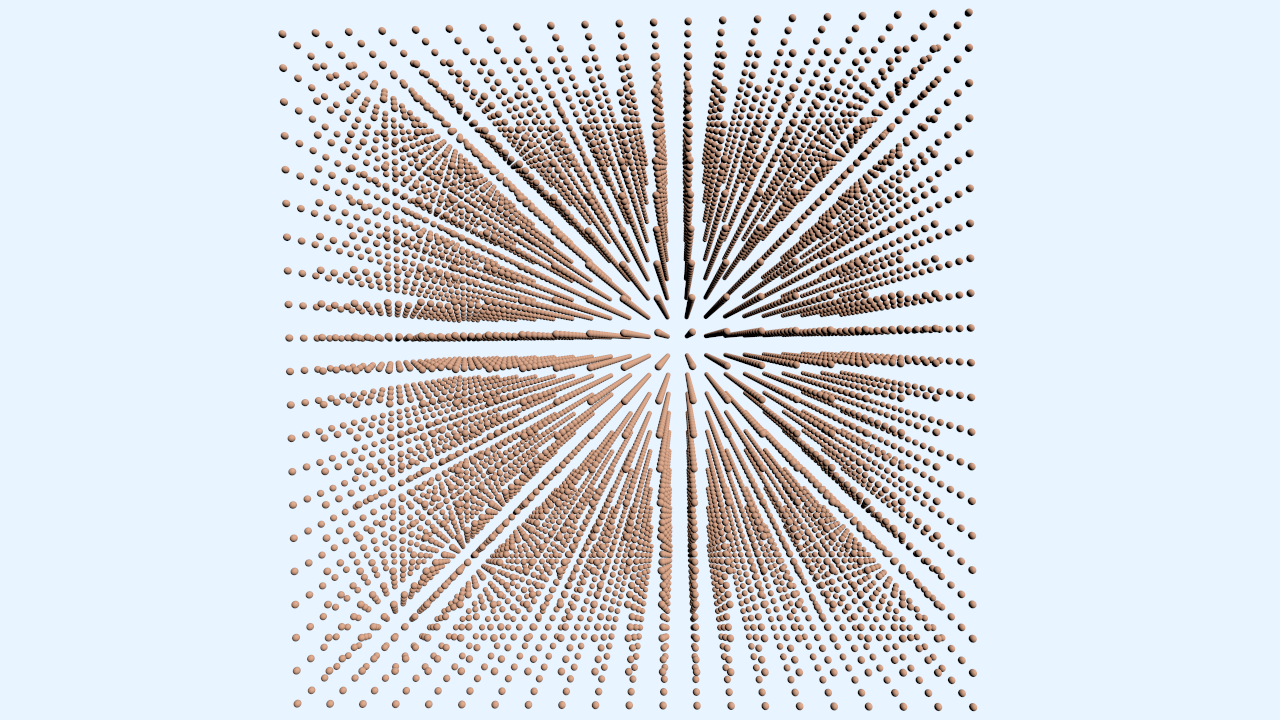} 
    \includegraphics[width=0.5\linewidth]{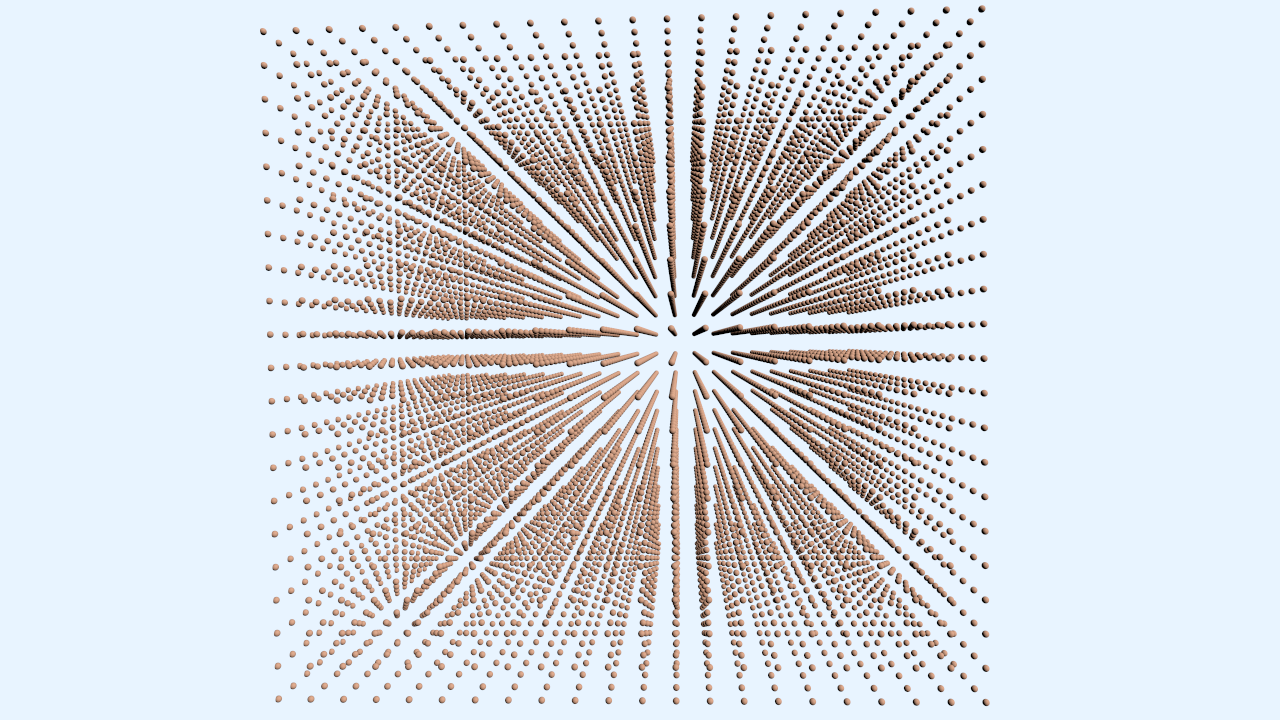}
    \end{tabular}
  \caption{A regular lattice of spherical particles is generated using two methods for the same-sized volume with the same number of particles. The left cloud is produced using periodic functions without grid discretization, using the Eq. \ref{eq:particles}, with a rendering time of 0.7 ms at 1193 fps. In contrast, the right cloud is generated using discrete grid methods based on Eq. \ref{eq:g-sdf} in the main document, resulting in a rendering time of 1.0 ms at 945 fps. The right cloud is certainly more expensive for evaluation compared to the left one because it requires many SDF evaluations to generate the particle at the grid cell, while the left cloud is generated using the evaluation of a single SDF at the grid cell. The time difference is small for a small particle cloud, but when we generate a particle cloud with a semi-infinite number of particles, the rendering time shows a significant difference. }
  \label{fig:correspondence}
\end{figure}

Fig.\ref{fig:correspondence} shows that the spherical particle clouds generated using the two methods periodic functions Eq. \ref{eq:particles} and discrete grids Eq. \ref{eq:g-sdf} in the main document. Almost both methods generate the same cloud, but using periodic functions is straightforward to generate the lattice of spherical particles. Using the Taylor series approximation on Eq. \ref{eq:g-sdf}, we show the correspondence between the periodic functions in Eq. \ref{eq:particles} and discrete grids SDF for spherical particles in Eq. 6 in the main document.

Starting with Eq. \ref{eq:particles} and including the frequency term $w_k$,

\begin{align}
SC_{\text{particles}}(\bm{p}) &=  \sin^2(w_kp_x) + \sin^2(w_kp_y) \notag \\
&\quad + \sin^2(w_kp_z) + w.
\end{align}

The function to expand in the Taylor series form is:

\begin{align}  \label{eq:particles}
SC(\theta, \phi, \psi) &= \sin^2(w_kp_x) + \sin^2(w_kp_y) \notag \\
&\quad + \sin^2(w_kp_z) + w.
\end{align}

The Taylor series expansion around the general point 
\( p_0 = (p_{x_0}, p_{y_0}, p_{z_0}) \) up to the second order is:

\begin{align}
f(p_x, p_y, p_z) &\approx f(p_0) 
+ \frac{\partial f}{\partial x} \bigg|_{p_0} (p_x - p_{x_0}) \notag \\
&\quad + \frac{\partial f}{\partial y} \bigg|_{p_0} (p_y - p_{y_0}) \notag \\
&\quad + \frac{\partial f}{\partial z} \bigg|_{p_0} (p_z - p_{z_0}) \notag \\
&\quad + \frac{1}{2} \frac{\partial^2 f}{\partial x^2} \bigg|_{p_0} (p_x - p_{x_0})^2 \notag \\
&\quad + \frac{1}{2} \frac{\partial^2 f}{\partial y^2} \bigg|_{p_0} (p_y - p_{y_0})^2 \notag \\
&\quad + \frac{1}{2} \frac{\partial^2 f}{\partial z^2} \bigg|_{p_0} (p_z - p_{z_0})^2.
\end{align}

The zeroth-order term is:

\begin{align}
f(p_{x_0}, p_{y_0}, p_{z_0}) &= \sin^2(w p_{x_0}) + \sin^2(w p_{y_0}) \notag \\
&\quad + \sin^2(w p_{z_0}) + w.
\end{align}

The first-order terms are:

\begin{align}
\frac{\partial f}{\partial x} \bigg|_{p_0} &= w \sin(2w p_{x_0}), \notag \\
\frac{\partial f}{\partial y} \bigg|_{p_0} &= w \sin(2w p_{y_0}), \notag \\
\frac{\partial f}{\partial z} \bigg|_{p_0} &= w \sin(2w p_{z_0}).
\end{align}

The second-order terms are:

\begin{align}
\frac{\partial^2 f}{\partial x^2} \bigg|_{p_0} &= 2w^2 \cos(2w p_{x_0}), \notag \\
\frac{\partial^2 f}{\partial y^2} \bigg|_{p_0} &= 2w^2 \cos(2w p_{y_0}), \notag \\
\frac{\partial^2 f}{\partial z^2} \bigg|_{p_0} &= 2w^2 \cos(2w p_{z_0}).
\end{align}

Substituting these into the Taylor expansion:

\begin{align}
f(p_x, p_y, p_z) &\approx \sin^2(w p_{x_0}) + \sin^2(w p_{y_0}) \notag \\
&\quad + \sin^2(w p_{z_0}) + w \notag \\
&\quad + w \sin(2w p_{x_0}) (p_x - p_{x_0}) \notag \\
&\quad + w \sin(2w p_{y_0}) (p_y - p_{y_0}) \notag \\
&\quad + w \sin(2w p_{z_0}) (p_z - p_{z_0}) \notag \\
&\quad + w^2 \cos(2w p_{x_0}) (p_x - p_{x_0})^2 \notag \\
&\quad + w^2 \cos(2w p_{y_0}) (p_y - p_{y_0})^2 \notag \\
&\quad + w^2 \cos(2w p_{z_0}) (p_z - p_{z_0})^2.
\end{align}

\begin{figure}
    \centering
     \begin{tabular}{@{}c@{}}
    \includegraphics[width=\linewidth]{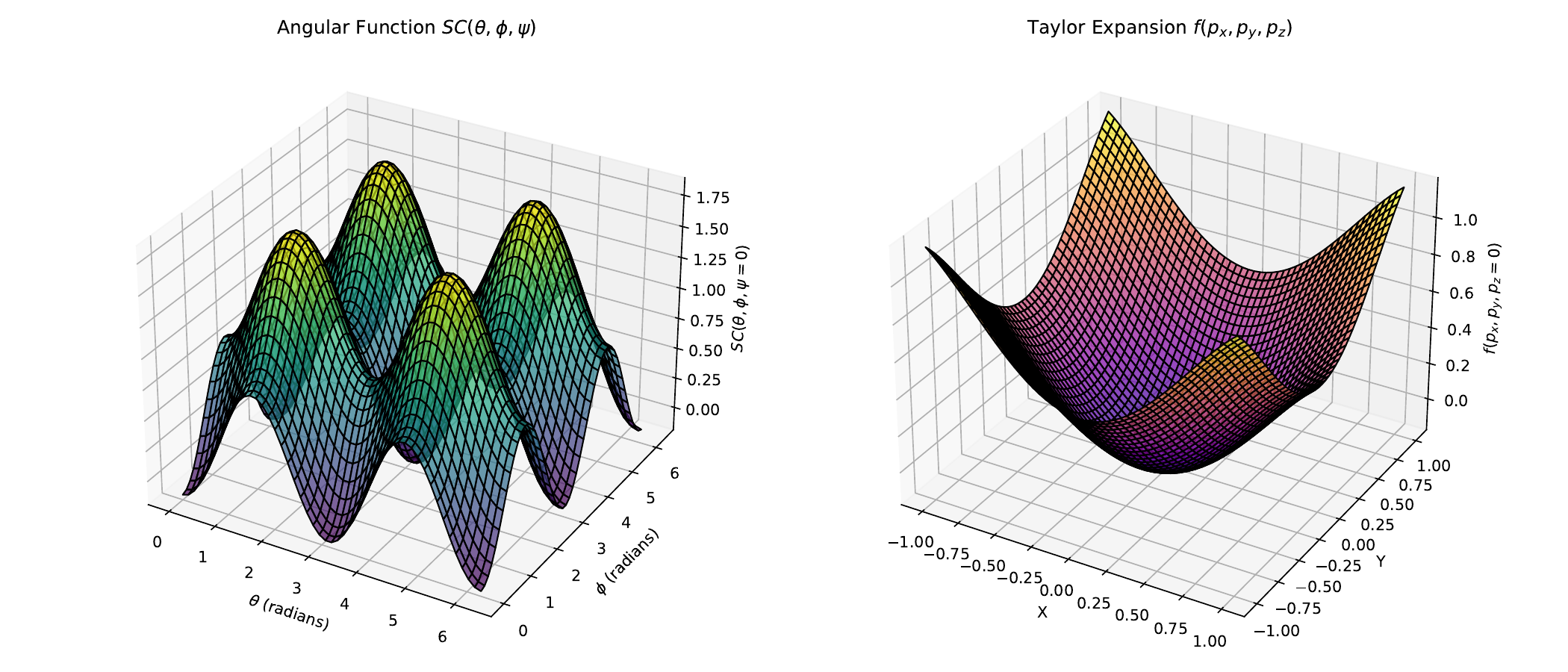}
    \end{tabular} 
  \caption{The left plot shows the visualization of the manifold from the implicit periodic function in \ref{eq:particles}, which can be used to generate the spherical particle cloud, and the right plot shows the visualization of the manifold from the polynomial function of a spherical primitive shape derived from the Taylor the series expansion of the left one, which is basically equal to the base primitive for the SDF in Eq. \ref{eq:g-sdf} in the main document. The diffeomorphism exists between the two manifolds, the transformation (or mapping between two manifolds) $\theta = w_kp_x, \phi = w_kp_y, \psi = w_kp_z$ (or it's inverse) likely preserves distances, implies that the transformation is an isometry. Since isometries are also diffeomorphisms.}
  \label{fig:coordinate-corres}
\end{figure}

The Taylor expansion of the periodic function in Eq.\ref{eq:particles} leads to the quadratic polynomial function representing the sphere, which means that the evaluation of the SDF of Eq.\ref{eq:particles} directly generates the spherical particle in each of the grid cell. which leads to the implicit function for the sphere, and it is the base primitive for the particle SDF in Eq. \ref{eq:g-sdf} in the main document. See Fig.\ref{fig:coordinate-corres} for the visualization of the mapping from the angular coordinates for the periodic implicit function to the polynomial generated from the Taylor expansion.
Now we consider the expansion around the Point \((0, 0, 0)\),thus, the expansion around \((0, 0, 0)\) is:
\[
f(p_x, p_y, p_z) \approx w + w^2 {p_x}^2 + w^2 {p_y}^2 + w^2 {p_z}^2.
\]

\section{Sphere Tracing Optimization}\label{sec:sto}

In this paper, we present an optimized sphere-tracing algorithm aimed at improving the performance of sphere tracing for multiscale grid variations. 
\clearpage
\begin{algorithm}[H]
\caption{Optimized sphere tracing with adaptive step size based on gradients of the SDF and the polynomial function $D(p)$}
\begin{algorithmic}[1]
\Function{SDF}{$p$}
    \State \Return $d$ \Comment{Signed distance at point $p$}
\EndFunction

\Function{LERP}{$a$, $b$, $t$}
    \State \Return $a + t \cdot (b - a)$
\EndFunction

\Function{ComputeGridScale}{$p$}
    \State $\varepsilon \gets 0.01$
    \State $dp_x \gets (\varepsilon, 0, 0)$
    \State $dp_y \gets (0, \varepsilon, 0)$
    \State $dp_z \gets (0, 0, \varepsilon)$
    
    \State $gp_x \gets |\text{SDF}(p + dp_x) - \text{SDF}(p - dp_x)|$
    \State $gp_y \gets |\text{SDF}(p + dp_y) - \text{SDF}(p - dp_y)|$
    \State $gp_z \gets |\text{SDF}(p + dp_z) - \text{SDF}(p - dp_z)|$
    
    \State $|\nabla d| \gets \sqrt{gp_x^2 + gp_y^2 + gp_z^2}$
    \State $scale \gets \text{clamp}(|\nabla d| \cdot 0.5, 0.0, 1.0)$
    \State \Return $scale$
\EndFunction

\Function{Sphere-tracing-multiscale-grid}{$ro$, $rd$, $t_{\min}$, $t_{\max}$}
    \State $t \gets t_{\min}$
    \State $d \gets \text{SDF}(ro + t \cdot rd)$
    \State $s \gets \text{sign}(d)$
    \State $lastScale \gets \text{ComputeGridScale}(ro + t \cdot rd)$

    \For{$i = 0$ to $n$}
        \If{$|d| < \text{precision} \cdot t$ \textbf{or} $t > t_{\max}$}
            \State \textbf{break}
        \EndIf

        \If{$D(p) = 1$} \Comment{Evaluate gradient only when polynomial checks $D(p)$ is true}
            \State $lastScale \gets \text{ComputeGridScale}(ro + t \cdot rd)$
        \EndIf

        \State $stepFactor \gets \text{LERP}(\delta_{\min}, 1.0, lastScale)$
        \State $t \gets t + s \cdot d \cdot stepFactor$
        \State $d \gets \text{SDF}(ro + t \cdot rd)$
    \EndFor
    \State \Return $t$
\EndFunction
\end{algorithmic}
\end{algorithm}

In the above algorithm, on line 28, we can use any other condition for reducing the gradient computations to improve the performance during sphere-tracing steps. For instance, we can simply check that using the following condition, where we check gradients to track the variations in grid scale for every m-step length. 
\begin{equation*}
    \text{If } i \mod m = 0, \text{ where } m < n
\end{equation*}

For instance, we use something like the following functions for the \( D(p) \) in the above algorithm to track the gradient changes. Essentially, we are sampling some sparse points throughout the volume to effectively track the grid scale variations. The polynomial conditional check \( D(p) \) is a way to reduce the gradient checks in sphere tracing iterations.

\begin{equation}
D(p) = 
\begin{cases}
1, & \text{if } \left| \mathrm{fmod}(p_y, n_1) \right| = 0.0 \\
   &  \lor \left( \left| \mathrm{fmod}(p_z, n_2) \right| = 0.0 \land \left| \mathrm{fmod}(p_x, n_3) \right| = 0.0 \right) \\
0, & \text{otherwise}
\end{cases}
\label{eq:fm-case}
\end{equation}

\begin{equation}
\begin{cases}
1, & \text{if } \begin{aligned}
       &\sin(p_x) \cos(p_y) + \sin(p_y) \cos(p_z) \\
       &+ \sin(p_z) \cos(p_x) = 0.5
     \end{aligned} \\
0, & \text{otherwise}
\end{cases}
\label{eq:sc-case}
\end{equation}

\begin{figure*}
    \centering
     \begin{tabular}{@{}c@{}c@{}c@{}}
    \includegraphics[width=0.33\linewidth]{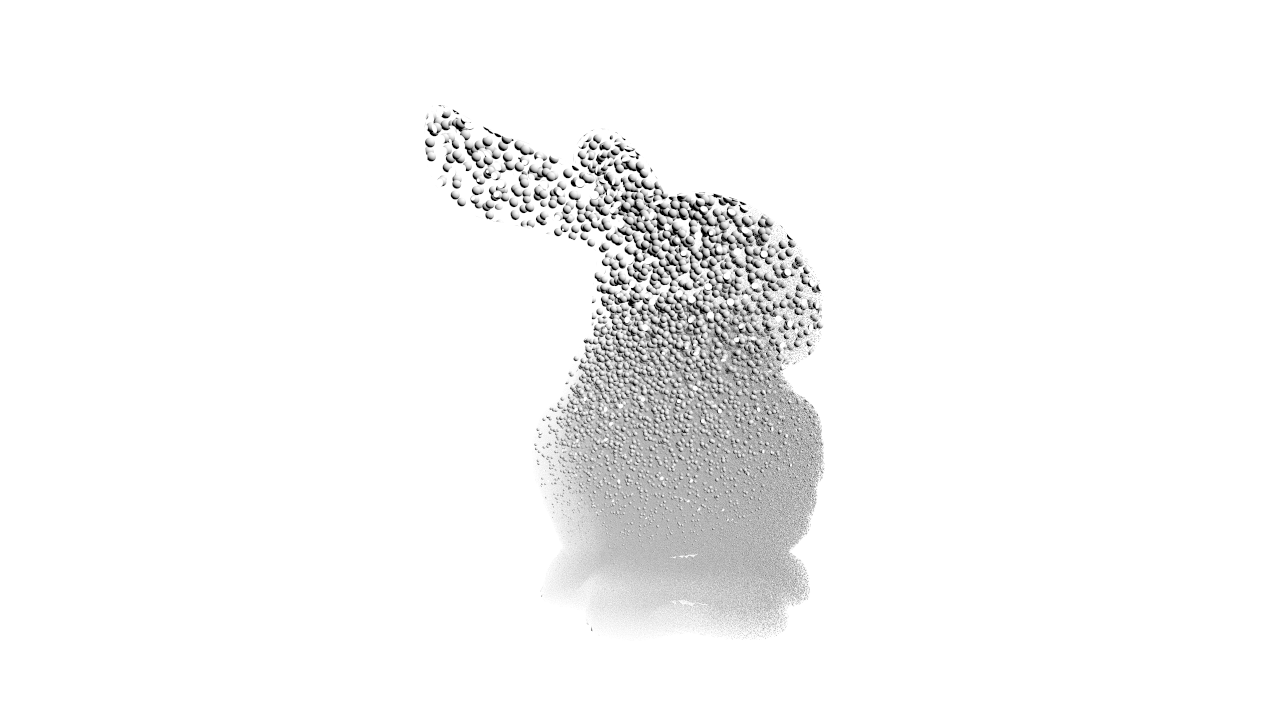}
    \includegraphics[width=0.33\linewidth]{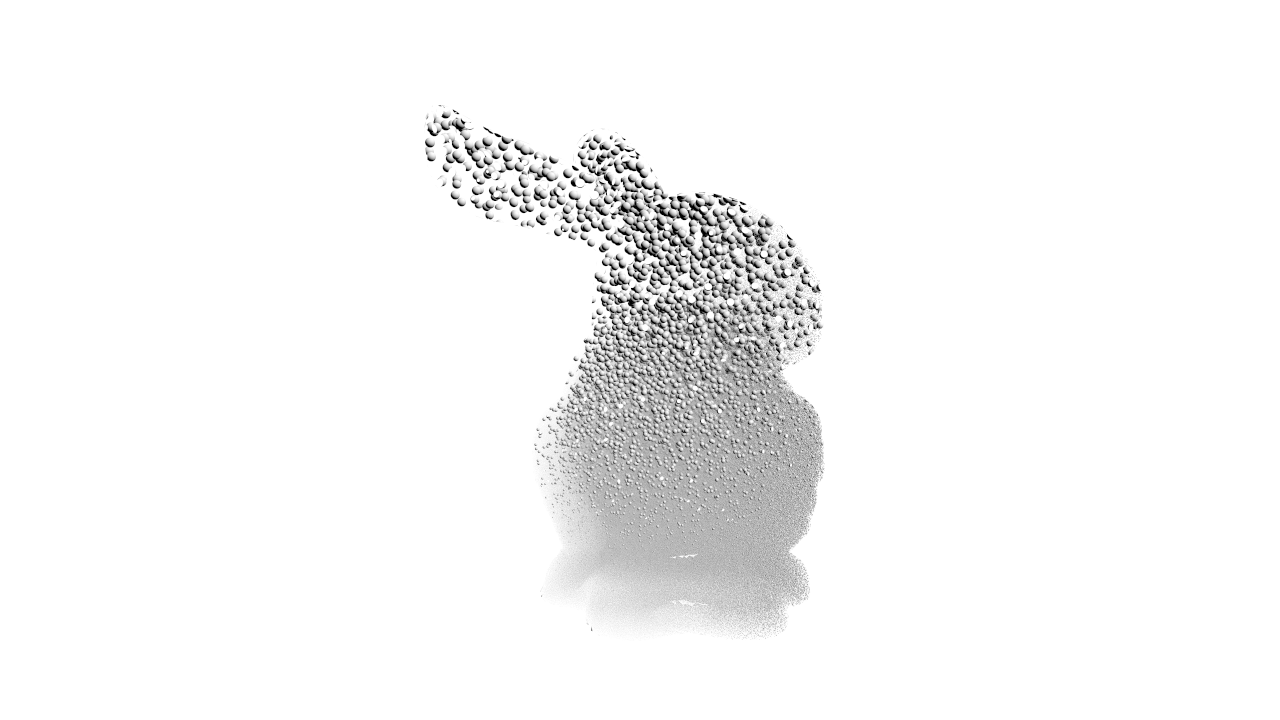}
    \includegraphics[width=0.33\linewidth]{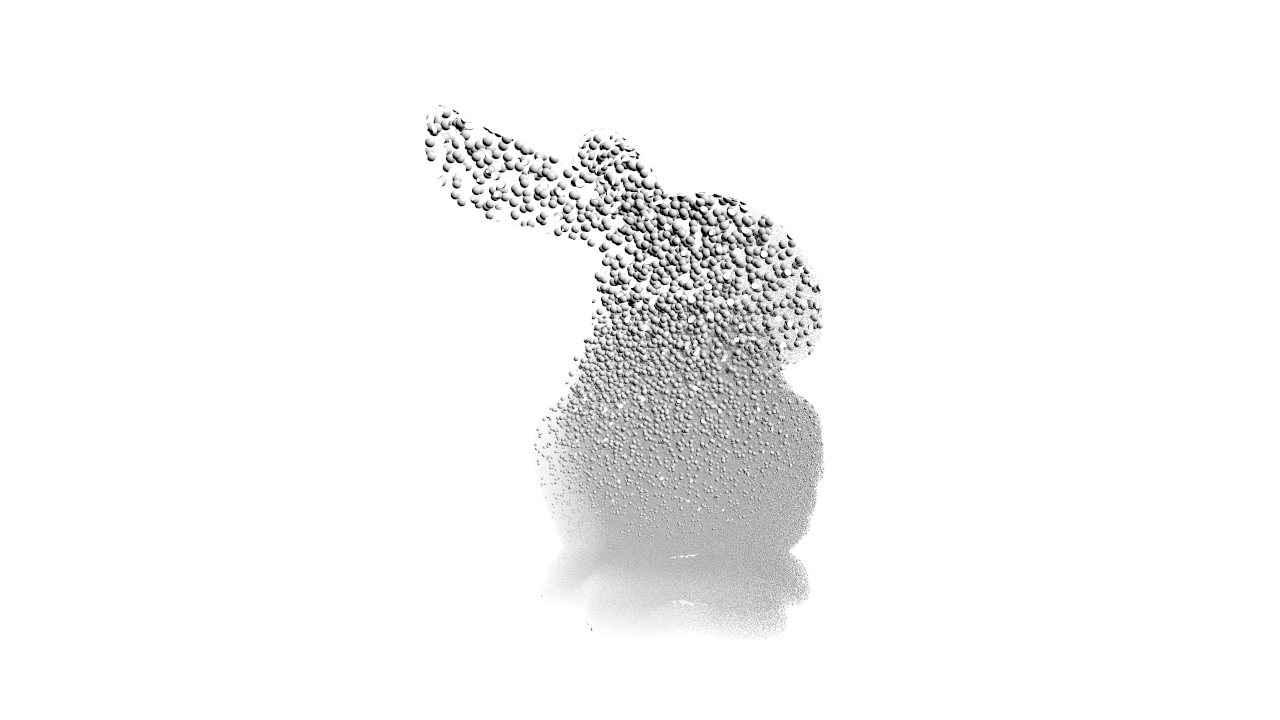}

    \end{tabular}
    
  \caption{Visualization of the rendered multiscale grid variational geometry with different step length methods. The results are shown for fixed step lengths, as well as for the two polynomial checks \(D(p)\) (as referenced in equations ~\ref{eq:fm-case} and ~\ref{eq:sc-case}) used for adaptive step lengths, respectively. The performance values for these three cases are as follows: for the fixed lengths case, the results are (RT - 8.6 ms, FPS - 112.3). For the polynomial-based gradient checks with the equation in ~\ref{eq:fm-case}, where the values of \(n1, n2, n3\) are 11, 5, and 7, the results are (RT - 7.0 ms, FPS - 140.1). With the equation in ~\ref{eq:sc-case}, the results are (RT - 6.9 ms, FPS - 141.6). It clearly shows that polynomial-based gradient checks for tracking the grid scale variations inside the sphere tracing algorithm improve the performance.}
  \label{fig:step-l1}
\end{figure*}
All of these methods can improve performance, but the use of polynomial checks to track variations in the grid scale is particularly effective. It allows for smaller step lengths in finer grid-scale regions and larger step lengths in coarser grid scales.

\section{Multiscale Geometry Reconstruction} \label{sec:mgr}

In this session, we address the reconstruction of the procedural microstructures introduced previously in Sec. ~\ref{sec:optimization} with two approaches: parameter fitting and analysis by synthesis. The second approach leverages a rendering pipeline to guide the optimization, while the first one consists of a direct fitting of the procedural SDF. For each approach, we outline the optimization problem and its solution, including the required data, the loss functions, and optimization algorithms. To validate these approaches as a proof of concept, we focus on synthetic microstructures, assuming the availability of either a signed distance field or an RGB image.


\subsection{Parameter Fitting}
\label{parameter_fitting_approach-add}

This approach formulates microstructure reconstruction as a model-fitting task, where the goal is to optimize the parameters of a procedural SDF to accurately match the microstructure's SDF ground truth. Given a set of spatial coordinates $\Omega = \{\mathbf{x} \mid \mathbf{x} \in \mathbb{R}^3\}$, a ground-truth SDF $\hat{S}:~\mathbb{R}^3 \rightarrow \mathbb{R}$ describing the surface at point $x$, a bounded parameter space $\Phi = [a_1, b_1] \times [a_2, b_2] \times \cdots \times [a_n, b_n]$ with $n$ parameters, and a loss function $\mathcal{L}: \mathbb{R} \times \mathbb{R} \rightarrow \mathbb{R}$, the problem is defined as minimizing the discrepancy between the fitted function $f$ and the ground truth $\hat{S}$ across the domain: 
\begin{equation}
\min_{\boldsymbol{\phi} \in \Phi} \int_{\Omega} \mathcal{L}(f(\mathbf{x}, \boldsymbol{\phi}), \hat{S}(\mathbf{x})) d\mathbf{x}.
\end{equation}

$f$, in our case, is a suitable representation for the given microstructure. Procedural SDFs allow efficient extrapolation of the microstructure beyond the optimization domain without modifying the fitted function. In addition, our compact parametrized microstructures are described with a relatively low number of parameters. We propose a loss function to evaluate the discrepancies in the frequency domain using the discrete Fourier transform (DFT) to consider the periodic behavior common in many microstructures. The loss function compares the amplitude and phase spectra of the Fourier coefficients, using the logarithm of the mean squared error (MSE) as an error metric:
%

\begin{multline}
\mathcal{L}_{FT} \left( f(\Omega, \boldsymbol{\phi}), \hat{S}(\Omega) \right) = \\
\log \biggl( \frac{1}{2N} \sum_{i=1}^{N} \Bigl( \left| A_i(\mathcal{F}(f(\Omega, \boldsymbol{\phi}))) - A_i(\mathcal{F}(\hat{S}(\Omega))) \right|^2 + \\
\left| \theta_i(\mathcal{F}(f(\Omega, \boldsymbol{\phi}))) - \theta_i(\mathcal{F}(\hat{S}(\Omega))) \right|^2 + \epsilon \Bigr) \biggr).
\label{FT_loss}
\end{multline}

For simplicity we use $\Omega$ to express a series of points, $\mathcal{F}$ being a DFT,  $A_i(\mathcal{F}(f(\Omega, \boldsymbol{\phi})))$ and $A_i(\mathcal{F}(\hat{S}(\Omega)))$ denoting the amplitudes, and \(\theta_i(f(\Omega, \boldsymbol{\phi}))\) and \(\theta_i(\hat{S}(\Omega))\) the phases of the Fourier coefficients derived from $f(\Omega, \boldsymbol{\phi})$ and $\hat{S}(\Omega)$, respectively, \(N\) representing the number of coefficients, and $\epsilon$ being a small positive value. Recall that $A_i(\mathcal{F}(x)) = |z_i| = \sqrt{\Re(z_i)^2 + \Im(z_i)^2}$, and $\theta_i(\mathcal{F}(x)) = \arctan\left(\frac{\Im(z_i)}{\Re(z_i)}\right)$, where $z_i$ is the $i$-th Fourier coefficient of $x$, and $\Re(z_i)$ and $\Im(z_i)$ are the real and imaginary parts of $z_i$, respectively. This loss is differentiable, enabling us to use gradient-based optimization algorithms as the operations involved are differentiable, including the DFT. Note this equation is closely related to the spectral density function commonly used in microstructure analysis~\cite{bostanabad2018computational}. 

For synthetic examples, ground-truth data generation consists of evaluating the corresponding target procedural SDF and its optimal parameters using a sampling method. Sampling points are generated within a transformed coordinate system as slices of a unit sphere in $\mathbb{R}^{D \times W \times H}$, aligned with the rotated basis vectors. These points are equidistant and centered at the origin, with Gaussian noise added to introduce small perturbations. Specifically, $i \in [1, D]$, $j \in [1, W]$, and $k \in [1, H]$ represent the indices corresponding to the generated points in the respective directions, and $\bar{\mathbf{x}}$ is a point inside the coordinates space with standard basis:

\begin{equation}
    \Omega = \left\{ \bar{\mathbf{x}}_{i,j,k} + \boldsymbol{\epsilon}_{i,j,k} \mid \boldsymbol{\epsilon}_{i,j,k} \sim \mathcal{N}(0, \sigma^2 \mathbf{I}) \right\}. 
\end{equation}
\label{point_generation_equation} 

Selecting a suitable optimization algorithm is critical for solving unconstraint global optimization problems. We employ two first-orded methods taking advantage of the gradient information: Basin-Hoppin(BH)~\cite{wales2003energy} and Simplicial Homology Global Optimization (SHGO)\cite{endres2018simplicial} algorithms. The BH algorithm is particularly effective for problems with funnel-like loss functions. Its key hyperparameters include the step size $s$ and the temperature $T$, which define the bounds of the uniformly distributed random perturbations and the acceptance probability for the candidate function, respectively. The temperature $T$ is empirically set to $T = 0.001$. Additionally, the Sequential Least-Squares Quadratic Programming (SLSQP) algorithm is used as a local minimizer in the second phase, which iteratively solves a QP subproblem derived from a second-order Taylor expansion of the loss function. The BH algorithm requires an initial $\boldsymbol{\phi}_0$. SHGO global optimization algorithm approximates locally convex subdomains using homology groups. The main hyperparameters for SHGO include the number of iterations  $l$ and the number of samples  $m$ for building the complex. Depending on problem complexity, we typically select $m \in [10,10000]  \cap \mathbb{N}$ and set $l = 1$. The Sobol sampling method and SLSQP are used for local search. Unlike the BH algorithm, SHGO does not require an initial population mean $\boldsymbol{\phi}_0$ but requires variable bounds.

As previously noted, the loss function is differentiable, but certain procedural SDFs are not differentiable, leading to non-differentiability of the final composite function $\mathcal{L} \circ f$. For gradient-free cases, we use the Covariance Matrix Adaptation Evolutionary Strategy (CMA-ES) algorithm~\cite{hansen2019pycma} as it is highly effective, in particular for rugged and non-differentiable problems. Its key hyperparameters are the population size $\mathcal{P}$ and the initial $\sigma_0$ of the covariance matrix, typically chosen depending on the complexity and dimensional of the problem. In our case, $\mathcal{P} \in [40, 400]$ and $\sigma_0 \in [70, 250]$. CMA-ES requires an initial population mean $\boldsymbol{\phi}_0$. 

\subsection{Analysis by Synthesis}
\label{analysis_by_synthesis_approach-add}

The main difference between the analytical optimization-based approach discussed in the first section and the analysis-by-synthesis method lies in the comparison process. Despite this difference, both methods aim to find an optimal SDF. In the latter approach, we employ a non-differentiable rendering pipeline to visualize the SDF and subsequently compare the resulting renders. Given a camera intrinsics $\mathbf{K} \in \mathbb{R}^{3\times 3}$ taken from a set of intrinsic matrices $\mathcal{K}$, the camera extrinsics $\mathbf{E} = [\mathbf{R} \mid \mathbf{t}] \in \mathbb{R}^{3\times 4}$  taken from a set of extrinsic matrices $\mathcal{E}$, a labeling function $\hat{S}:~\mathbb{R}^3 \rightarrow \mathbb{R}$, a bounded parameter space $\Phi = [a_1, b_1] \times [a_2, b_2] \times \cdots \times [a_n, b_n]$ with $n$ parameters, a signed distance fitting function $f: \mathbb{R}^3 \times \Phi \rightarrow \mathbb{R}$, a loss function $\mathcal{L}: \mathbb{R}^{W\times H \times 3} \times \mathbb{R}^{W\times H \times 3} \rightarrow \mathbb{R}$, with $W, H$ being width and the height respectively, a set of functions $\mathcal{A}$, and finally the rendering function $\mathcal{R}:~\mathcal{A}\times \mathcal{K} \times \mathcal{E} \rightarrow \mathbb{R}^{W\times H \times 3}$, we can express the problem as follows:

\begin{equation}
    \min_{\boldsymbol{\phi} \in \Phi} \mathcal{L}\left( \mathcal{R}\left( f\left( \cdot, \boldsymbol{\phi}\right), \mathbf{K}, \mathbf{E}\right), \mathcal{R} ( \hat{S}\left( \cdot \right), \mathbf{K}, \mathbf{E} ) \right).
\label{analysis_by_synthesis_eq-add}
\end{equation}

$\mathcal{R}$ renders the microstructure bounded by a unit sphere from the given view with the camera parameters using the sphere-tracing algorithm~\cite{hart1996sphere} in the CUDA-based framework OptiX \cite{parker2010optix}. Similar to the parameter fitting approach, we can set $f := \hat{f}$. To generate the ground-truth data for the analysis-by-synthesis approach, we must provide a target render image. This involves rendering the microstructure using the $\mathcal{R}\Bigl(\hat{S}(\cdot), \mathbf{K}, \mathbf{E} \Bigr)$ with a fixed $\mathbf{K}, \mathbf{E}$ later used for the optimization.\\ 

For the loss function, we use the two-dimensional DFT, denoted as $\mathcal{F}_{2D}$, to transform the image signal for comparison in the frequency domain expressed as:
%
%
\begin{align}
\mathcal{L}_{2DFT} &\left( \mathcal{R}(f(\cdot, \boldsymbol{\phi}), \mathbf{K}, \mathbf{E}), \mathcal{R}(\hat{S}(\cdot), \mathbf{K}, \mathbf{E}) \right) = \notag \\
&\log \biggl( \frac{1}{N} \sum_{i=1}^{N} \left| A_i\Bigl(\mathcal{F}_{2D}(\mathcal{R}(f(\cdot, \boldsymbol{\phi}), \mathbf{K}, \mathbf{E})) \Bigr) - \right. \notag \\
&\quad \left. A_i \Bigl(\mathcal{F}_{2D}(\mathcal{R}(\hat{S}(\cdot), \mathbf{K}, \mathbf{E})) \Bigr) \right|^2 + \epsilon \biggr),  
\label{FT_loss_image}
\end{align}

where $A_i\Bigl(\mathcal{F}_{2D}(\mathcal{R}(f(\cdot, \boldsymbol{\phi}), \mathbf{K}, \mathbf{E}))\Bigr) $ and $A_i\Bigl(\mathcal{F}_{2D} (\mathcal{R}(\hat{S}(\cdot), \mathbf{K}, \mathbf{E}) ) \Bigr)$ denote the amplitudes of the 2D Fourier coefficients for the rendered image and the target image, respectively. $N$ represents the total number of coefficients, and $\epsilon$ is a small constant. By focusing on the amplitudes of the Fourier coefficients, we capture the key periodic features of the image, which proved to be crucial for effective global optimization in our experiments. While multi-view optimization is theoretically possible, we opted for single-view optimization due to its computational overhead, as rendering multiple views would be even more time-consuming.\\ 

In this approach, the rendering function $\mathcal{R}$ is non-differentiable, forcing us to only use gradient-free optimizers. The first optimizer we consider is again the evolutionary algorithm CMA-ES \cite{hansen2019pycma} as it is gradient-free, adaptive, robust to noise, and effective in complex landscapes. As the evaluation of the loss function in Equation~\ref{FT_loss_image} is computationally expensive due to the overhead of $\mathcal{R}$, we decide to decrease the population size to $\mathcal{P} \in [5, 10] \cap \mathbb{N}$. For our microstructure experiments, we maintain the initial standard deviation at $\sigma_0 \in~[60, 80]$. The second optimization algorithm that we considered is the slightly modified Powell method \cite{press1988numerical, powell1964efficient}, a local optimizer without hyperparameters that can be proven to be effective when a reasonable close initialization is provided.

\subsection{Synthetic Dataset}
\label{synthetic_dataset}

\begin{figure}[ht]
    \centering
    \begin{adjustbox}{max width=\textwidth, max height=0.17\textheight}
    \setlength{\tabcolsep}{1.5pt} 
    \renewcommand{\arraystretch}{0.3} 
    \begin{tabular}{c c c}
        \includegraphics[width=0.15\textwidth]{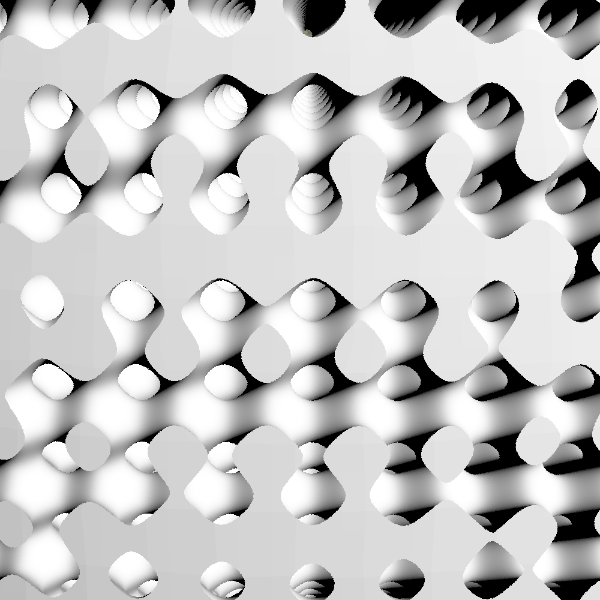} &
        \includegraphics[width=0.15\textwidth]{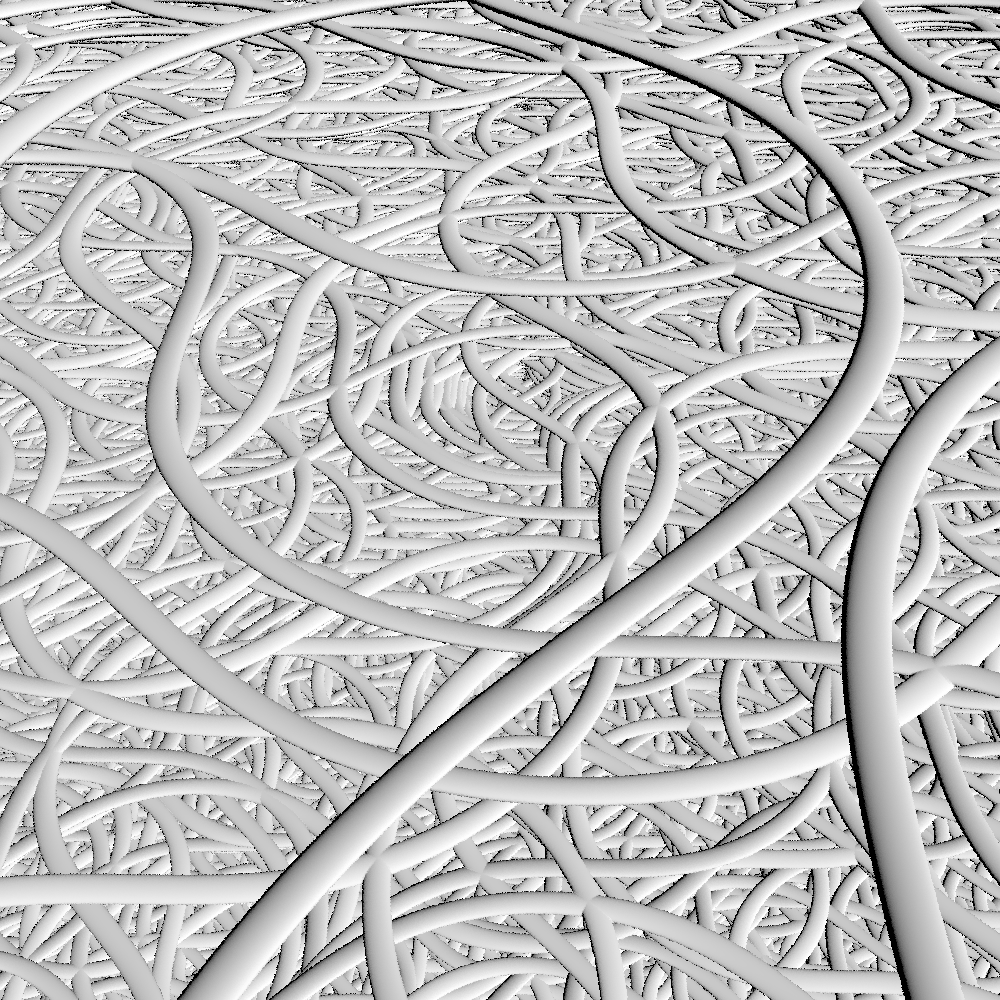} &
        \includegraphics[width=0.15\textwidth]{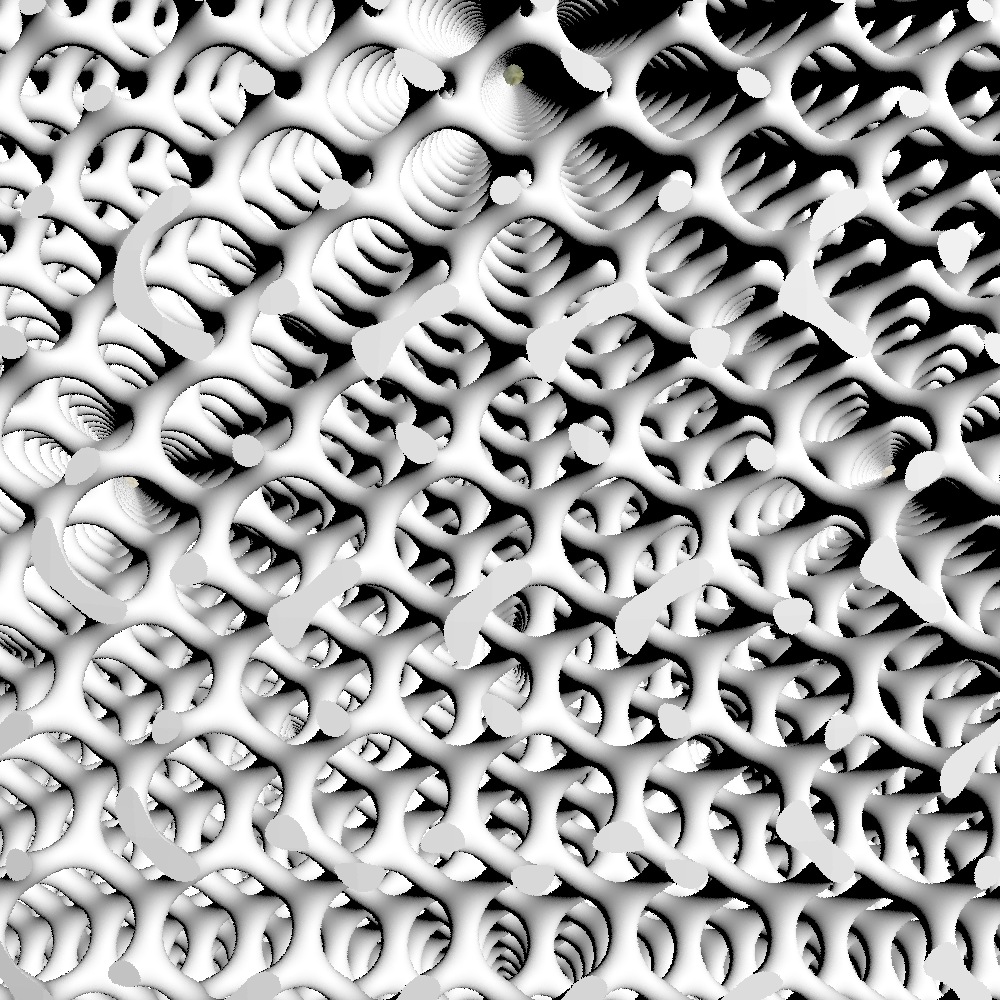}\\
        \includegraphics[width=0.15\textwidth]{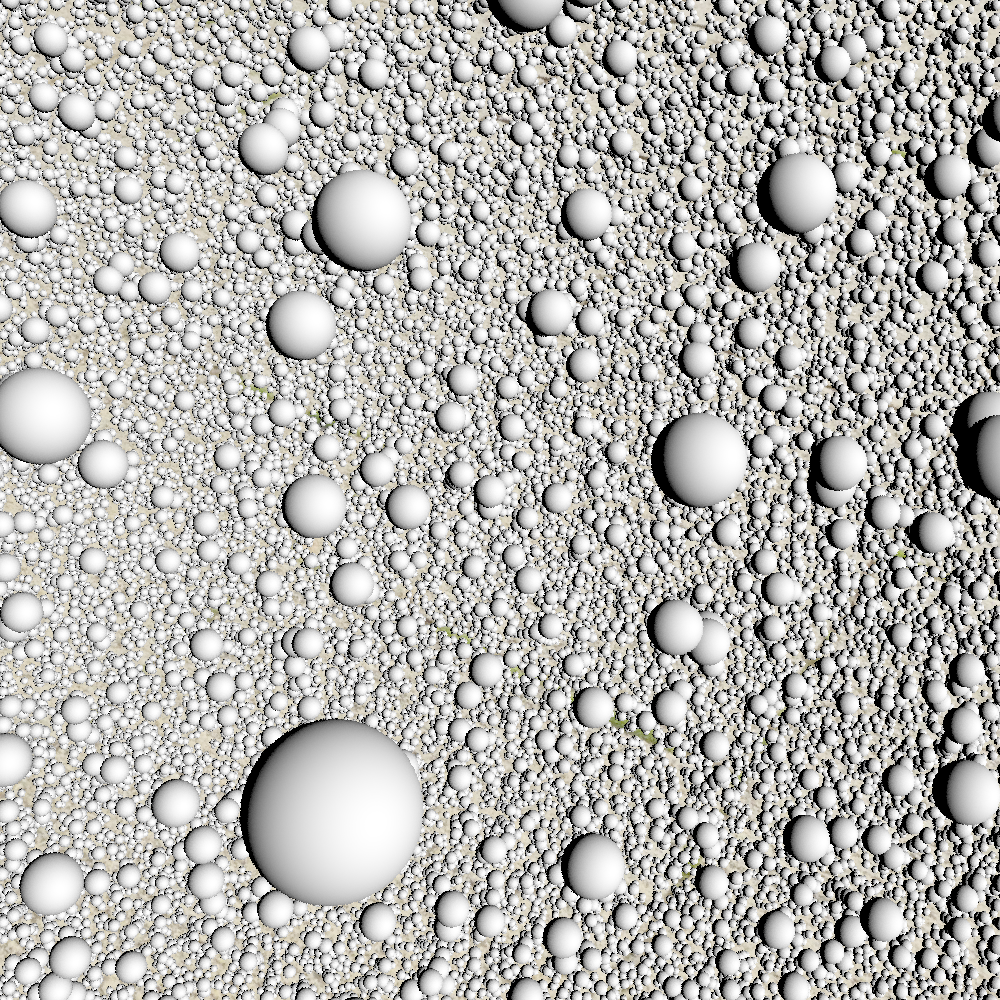} &
        \includegraphics[width=0.15\textwidth]{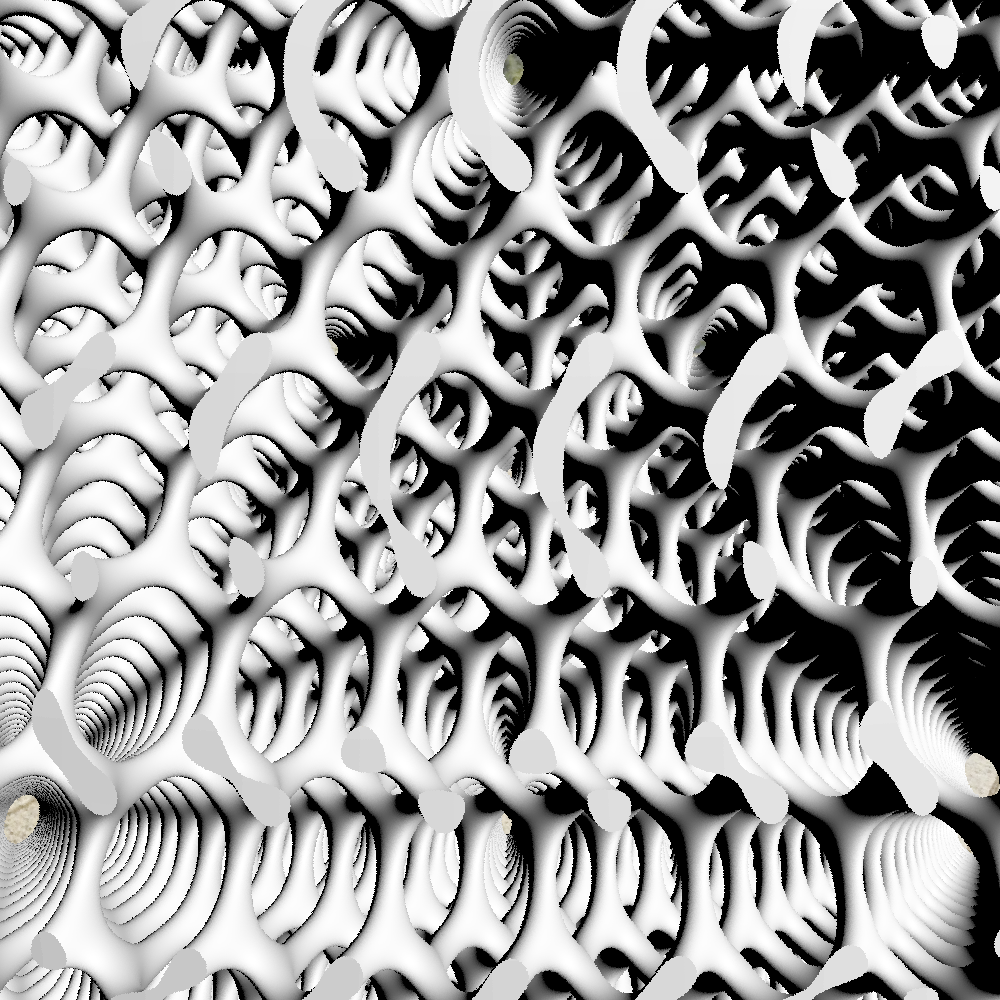} &
        \includegraphics[width=0.15\textwidth]{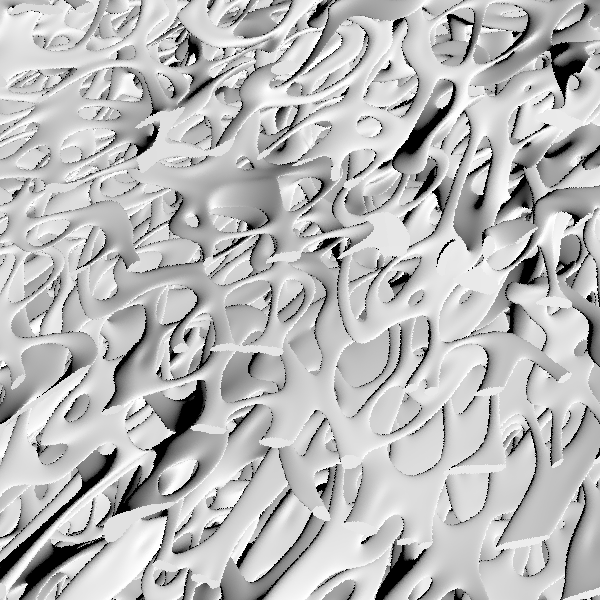} 
    \end{tabular}
    \end{adjustbox}
    \caption{Instances of exemplar renders (\textit{top row}) Gyroid~[1D] \ref{gyroid1d}, Fibers~[2D] \ref{fibers_sdf}, Gyroid~[3D] \ref{gyroid3d} respectively. Rest of the renders (\textit{bottom row}) are Spheres~[2D]~\ref{sphere_microstructure}, Gyroid~[5D] \ref{gyroid5d}, Porous~[28D]~\ref{porous_material_sdf} respectively.}
    \label{tab:renders_exemplars}
\end{figure}

For the proof of concept and more control over the experiments, we test our algorithms on representative synthetic examples that closely resemble SEM/TEM images. Let us now describe our synthetic dataset of microstructures. The dataset consists of six microstructures. Each microstructure is defined by a SDF $\hat{f}$ with $n$ parameters, denoted as $\boldsymbol{\phi}$ in Approaches~\ref{parameter_fitting_approach} and~\ref{analysis_by_synthesis_approach}. We include the number of parameters inside square brackets in the naming conventions to reflect both the degrees of freedom and to provide an indication of the optimization complexity. All ground-truth parameter values presented in the dataset were selected for their convenience and to ensure they fit in the defined domain range.

\subsubsection{Gyroid Family}  

The first family of microstructures are the gyroid microstructures. These consist of periodic fully deterministic patterns created using trigonometric functions. Our dataset includes three such microstructures with 1, 3, 5 degrees of freedom respectively. 

The simplest example is the Gyroid [1D]. The only parameter is a scalar variable $\eta \in \mathbb{R}$. We call $\eta$ the noise scale. The $\eta$ variable controls the density of the microgeometry. We create the ground-truth using $\eta = 100$:

\begin{equation}
\hat{f}(\mathbf{x} \mid \eta) = \sum_i^3{\sin(x_i \cdot \eta )} / \eta.
\label{gyroid1d}
\end{equation}

The second element from the set of gyroid functions, Gyroid [3D], is guided by $\eta \in \mathbb{R}$, and $k = 2 \cdot \pi / a \in \mathbb{R}, t \in \mathbb{R}$. The $a$ represents the cell size of the gyroid structure, $t$ represents the wall thickness. 
We create the ground-truth using the setting of $\eta = 100.0$, $a = 7.0$, $t = 1.2$ as follows
\begin{gather}
 \label{gyroid3d}
 \hat{f}(\mathbf{x} \mid \eta, k, t  ) = g(\mathbf{x} \cdot \eta \mid k, t  ) / \eta ,\\
 \notag\\
g(\mathbf{x} \mid k, t ) = \sin(k x_0)\cos(k x_1)+ \\ \sin(k x_1)\cos(kx_2)+\sin(kx_2)\cos(kx_0)+t.
\end{gather}
Finally, the last exemplar, the Gyroid [5D], closely resembles the Gyroid [3D] problem. The only difference is the division of the $\mathbf{a} \in \mathbb{R}^3$ parameter into three variables $k_i = 2 \cdot \pi / a_i$, each controlling the cell size of the gyroid in the $x$, $y$, and $z$ directions. We create the ground-truth using the setting $\eta = 100.0$, $\mathbf{a} = (7.0, 10.0, 15.0)$, $t = 1.2$. The fitting is again thus: 
\begin{gather}
\label{gyroid5d}
\hat{f}(\mathbf{x} \mid \eta, \mathbf{k}, t ) = r(\mathbf{x} \cdot \eta \mid \mathbf{k}, t  ) / \eta, \\
\notag\\
r(\mathbf{x} \mid \mathbf{k}, t ) = \sin(k_1 x_0)\cos(k_2 x_1)+ \\ \sin(k_2 x_1)\cos(k_3 x_2)+\sin(k_3 x_2)\cos(k_1 x_0)+t,
\end{gather}
Note that the $\hat{f}$ is, in all cases, differentiable, thus enabling the use of first-order optimization. 

\subsubsection{Spherical Microstructure}

Our next microstructure consists of a signed distance function for randomly placed spheres. We generate a granular material microgeometry; to achieve this, we use a space-filling grid-based random variable of points. A grid cell index, which is computed from its floored location, is used to generate a seed $t$ for PRNG. A multilinear hash function defining the seed looks as follows:

\begin{equation}
t_l(\mathbf{x}) = M \sum_{i=0}^1 \sum_{j=0}^1 \sum_{k=0}^1 a_{ijk} q_l(x_0)^i q_l(x_1)^j q_l(x_2)^k,    
\end{equation}

where $(q_l(x_0), q_l(x_1), q_l(x_2))$ are grid indices generated in eight possible directions by adding binary-encoded offsets, dependent on $l$, to the base grid coordinates computed from the floored location $\eta \cdot \mathbf{x}$ which is noise-scale $\eta$ dependent. $M$ and all $a_{ijk}$ were selected such that regularity artifacts are avoided.  
Seed $t_l$ is later used as an input to PRNG, generating a pseudo-random number in the interval~$[0, 1)$. We then add the randomly generated number to the grid position to obtain the center of a random sphere in the current grid. We repeat the described procedure in total 8 times (once for each potential overlapping grid cell). Finally, the signed distance is calculated by finding a minimum distance to all spheres taking into account their radius. The variables are the density-controlling $\eta \in \mathbb{R}$ and the radius of the spheres $r \in \mathbb{R}$. Mathematically, the procedure looks as follows
\begin{gather}   
\label{sphere_microstructure}
\hat{f}(\boldsymbol{x} \mid \eta, r ) = \Bigl( \min_{l=1}^8 || c_l(\eta \cdot\mathbf{x}, t_l(\eta \cdot\mathbf{x})) - \mathbf{x} ||_2  - r \Bigr) / \eta,    \\
c_l(\eta \cdot\mathbf{x}, t_l(\eta \cdot\mathbf{x})) = \mathbf{q}_l(\eta \cdot\mathbf{x}) + \text{PRNG}(t_l(\eta \cdot\mathbf{x})), \\
\mathbf{q}_l(\eta \cdot \mathbf{x}) = \lfloor \eta \cdot \mathbf{x} \rfloor + \mathcal{U}_l, 
\end{gather}

where $\mathbf{x}, t_i(\mathbf{x}), c(\mathbf{x}, t_i(\mathbf{x}))$, $\mathcal{U}_l$ are a point of evaluation, generated random seed, generated center of a random sphere, and $l$-th element of the unit cube vertices $\mathcal{U} = \{ (v_1, v_2, v_3)  \mid v_i \in \{0, 1\}\}$, respectively. We use the following setting to create the ground-truth $\eta = 30.0$, $r = 0.08$.

The $\mathbf{q}_l(\eta \cdot \mathbf{x})$ is computed by flooring the point coordinates, which eliminates the gradient with respect to $\eta$ entirely. Additionally, the center is calculated by adding a random value $\text{PRNG}(t_l(\eta \cdot\mathbf{x}))$ to the grid. However, the PRNG is not differentiable, making $\hat{f}$ a non-differentiable function.

To maintain consistency between the Python implementation of the signed distance field generation, defined by the Equation~\ref{sphere_microstructure},  and the rendering of the microstructure, given by the same equation and processed within a CUDA shader, the CUDA PRNG is called directly from Python. This ensures both environments use the same random number generation process, avoiding potential differences in results.
 
\subsubsection{Fibers Microstructure}

Next, we describe a procedure to obtain a union of two rotated sets of fibers parallel to the y-axis arranged in layers. To achieve this, we need a rotation transformation matrix $T_\varphi$ along the axis y by an angle $\varphi \in [0, 2\pi]$, which is a parameter. Thanks to the symmetry of the problem, we can constrain the $\varphi \in [0, \pi]$.  We also need a function $u : \mathbb{R}^3 \rightarrow \mathbb{R}$ defining the fibers. The last parameter is the $\eta \in \mathbb{R}$ controlling the density of the fibers. The microstructure Fibers [2D] is defined by
\begin{gather}
\label{fibers_sdf}
 \hat{f}(\mathbf{x} | \eta, \phi ) =  h(\mathbf{x} \cdot \eta | \phi) / \eta, \\
 \notag\\
 h(\mathbf{x} | \phi ) = \min \Bigl( u(\mathbf{x}) + u((x_1, x_1, x_1)) , u(T_\phi(\mathbf{x})) + u(T_\phi(x_1, x_1, x_1)) \Bigr), \\
  \notag\\
u(x) = (\sin(x_0)\cos(x_1)+\sin(x_1)\cos(x_2)+\sin(x_2)\cos(x_0))^2 .   
\end{gather}

Note that all functions used are periodic functions and linear transformations, all of which are differentiable, making the $\hat{f}$ also differentiable. To create the ground truth, we set $\eta = 100, \varphi = \pi / 4$.

\subsubsection{Porous Microstructure}
\label{porous_material}

Lastly, we present the function defining a porous microstructure named Porous [28D]. The first parameter, $\eta \in \mathbb{R}$, controls the density of the material. The SDF is composed of a summation of various sine and cosine waves with different amplitudes and frequencies, resulting in a complex material structure. The frequencies are determined by three $3\times3$ parameter transformation matrices, simplified as $\mathbf{T} \in \mathbb{R}^{3\times 3 \times 3}$. The final distance is then given by
\begin{gather}
\label{porous_material_sdf}
\hat{f}(\mathbf{x} | \eta, \mathbf{T}) = \min \Bigl(\epsilon, v(\eta\mathbf{T}_1 \mathbf{x}) + w(\eta\mathbf{T}_2 \mathbf{x}) \cdot v(\eta\mathbf{T}_3 \mathbf{x}) \Bigr) / \eta, \\
v(\mathbf{x}) = \Bigl( \sin(x_0) \cos(x_1) + \sin(x_1) \cos(x_2) + \sin(x_2) \cos(x_0) \Bigr)^2, \\
w(\mathbf{x}) = \Bigl( \sin(x_0) \sin(x_1) + \sin(x_1) \sin(x_2) + \sin(x_2) \sin(x_0) \Big)^2,
\end{gather}

where $\epsilon$ is a small positive number. The $\hat{f}$ is again differentiable thanks to a differentiability of composition of linear transformations and periodic functions. The ground truth is created with a fixed $\eta = 30$ and the $\mathbf{T} \in [-4, 7]^{3\times 3 \times 3}$.

\subsection{Results}

\definecolor{darkorange}{RGB}{255,140,0}  

\begin{table*}[ht]
\begin{adjustbox}{max width=\textwidth, max height=0.2\textheight, center}
\begin{tabular}{l r | >{\centering\arraybackslash}m{2.5cm} >{\centering\arraybackslash}m{2.5cm} >{\centering\arraybackslash}m{2.5cm} >{\centering\arraybackslash}m{2.5cm} >{\centering\arraybackslash}m{2.5cm}} 
Microstructure & & SHGO-SDF & BH-SDF & CMA-ES-SDF & CMA-ES-S & Powell-S\\
\toprule
    \multirow{5}{*}{Gyroid [1D]} & Val. Error$\downarrow$ & \cellcolor{orange!25}$\mathbf{0.00}$ & \cellcolor{orange!25}$\mathbf{0.00}$ & \cellcolor{orange!25}$\mathbf{0.00}$ & $10^{-4}$ & \cellcolor{yellow!25}$10^{-6}$\\ 
& LPIPS$\downarrow$ & \cellcolor{orange!25}\textbf{0.00} & \cellcolor{orange!25}\textbf{0.00}& \cellcolor{orange!25}\textbf{0.00} & \cellcolor{yellow!25}0.02 & \cellcolor{orange!25}\textbf{0.00}\\ 
& \FLIP$\downarrow$ &  \cellcolor{yellow!25}$3\times10^{-7}$  & \cellcolor{orange!25}$\mathbf{1\times10^{-7}}$ &  $1\times10^{-3}$  &  $2\times10^{-3}$ &  $2\times10^{-3}$\\ 
& Time$\downarrow$ & \cellcolor{orange!25}\textbf{2.85s}  & \cellcolor{yellow!25}16.67s & \cellcolor{orange!25}\textbf{2.23s}  & 20m & 7m\\ 
& $\Delta$-loss$\uparrow$ & 23.67 & 23.87 & 19.25  & 4.55 & 8.12 \\
\\
\multirow{5}{*}{Fibers [2D]} & Val. Error$\downarrow$ & \cellcolor{yellow!25}0.34 & \cellcolor{orange!25}$\mathbf{1\times10^{-3}}$ & 0.65 & 1.00 & 1.00\\ 
& LPIPS$\downarrow$ & \cellcolor{yellow!25}0.41 & \cellcolor{orange!25}\textbf{0.13} & 0.54 & 0.56 & 0.56   \\ 
& \FLIP$\downarrow$ &  \cellcolor{yellow!25}0.3  & \cellcolor{orange!25}\textbf{0.14} & 0.42  &  0.40 &  0.43 \\ 
& Time$\downarrow$ & 13m  & 20m & \cellcolor{yellow!25}6m  & 20m & \cellcolor{orange!25}\textbf{5m}\\ 
& $\Delta$-loss$\uparrow$   & 2.17 & 7.58 & 1.46  & 0.08 & 0.05 \\ 
\\
\multirow{5}{*}{Gyroid [3D]} & Val. Error$\downarrow$ & \cellcolor{orange!25}$\mathbf{1\times10^{-3}}$ & $6\times10^{-3}$ & 1.02& 0.15 & \cellcolor{yellow!25}$2\times10^{-3}$\\ 
& LPIPS$\downarrow$ & \cellcolor{orange!25}\textbf{0.01} & 0.04 & 0.65 & 0.35 & \cellcolor{yellow!25}0.02\\ 
& \FLIP$\downarrow$ &  \cellcolor{orange!25}\textbf{0.01}  & 0.05 &  0.61 &  0.42 &  \cellcolor{yellow!25}0.03 \\ 
& Time$\downarrow$ & \cellcolor{orange!25}\textbf{22.7s} & 20m & \cellcolor{yellow!25}7m & 14m & 20m\\ 
& $\Delta$-loss$\uparrow$   & 18.2 & 18.88 & 0.72& 1.64 & 4.66 \\ 
\\
\multirow{5}{*}{Spheres [2D]} & Val. Error$\downarrow$ & \sout{N/A} & \sout{N/A} & \cellcolor{orange!25}$\mathbf{1\times10^{-5}}$ & 1.01 & 1.01\\ 
& LPIPS$\downarrow$ & \sout{N/A} & \sout{N/A} & \cellcolor{orange!25}\textbf{0.00} & 0.53 & 0.53\\ 
& \FLIP$\downarrow$ &  \sout{N/A}& \sout{N/A} & \cellcolor{orange!25}$\mathbf{3\times10^{-7}}$  &  0.55 & 0.55 \\ 
& time$\downarrow$ &\sout{N/A} & \sout{N/A} & \cellcolor{orange!25}\textbf{4m}  & 20m & 5m\\ 
& $\epsilon$-loss$\uparrow$  & \sout{N/A} &\sout{N/A} & 11.34  & 0.63 & 0.61 \\ 
\\
\multirow{5}{*}{Gyroid [5D]} & Val. Error$\downarrow$ & 0.66 & 0.65 & \cellcolor{orange!25}\textbf{$\mathbf{3\times10^{-4}}$} & \cellcolor{yellow!25}0.03 & 0.44 \\ 
& LPIPS$\downarrow$ & 0.42 & 0.57 & \cellcolor{orange!25}\textbf{0.00} & \cellcolor{yellow!25}0.21 & 0.54 \\ 
& \FLIP$\downarrow$ &  0.42  & 0.62 &  \cellcolor{orange!25}\textbf{$\mathbf{6\times10^{-3}}$}  &  \cellcolor{yellow!25}0.28 &  0.60\\ 
& Time$\downarrow$ & \cellcolor{orange!25}\textbf{2m} & 15m & \cellcolor{yellow!25}11m  & 20m & 15m \\ 
& $\Delta$-loss$\uparrow$  & 1.92 & 1.75 & 20.34  & 2.14 & 1.60\\ 
\\
\multirow{5}{*}{Porous [28D]} & Val. Error$\downarrow$ & \sout{N/A} & \cellcolor{orange!25}\textbf{1.00} & \cellcolor{orange!25}\textbf{1.00} & \cellcolor{orange!25}\textbf{1.00} & \cellcolor{orange!25}\textbf{1.00}\\ 
& LPIPS$\downarrow$ & \sout{N/A} & \cellcolor{yellow!25}0.64 & \cellcolor{yellow!25}0.64 & \cellcolor{yellow!25}0.64 & \cellcolor{orange!25}\textbf{0.61}\\ 
& \FLIP$\downarrow$ &  \sout{N/A} & \cellcolor{orange!25}\textbf{0.51} & \cellcolor{yellow!25}0.52 & \cellcolor{yellow!25}0.52 & 0.59\\ 
& Time$\downarrow$ &  \sout{N/A} & \cellcolor{orange!25}\textbf{4m} & \cellcolor{yellow!25}8m & 20m & 20m\\ 
& $\Delta$-loss$\uparrow$  & \sout{N/A} & 0.88 & 0.01 & 1.6 & 0.23\\ 
\end{tabular}
\end{adjustbox}
\caption{Fitted Functions. Table of comparison of the global optimization algorithms used on our synthetic dataset~\ref{synthetic_dataset}. We show for each problem the following: \textbf{a)} Validation Error, \textbf{b)} LPIPS Perceptual loss, \textbf{c)} \FLIP error, \textbf{d)} Time of the optimization, \textbf{e)} The difference between the loss at the beginning and at the end of the optimization process is denoted as $\Delta$-loss. The time limit for all examples was set to 20 minutes. We consider a numerical~0 to be a number below $1\times10^{-7}$. The \sout{N/A} implies that the method could not be used for the problem. }
\label{optimization_table_results-add}
\end{table*}

\definecolor{darkorange}{RGB}{0,0,0}  
\begin{table*}[ht]
\begin{adjustbox}{max width=\textwidth, max height=0.4\textheight}
\begin{tabular}{ |c c c||c|c|c|c|c| } 
\hline
Problem & Ground Truth & Init. & SHGO-A  & BH-A & CMA-ES-A & CMA-ES-S & Powell-S\\
\hline
\hline
\multirow{1}{*}{Gyroid [1D]} & $\eta$: 100.0 & 300.0 & \cellcolor{orange!25}\textbf{100.000} & \cellcolor{orange!25}\textbf{100.000} & \cellcolor{orange!25}\textbf{100.000} & 100.001 & \cellcolor{orange!25}\textbf{100.000}\\ 
\hline
\multirow{2}{*}{Fibers [2D]} & $\eta$: 100.0 & 200.0 & \cellcolor{orange!25}\textbf{100.000} & \cellcolor{orange!25}\textbf{100.000} & 100.100 & 175.159 &  133.202\\ 
& $\theta: \pi / 4$  &  $\pi$ & 0.142 &  \cellcolor{orange!25}\textbf{0.785} &  2.825 &  3.126 &  0.831\\
\hline
\multirow{3}{*}{Gyroid [3D]} & 
$\eta$: 100.0 & 300.0 &  \cellcolor{orange!25}\textbf{116.633} &  213.469 & 32.000 & 186.593 & 86.780 \\ 
& $a: 7.0$  & 6.0 &  8.164 &  14.942 &  11.010 &  12.958 &  \cellcolor{orange!25}\textbf{6.075}\\
& $t: 1.2$  & 0.0 &  \cellcolor{orange!25}\textbf{1.200} &  \cellcolor{orange!25}\textbf{1.200}  &  1.196 &  1.173 & 1.198 \\
\hline
\multirow{2}{*}{Spheres [2D]} & 
$\eta$: 30.0 & 150.0 & \sout{N/A} & \sout{N/A} & \cellcolor{orange!25}\textbf{29.999} & 291.739 & 300.964  \\ 
& $r: 0.08$  & 0.2 &  \sout{N/A} &  \sout{N/A} &  \cellcolor{orange!25}\textbf{0.079} &  0.093 &  0.086\\
\hline
\multirow{5}{*}{Gyroid [5D]} & 
$\eta$: 100.0 & 200.0 &  114.983 & 142.570 & \cellcolor{orange!25}\textbf{97.668} & 74.162 & 89.104 \\ 
& $a: 7.0$  & 6.0 &  8.051 &  9.979 &  \cellcolor{orange!25}\textbf{6.836} &  5.180 &  6.319\\
& $t: 10.0$  & 6.0 &  11.534 &  14.162  &  \cellcolor{orange!25}\textbf{9.766} &  7.376 & 9.228 \\
& $t: 15.0$  & 6.0 &  \cellcolor{orange!25}\textbf{14.961} &  14.778  &  14.650 &  11.062 & 13.125 \\
& $t: 1.2$  & 6.0 &  1.180 &  1.199  &  \cellcolor{orange!25}\textbf{1.200} &  1.165 & 1.151 \\
\hline
\multirow{1}{*}{Porous [28D]} & 
$\eta$: 30.0 & 50.0 &  \sout{N/A} & 38.586 & 42.796 & \cellcolor{orange!25}\textbf{28.268}  & 43.272 \\ 
\hline
\end{tabular}
\end{adjustbox}
    \caption{Optimal function parameters $\boldsymbol{\phi}$ Table. Optimal parameters found by the specified global optimizers. These results refer to the Table~\ref{optimization_table_results}. The parameters were rounded to 3 decimal points. We also specify the ground truth and the initialization of the variables.}
\label{optimization_table_parameters}
\end{table*}

In Table \ref{optimization_table_results}, we report the performance of different algorithms for different microstructures. In Fig. \ref{fig:reconstruction_renders}, we show the corresponding renders. We set a maximum time limit for the optimization of approximately 20 minutes, as results typically do not improve significantly beyond this point. The results show that no single optimization algorithm consistently outperforms the others across all microstructures. SHGO-SDF and BH-SDF perform well for small-variable problems but struggle with higher-dimensional cases (e.g., Gyroid [5D], Porous [28D]) and cannot be applied to certain discrete problems (e.g., Spheres [2D]). CMA-ES-SDF demonstrates solid performance across different microstructures, including discrete ones like Spheres [2D]. As expected, parameter fitting methods outperform analysis-by-synthesis in both accuracy and efficiency, given the larger complexity of the latter. Interestingly, CMA-ES-S resemble more Porous [28D] more than the other methods, highlighting its potential for high-dimensional reconstructions.\\

In Table \ref{optimization_table_results}, we report the performance of different algorithms for different microstructures. In Fig. \ref{fig:reconstruction_renders}, we show the corresponding renders. We set a maximum time limit for the optimization of approximately 20 minutes, as results typically do not improve significantly beyond this point. The results show that no single optimization algorithm consistently outperforms the others across all microstructures. SHGO-SDF and BH-SDF perform well for small-variable problems but struggle with higher-dimensional cases (e.g., Gyroid [5D], Porous [28D]) and cannot be applied to certain discrete problems (e.g., Spheres [2D]). CMA-ES-SDF demonstrates solid performance across different microstructures, including discrete ones like Spheres [2D]. As expected, parameter fitting methods outperform analysis-by-synthesis in both accuracy and efficiency, given the larger complexity of the latter. Interestingly, CMA-ES-S resamble more Porous [28D] more than the other methods, highlighting its potential for high-dimensional reconstructions.\\

\definecolor{darkorange}{RGB}{255,140,0}  
\begin{table}[ht]
\begin{adjustbox}{max width=\textwidth, max height=0.17\textheight, center}
\begin{tabular}{l r | c c c c c} 
Microstructure & Metrics & SHGO-SDF & BH-SDF & CMA-ES-SDF & CMA-ES-S & Powell-S\\
\toprule
    \multirow{4}{*}{Gyroid [1D]} & Val. Error$\downarrow$ & \cellcolor{orange!25}$\mathbf{0.00}$ & \cellcolor{orange!25}$\mathbf{0.00}$ & \cellcolor{orange!25}$\mathbf{0.00}$ & $10^{-4}$ & \cellcolor{yellow!25}$10^{-6}$\\ 
& LPIPS$\downarrow$ & \cellcolor{orange!25}\textbf{0.00} & \cellcolor{orange!25}\textbf{0.00}& \cellcolor{orange!25}\textbf{0.00} & \cellcolor{yellow!25}0.02 & \cellcolor{orange!25}\textbf{0.00}\\ 
& \FLIP$\downarrow$ &  \cellcolor{yellow!25}$3\times10^{-7}$  & \cellcolor{orange!25}$\mathbf{1\times10^{-7}}$ &  $1\times10^{-3}$  &  $2\times10^{-3}$ &  $2\times10^{-3}$\\ 
& Time$\downarrow$ & \cellcolor{orange!25}\textbf{2.85s}  & \cellcolor{yellow!25}16.67s & \cellcolor{orange!25}\textbf{2.23s}  & 20m & 7m\\ 
\\
\multirow{4}{*}{Fibers [2D]} & Val. Error$\downarrow$ & \cellcolor{yellow!25}0.34 & \cellcolor{orange!25}$\mathbf{1\times10^{-3}}$ & 0.65 & 1.00 & 1.00\\ 
& LPIPS$\downarrow$ & \cellcolor{yellow!25}0.41 & \cellcolor{orange!25}\textbf{0.13} & 0.54 & 0.56 & 0.56   \\ 
& \FLIP$\downarrow$ &  \cellcolor{yellow!25}0.3  & \cellcolor{orange!25}\textbf{0.14} & 0.42  &  0.40 &  0.43 \\ 
& Time$\downarrow$ & 13m  & 20m & \cellcolor{yellow!25}6m  & 20m & \cellcolor{orange!25}\textbf{5m}\\ 
\\
\multirow{4}{*}{Gyroid [3D]} & Val. Error$\downarrow$ & \cellcolor{orange!25}$\mathbf{1\times10^{-3}}$ & $6\times10^{-3}$ & 1.02& 0.15 & \cellcolor{yellow!25}$2\times10^{-3}$\\ 
& LPIPS$\downarrow$ & \cellcolor{orange!25}\textbf{0.01} & 0.04 & 0.65 & 0.35 & \cellcolor{yellow!25}0.02\\ 
& \FLIP$\downarrow$ &  \cellcolor{orange!25}\textbf{0.01}  & 0.05 &  0.61 &  0.42 &  \cellcolor{yellow!25}0.03 \\ 
& Time$\downarrow$ & \cellcolor{orange!25}\textbf{22.7s} & 20m & \cellcolor{yellow!25}7m & 14m & 20m\\ 
\\
\multirow{4}{*}{Spheres [2D]} & Val. Error$\downarrow$ & \sout{N/A} & \sout{N/A} & \cellcolor{orange!25}$\mathbf{1\times10^{-5}}$ & 1.01 & 1.01\\ 
& LPIPS$\downarrow$ & \sout{N/A} & \sout{N/A} & \cellcolor{orange!25}\textbf{0.00} & 0.53 & 0.53\\ 
& \FLIP$\downarrow$ &  \sout{N/A}& \sout{N/A} & \cellcolor{orange!25}$\mathbf{3\times10^{-7}}$  &  0.55 & 0.55 \\ 
& time$\downarrow$ &\sout{N/A} & \sout{N/A} & \cellcolor{orange!25}\textbf{4m}  & 20m & 5m\\ 
\\
\multirow{4}{*}{Gyroid [5D]} & Val. Error$\downarrow$ & 0.66 & 0.65 & \cellcolor{orange!25}\textbf{$\mathbf{3\times10^{-4}}$} & \cellcolor{yellow!25}0.03 & 0.44 \\ 
& LPIPS$\downarrow$ & 0.42 & 0.57 & \cellcolor{orange!25}\textbf{0.00} & \cellcolor{yellow!25}0.21 & 0.54 \\ 
& \FLIP$\downarrow$ &  0.42  & 0.62 &  \cellcolor{orange!25}\textbf{$\mathbf{6\times10^{-3}}$}  &  \cellcolor{yellow!25}0.28 &  0.60\\ 
& Time$\downarrow$ & \cellcolor{orange!25}\textbf{2m} & 15m & \cellcolor{yellow!25}11m  & 20m & 15m \\ 
\\
\multirow{4}{*}{Porous [28D]} & Val. Error$\downarrow$ & \sout{N/A} & \cellcolor{orange!25}\textbf{1.00} & \cellcolor{orange!25}\textbf{1.00} & \cellcolor{orange!25}\textbf{1.00} & \cellcolor{orange!25}\textbf{1.00}\\ 
& LPIPS$\downarrow$ & \sout{N/A} & \cellcolor{yellow!25}0.64 & \cellcolor{yellow!25}0.64 & \cellcolor{yellow!25}0.64 & \cellcolor{orange!25}\textbf{0.61}\\ 
& \FLIP$\downarrow$ &  \sout{N/A} & \cellcolor{orange!25}\textbf{0.51} & \cellcolor{yellow!25}0.52 & \cellcolor{yellow!25}0.52 & 0.59\\ 
& Time$\downarrow$ &  \sout{N/A} & \cellcolor{orange!25}\textbf{4m} & \cellcolor{yellow!25}8m & 20m & 20m\\ 
\end{tabular}
\end{adjustbox}
\caption{Reconstruction performance for different optimization algorithms (columns) and synthetic procedural microstructures (rows). We report for each microstructure the \textbf{a)} Validaton Error, \textbf{b)} LPIPS Perceptual loss, \textbf{c)} \FLIP error, \textbf{d)} Time of the optimization. We consider a numerical~0 to be a number below $1\times10^{-7}$. The \sout{N/A} implies that the optimization method could not be applied for the corresponding microstructure. The best results are
highlighted in orange, while the second-best results are marked in yellow. }
\label{optimization_table_results-add1}
\end{table}

\begin{figure*}
    \centering
     \begin{tabular}{@{}c@{}c@{}c@{}}
    \includegraphics[width=0.33\linewidth]{figures/optimization_results/fitting_render-0_resized.jpg}
    \includegraphics[width=0.33\linewidth]{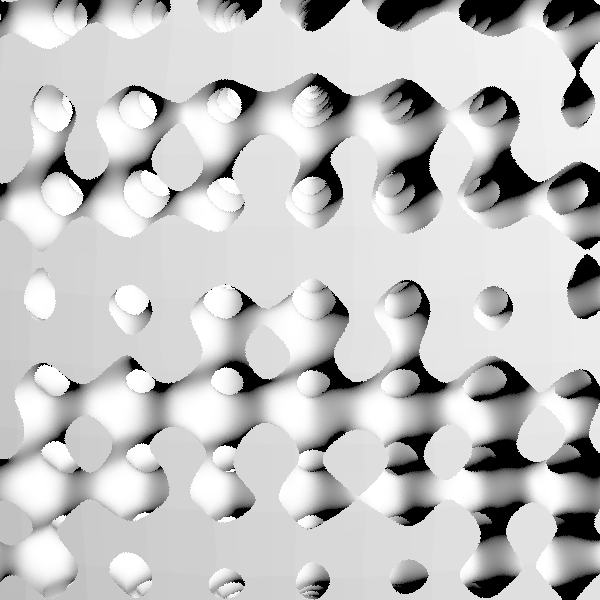}
    \includegraphics[width=0.33\linewidth]{images/gyroid1d_siren_volume_512.png}

    \end{tabular}
     Reference\hspace{2.5cm} SIREN-VOL-512 (0.01) \hspace{2.5cm}  SIREN-VOL-256 (0.02) \\[-0.5ex]
    
  \caption{Resulst for the Gyroid~[1D] with neural networks. Neural networks perform reasonably well with periodic function optimization.}
  \label{fig:gyroid1D}
\end{figure*}

\begin{figure*}
    \centering
     \begin{tabular}{@{}c@{}c@{}c@{}}
    \includegraphics[width=0.33\linewidth]{figures/optimization_results/fitting_render-1_resized.jpg}
    \includegraphics[width=0.33\linewidth]{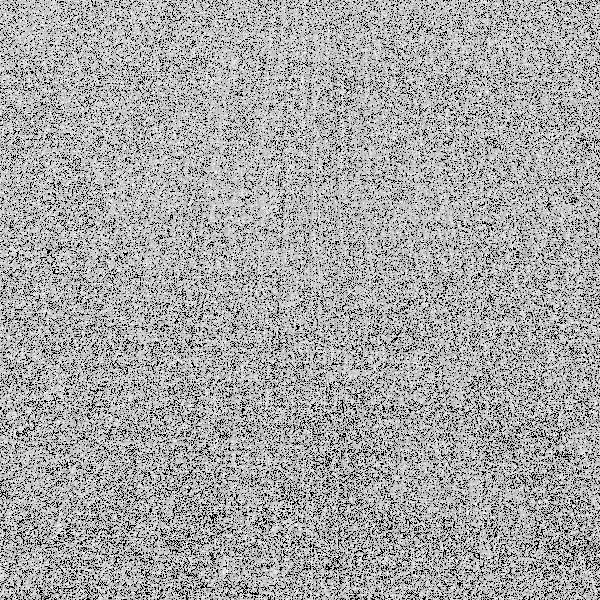}
    \includegraphics[width=0.33\linewidth]{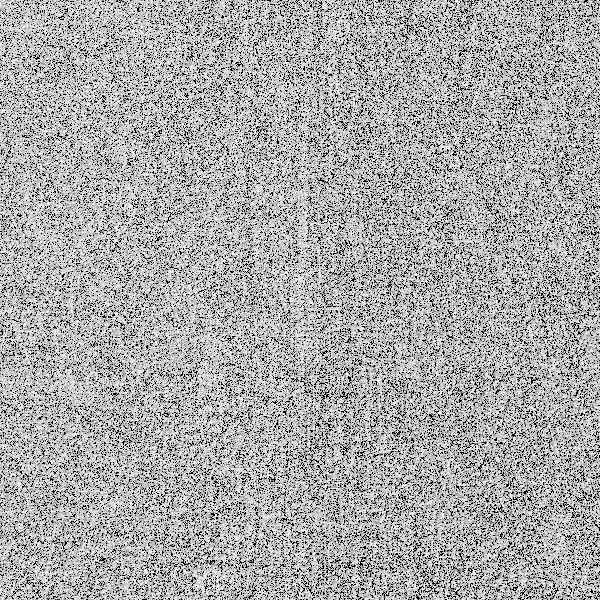}

    \end{tabular}
    Reference\hspace{2.5cm} MLP-PE-VOL-256 (0.39) \hspace{2.5cm}  MLP-PE-VOL-512 (0.43) \\[-0.5ex]
    
  \caption{Resulst for the Fibers~[2D] with neural networks. While the other two types of methods, parameter fitting and analysis by synthesis, perform well in learning space-filling primitive distributions, neural networks are struggling. Further exploration is needed in neural methods.}
  \label{fig:fibers2d}
\end{figure*}

\begin{figure*}
    \centering
     \begin{tabular}{@{}c@{}c@{}c@{}}
    \includegraphics[width=0.33\linewidth]{figures/optimization_results/fitting_render-3_resized.jpg}
    \includegraphics[width=0.33\linewidth]{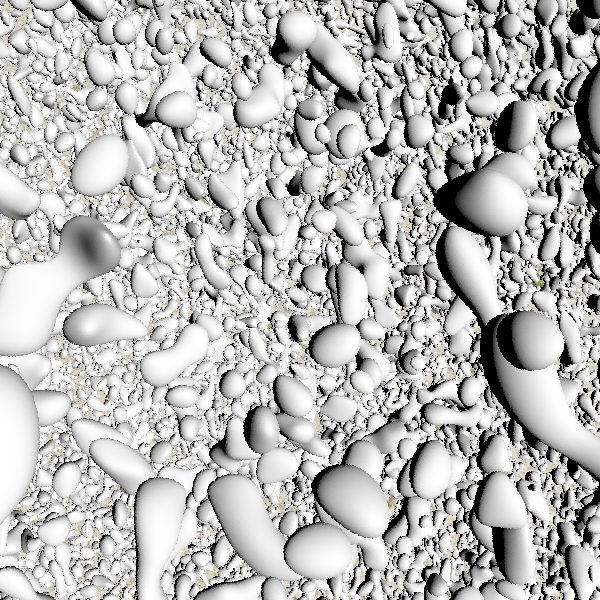}
    \includegraphics[width=0.33\linewidth]{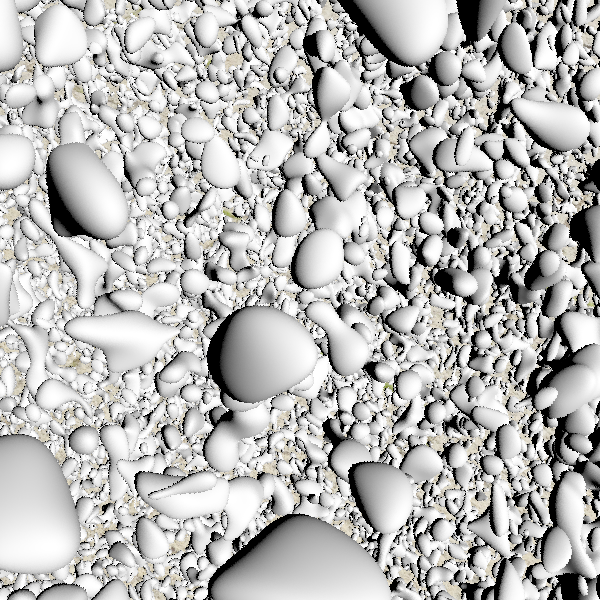}

    \end{tabular}
   Reference\hspace{2.5cm}SIREN-VOL-512 (0.36) \hspace{2.5cm}  SIREN-VOL-256 (0.46) \\[-0.5ex]
  \caption{Resulst for the Spheres~[2D] with neural networks. While the other two types of methods, parameter fitting and analysis by synthesis, perform well in learning space-filling primitive distributions, neural networks are struggling. Further exploration is needed in neural methods.}
  \label{fig:spheres2d}
\end{figure*}

\begin{figure*}
    \centering
     \begin{tabular}{@{}c@{}c@{}c@{}}
    \includegraphics[width=0.33\linewidth]{figures/optimization_results/fitting_render-4_resized.jpg}
    \includegraphics[width=0.33\linewidth]{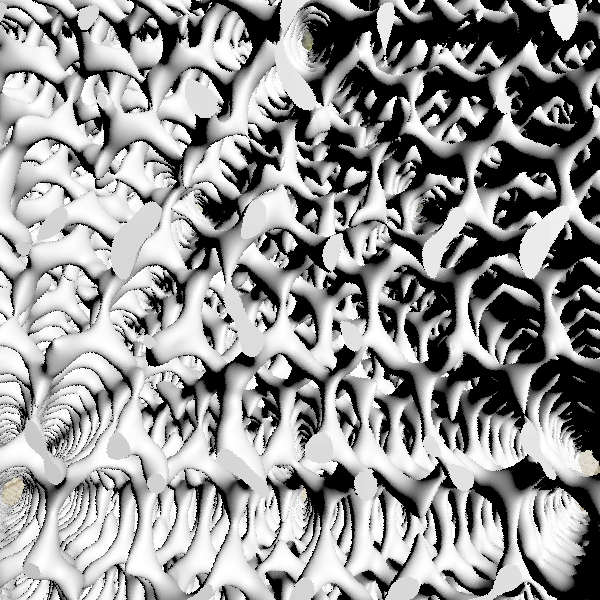}
    \includegraphics[width=0.33\linewidth]{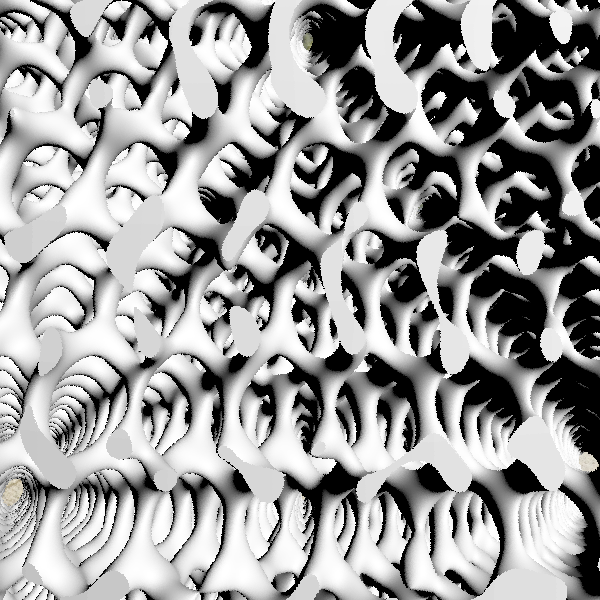}

    \end{tabular}
    Reference\hspace{2.5cm} SIREN-VOL-512 (0.04) \hspace{2.5cm} SIREN-VOL-256 (0.03) \\[-0.5ex]
   \caption{Resulst for the Gyroid~[5D] with neural networks. Neural networks perform reasonably well with periodic function optimization.}
  \label{fig:gyroid5d}
\end{figure*}

\begin{figure*}
    \centering
     \begin{tabular}{@{}c@{}c@{}c@{}}
    \includegraphics[width=0.33\linewidth]{figures/optimization_results/fitting_render-5.png}
    \includegraphics[width=0.33\linewidth]{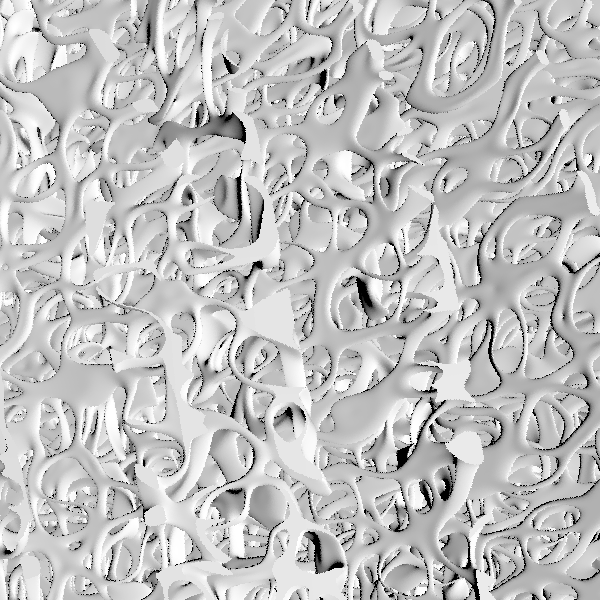}
    \includegraphics[width=0.33\linewidth]{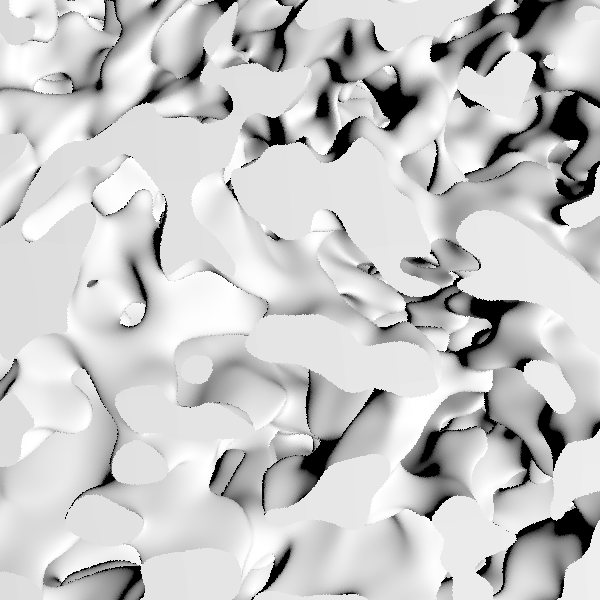}

    \end{tabular}
     Reference\hspace{2.5cm}CMA-ES-S (1.00) \hspace{3.0cm} SIREN-VOL-512 (0.7) \\[-0.5ex]
   \caption{Results for the Porous~[28D] with parameter fitting (CMA-ES-S) and neural network methods indicate that neural networks are struggling to learn the volumetric geometry from exemplar SDFS. Further exploration is needed for neural-based approaches.}
  \label{fig:porous28d}
\end{figure*}

\end{document}